\documentclass[12pt, a4paper]{article}

\pdfoutput=1

\usepackage{amsmath}
\usepackage{amsfonts}
\usepackage{amssymb}
\usepackage{bbm}
\usepackage{verbatim}
\usepackage{dsfont}
\usepackage{booktabs}
\usepackage{slashed}

\usepackage{epsfig}
\usepackage{color}
\usepackage[table]{xcolor}
\usepackage{graphicx}
\usepackage[]{caption}
\usepackage[listofformat=empty,subrefformat=empty]{subfig}
\usepackage{float}
\usepackage{overpic}

\usepackage{enumerate}
\usepackage{hhline}
\usepackage{multirow}

\usepackage{cite}
\usepackage{xspace}
\usepackage{setspace}

\usepackage[pdftitle={Chiral fermions and anomaly cancellation on orbifolds with Wilson lines and flux},
  pdfauthor={Wilfried Buchmuller, Markus Dierigl, Fabian Ruehle, Julian Schweizer},
  pdfsubject={orbifolds, Wilson lines, flux, anomalies},
  bookmarksopen, bookmarksnumbered, bookmarksopenlevel=2, colorlinks=false, linkcolor=blue, citecolor=blue, urlcolor=blue]{hyperref}

\newcommand{\comma}{\ensuremath{\enspace , \enspace}}

\renewcommand{\tfrac}{\genfrac{}{}{}1}
\newcommand{\dc}{\partial}
\newcommand{\db}{\overline{\partial}}

\begin{document}

\thispagestyle{empty}

\begin{flushright}
DESY-18-167\\
\end{flushright}
\vskip .8 cm
\begin{center}
{\Large {\bf Magnetized orbifolds and localized flux}}\\[10pt]
%{\bf{Yoshiyuki Tatsuta}\footnote{E-mail: yoshiyuki.tatsuta@desy.de}}
\bigskip
\bigskip 
{
{\bf{Wilfried Buchmuller$^{a}$}\footnote{E-mail: wilfried.buchmueller@desy.de}},
{\bf{Markus Dierigl$^{b,c}$}\footnote{E-mail: m.j.dierigl@uu.nl}},  
{\bf{Yoshiyuki Tatsuta$^{a}$}\footnote{E-mail: yoshiyuki.tatsuta@desy.de}}
\bigskip}\\[0pt]
\vspace{0.23cm}
{\it $^{a}$ Deutsches Elektronen-Synchrotron DESY, 22607 Hamburg, Germany \\ \vspace{0.2cm}
$^b$ Institute for Theoretical Physics, Utrecht University, 3584 CC Utrecht,\\ The Netherlands \\ \vspace{0.2cm}
$^c$ Institute of Physics, University of Amsterdam, 1098 XH Amsterdam,\\ The Netherlands}\\[20pt] 
\bigskip
\end{center}

\begin{abstract}
\noindent
Magnetized orbifolds play an important role in compactifications of
string theories and higher-dimensional field theories to four
dimensions. Magnetic flux leads to chiral fermions, it can be a source
of supersymmetry breaking and it is an important ingredient of moduli stabilization.
Flux quantization on orbifolds is subtle due to the orbifold
singularities. Generically, Wilson line integrals around these singularities are
non-trivial, which can be interpreted as localized flux. As a
consequence, flux densities on orbifolds can take the same values as
on tori. We determine the transition functions for the
flux vector bundle on the orbifold $T^2/\mathbb{Z}_2$ and the related
twisted boundary conditions of zero-mode wave functions. We also
construct ``untwisted'' zero-mode functions that are obtained for
singular vector fields related to the Green's function on a torus,
and we discuss the connection between zeros of the wave functions and localized flux.
Twisted and untwisted zero-mode functions are related by a singular
gauge transformation.
\end{abstract}

\newpage 
\setcounter{page}{2}
\setcounter{footnote}{0}
\tableofcontents

{\renewcommand{\baselinestretch}{1.3}

\section{Introduction}
\label{sec:Introduction} 

Orbifold compactifications play an important role in string theory
\cite{Dixon:1985jw,Dixon:1986jc}. They partially break supersymmetry and lead to chiral fermion spectra
in four dimensions. The same effects can be achieved by compactifications on magnetized tori
\cite{Witten:1984dg,Bachas:1995ik}. Particularly interesting are magnetized D-branes wrapping tori or toroidal
orbifolds, which were studied in compactifications of type-I string theory
\cite{Blumenhagen:2000wh,Angelantonj:2000hi,Aldazabal:2000dg}. The interplay of these ideas 
led to a class of four-dimensional chiral gauge theories, constructed as type-I or type-II string vacua
with D-branes and orientifolds, which capture the main features of the
Standard Model and its supersymmetric extension (for reviews see, for
example \cite{Angelantonj:2002ct,Blumenhagen:2006ci,
Ibanez:2012zz}).

In view of the complexity of string compactifications, especially 
 the need to stabilize all moduli
fields of the theory, also compactifications of higher-dimensional
field theories have been considered as an intermediate step towards a
solution of the full problem. In particular orbifold grand unified
models (GUTs) in five and six dimensions can successfully account for
the doublet-triplet splitting, the breaking of GUT gauge groups and flavour physics
\cite{Kawamura:2000ev,Hall:2001pg,
  Hebecker:2001wq,Asaka:2001eh,Kim:2002im,
deAnda:2018oik}, as well as the stabilization of moduli
\cite{Ponton:2001hq,Ghilencea:2005vm,Buchmuller:2009er}.
Orbifold GUTs are strongly inspired by
heterotic string compactifications (for reviews see, for example \cite{Nilles:2014owa,Raby:2017ucc}).

A complementary approach are
compactifications on magnetized tori, which were thoroughly studied in
\cite{Cremades:2004wa} and extended to magnetized orbifolds in \cite{Braun:2006se,Abe:2008fi,
Abe:2013bca,Abe:2014noa,Buchmuller:2015eya}. Since magnetic flux can provide
both, chiral fermions and supersymmetry breaking, one can obtain
extensions of the Standard Model where supersymmetry is broken at a
high scale
that is related to the size of the compact dimensions
\cite{Buchmuller:2015jna}. Moreover, magnetized orbifold compactifications have
interesting implications for flavour physics \cite{Abe:2015yva,Kobayashi:2016qag,
Buchmuller:2017vho,Buchmuller:2017vut,Abe:2018qbp}, and in this simple setup
the interplay
of flux and nonperturbative effects at the orbifold fixed points
allows to stabilize all moduli in Minkowski or de Sitter vacua 
\cite{Braun:2006se,Buchmuller:2016dai}. Particularly interesting is the effect of flux on quantum
corrections to scalar masses. In models of gauge-Higgs unification
magnetic flux can keep Wilson line scalars at zero mass due to symmetries of the
higher-dimensional theory \cite{Buchmuller:2016gib,Ghilencea:2017jmh,
Buchmuller:2018eog}.

Compared to compactifications on magnetized tori, flux
compactifications on orbifolds are subtle due to the
orbifold singularities. In particular, there is a puzzle concerning the allowed flux
densities. For instance, it has been argued that for a quantized flux
density $f = 2\pi M$, $M \in \mathbb{Z}$ on a torus $T^2$, the allowed
flux density on an orbifold $T^2/\mathbb{Z}_2$ is $f=4\pi M$
\cite{Bachas:1995ik,Braun:2006se,Buchmuller:2015eya}. Considering a
closed path on the orbifold and using naively Stokes' theorem, this follows immediately from the
fact that the area of the orbifold is half the area of the torus. On
the other hand, zero-mode wave functions have been constructed on
magnetized orbifolds for flux densities $f = 2\pi M$ without any signs
of inconsistency \cite{Abe:2008fi,Abe:2013bca}. One of the main goals of this paper is to
clarify this puzzle. This will be achieved by carefully discussing the
flux vector bundle on the orbifold. As we shall see, the non-trivial
transition functions on the orbifold will lead to non-trivial
Wilson line integrals around orbifold fixed points, which makes flux
densities $f = 2\pi M$ indeed consistent.

The non-trivial Wilson lines around orbifold fixed points
can be interpreted as localized magnetic flux. Localized flux has
previously been considered in connection with fixed-point anomalies
\cite{Scrucca:2004jn,vonGersdorff:2006nt} as well as localized Fayet-Illiopoulos terms \cite{Lee:2003mc}. 
In the latter case zero-mode wave functions of charged bulk fields have been
constructed by means of torus Green's functions whose singularities
are localized at orbifold fixed points. We shall extend this
construction to magnetized orbifolds and show that these ``untwisted''
zero-mode functions are closely related to the standard ``twisted''
zero-mode functions with boundary conditions of Scherk-Schwarz type
\cite{Scherk:1978ta}. As we shall see, untwisted wave functions can be
mapped to twisted wave functions by means of a singular gauge transformation.

The paper is organized as follows. In Section~\ref{sec:orbifoldGT} we
first briefly review  1-cycles on orbifolds. We then
discuss the vector bundle for magnetic flux in Landau gauge and
derive the non-trivial transition functions and the related twisted
boundary conditions. Subsequently, we consider the singular vector
fields obtained from the Green's function of the bosonic string on the torus.
Section~\ref{sec:Wavefunctions} is devoted to zero-modes
on the orbifold. We first consider the regular flux vector bundle and
briefly recall the derivation of ``twisted'' wave functions in terms of Jacobi
theta-functions. We then discuss the pattern of zeros of the wave
functions for odd and even flux densities. Finally, untwisted
zero-mode functions are constructed for singular  vector fields
and compared with the corresponding twisted wave functions. Our
results are summarized in Section~\ref{sec:Conclusion}. In the
appendices we collect some formulae for Jacobi theta-functions
and 6d gamma-matrices, which are used in the calculations presented
in the main text.

\section{Gauge theories on orbifolds}
\label{sec:orbifoldGT}

In our discussion of gauge theories on orbifolds 1-cycles around
orbifold fixed points play a crucial role. We therefore briefly recall
their relation to the standard torus one-cycles. We then discuss the
vector bundles related to magnetic flux and Wilson lines and compute
the corresponding transition functions. They are compared with
singular vector fields obtained from the torus Green's function, which
are invariant under torus translations and for which the transition
functions are trivial. 

\subsection{One-cycles on orbifolds}

Consider  a six-dimensional theory compactified
on a torus $T^2$ to four-dimensional Minkowski space.   
The torus is obtained from the covering space $\mathbb{R}^2$ by
modding out a two-dimensional lattice,
\begin{align}
y \sim t(y) = y + \lambda \,,
\end{align}
where $\lambda = n_1 \lambda_1 + n_2 \lambda_2$ is a linear
combination of two lattice vectors $\lambda_{1,2}$ with integer coefficients.
They correspond to two basic translations $t_{1,2}$ and define the fundamental domain of the
torus. The dimensionless coordinates $y$ are related to
physical coordinates by $y=(y_1,y_2)= (x^5, x^6)/L$, where $L$ denotes a fixed physical length scale.

The shape of the torus is parametrized by two real moduli $\tau_{1,2}$ in the two-dimensional metric $(g_2)_{mn}$,
\begin{align}\label{torusmetric}
(g_2)_{mn} = \frac{1}{\tau_2} \begin{pmatrix} 1 & \tau_1 \\ \tau_1 & \tau_1^2 + \tau_2^2 \end{pmatrix} \,,
\end{align}
and the physical volume of the torus is $V_{T^2} = L^2$. A basis of 
1-cycles and the fundamental domain of the torus are
depicted in Fig.~\ref{toruslattice}.
\begin{figure}
\centering
\begin{overpic}[scale = .4, tics=10]{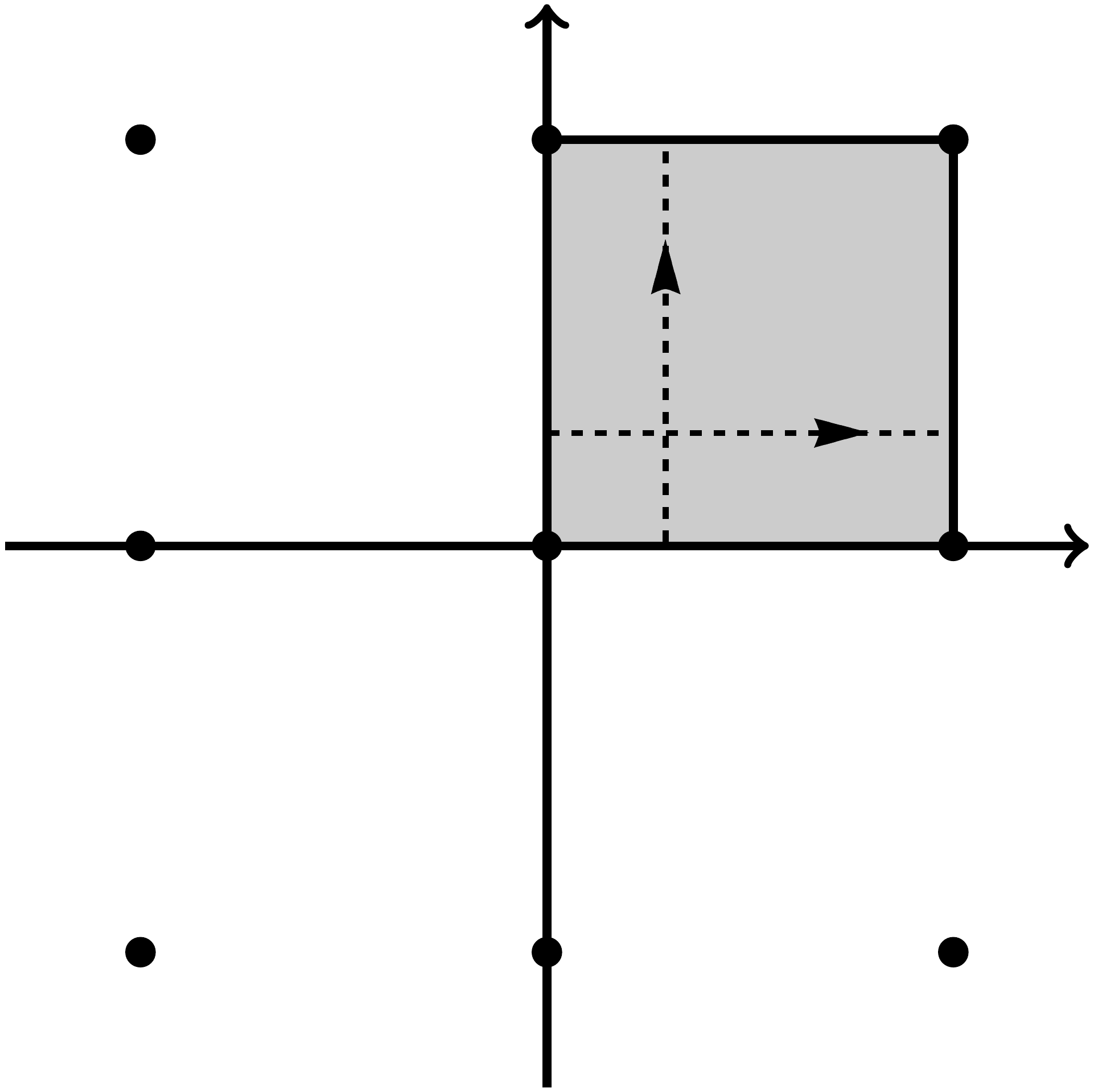}
	\put(96, 43){$y_1$}
	\put(42, 96){$y_2$}
	\put(75, 65){$\mathcal{T}_1$}
	\put(65, 75){$\mathcal{T}_2$}
	\put(7, 43){$-1$}
	\put(40, 10){$-1$}
	\put(86, 43){$1$}
	\put(44, 43){$0$}
	\put(43, 86){$1$}
\end{overpic}
\caption{A torus lattice $\mathbb{Z}^2$ and its fundamental domain (gray); the basis of 1-cycles $\mathcal{T}_{1,2}$ is depicted by dashed arrows.}
\label{toruslattice}
\end{figure}
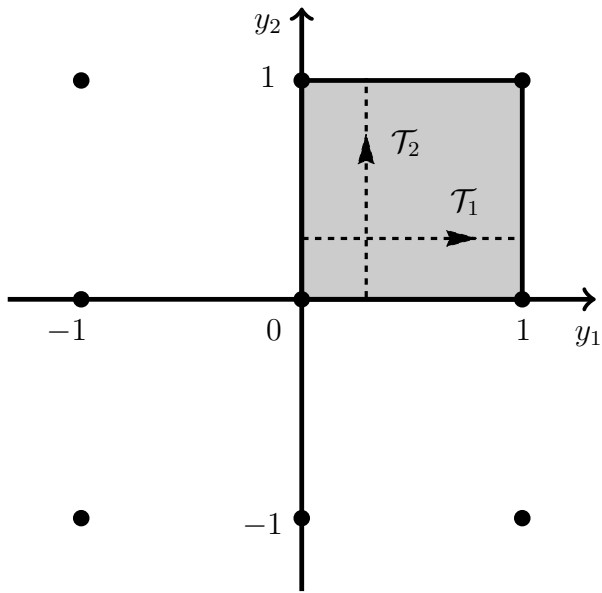
Modding out a rotational $\mathbb{Z}_2$ symmetry,
\begin{align}
 y \sim p(y) = -y \,,
\end{align}
one obtains the orbifold $T^2/\mathbb{Z}_2$. Its fundamental domain
can be chosen as $y_1 \in
[0, 1/2)$, $y_2 \in [0, 1)$, or alternatlively as $y_1 \in[0, 1)$, $y_2 \in [0, 1/2)$.
Hence, its volume is $V_{T^2/\mathbb{Z}_2} = L^2/2$. The
transformations $\{t_1, t_2, p\}$ generate the so-called space group. Modding
out its action from the covering space $\mathbb{R}^2$ yields the
orbifold $T^2/\mathbb{Z}_2$.
The space group does not act freely, but there are fixed points located at
\begin{align}
\zeta_1 = (0, 0) \comma \zeta_2 = (1/2, 0) \comma \zeta_3 = (0, 1/2) \comma \zeta_4 = (1/2, 1/2) \,.
\end{align}
The orbifold has the topology of a sphere with four points
removed. At each fixed point there is a conical singularity with
deficit angle $\pi$, corresponding to singular curvature. The bulk away from the fixed points is flat.

As shown in \cite{Buchmuller:2015eya}, it is often
convenient to decompose the bulk 1-cycles $\mathcal{T}_{1,2}$ in
terms of the ``canonical'' 1-cycles $\mathcal{C}_{i}$, $i=1,\dots, 4$, encircling the
orbifold fixed point, see~\ Fig.~\ref{fundamentaldomain}. 
The $\mathbb{Z}_2$ operator $p$ corresponds to the 1-cycle
$\mathcal{C}_1$. 
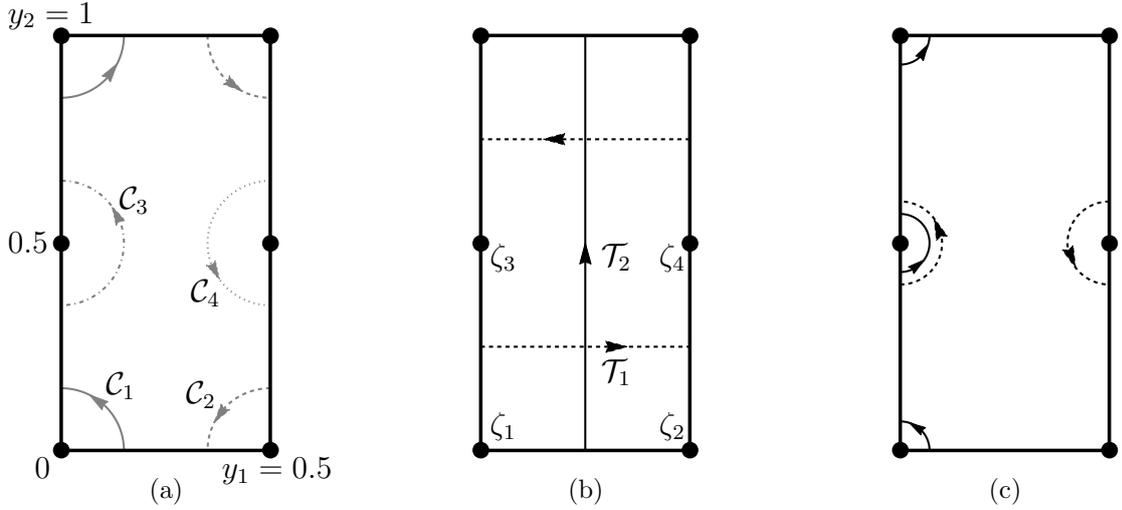
\begin{figure}
\centering
	\subfloat[\label{fundamentaldomain}]{
		\begin{overpic}[width= 0.2\textwidth]{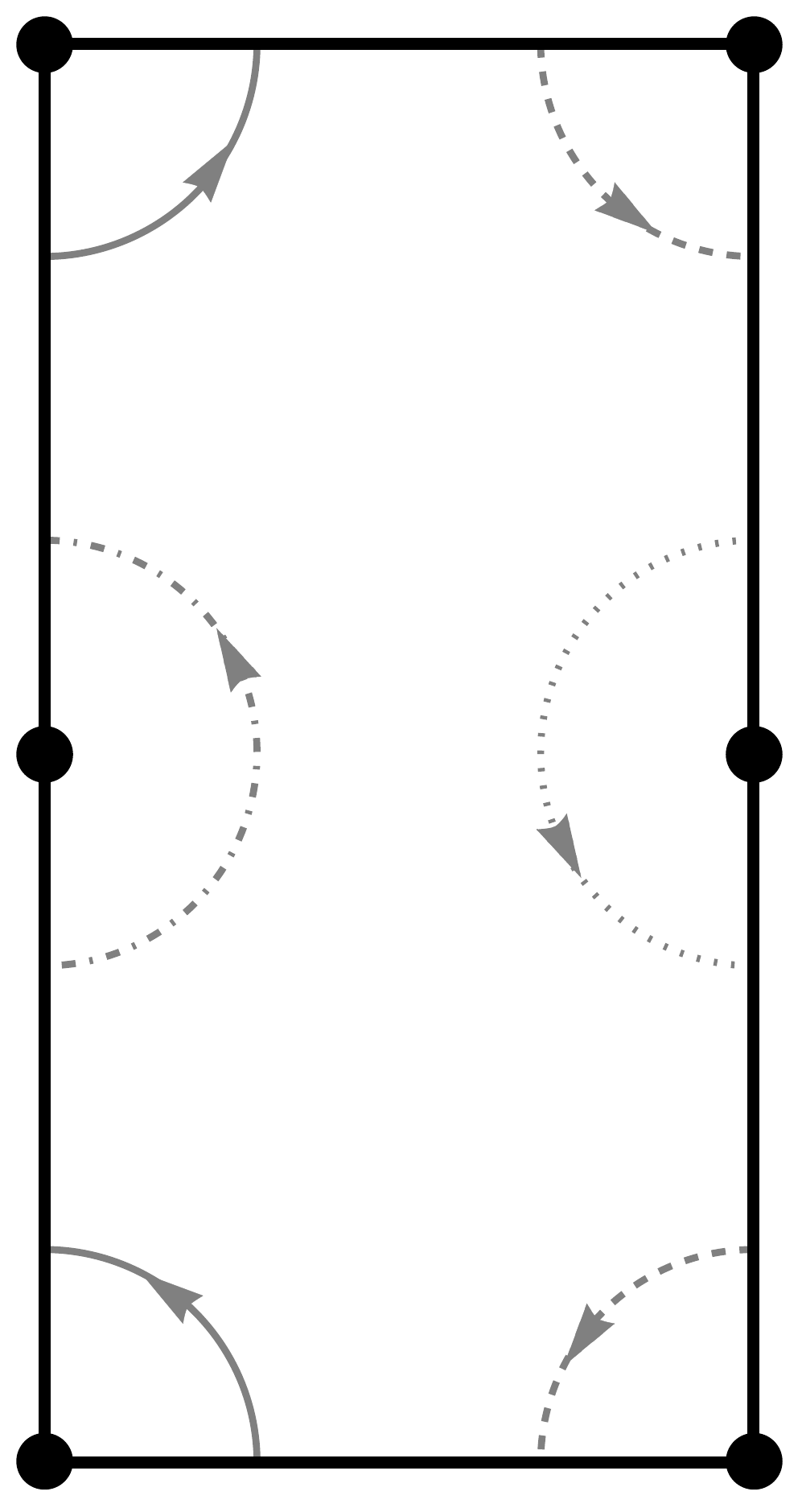}
			\put(13,16){$\mathcal{C}_1$}
			\put(31,14){$\mathcal{C}_2$}
			\put(16,58){$\mathcal{C}_3$}
			\put(32,37){$\mathcal{C}_4$}
			\put(-3,-3){$0$}
			\put(-9,48){$0.5$}
			\put(-9,100){$y_2 = 1$}
			\put(39,-3){$y_1 = 0.5$}
		\end{overpic}
	}
	\hspace{2cm}
	\subfloat[\label{projection}]{
		\begin{overpic}[width=0.2 \textwidth]{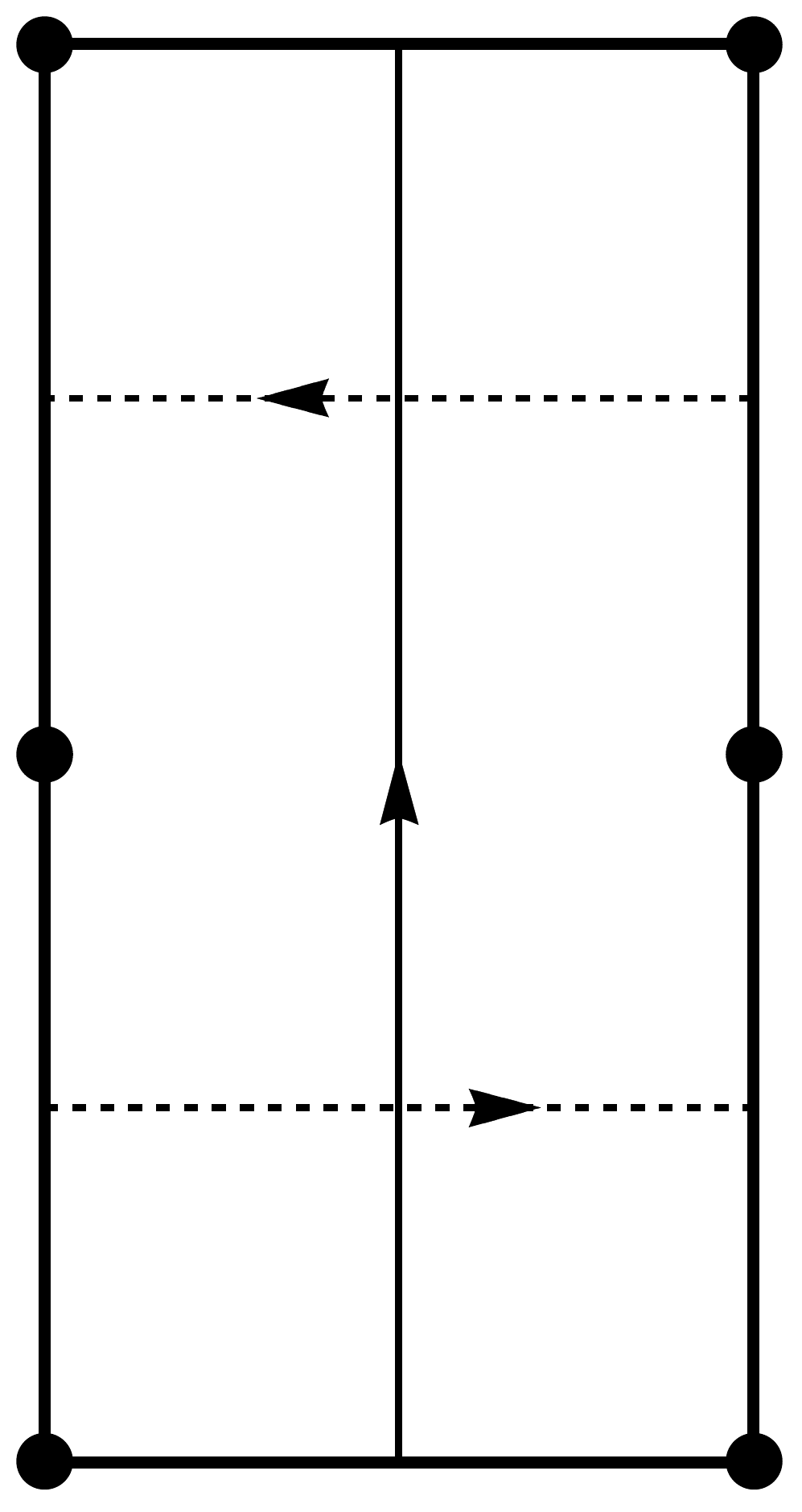}
			\put(5,7){$\zeta_1$}
			\put(43,7){$\zeta_2$}
			\put(5,45){$\zeta_3$}
			\put(43,45){$\zeta_4$}
			\put(30,45){$\mathcal{T}_2$}
			\put(30,19){$\mathcal{T}_1$}
	\end{overpic}
	}
	\hspace{2cm}
	\subfloat[\label{projectiondeformation}]{
		\begin{overpic}[width= 0.2\textwidth]{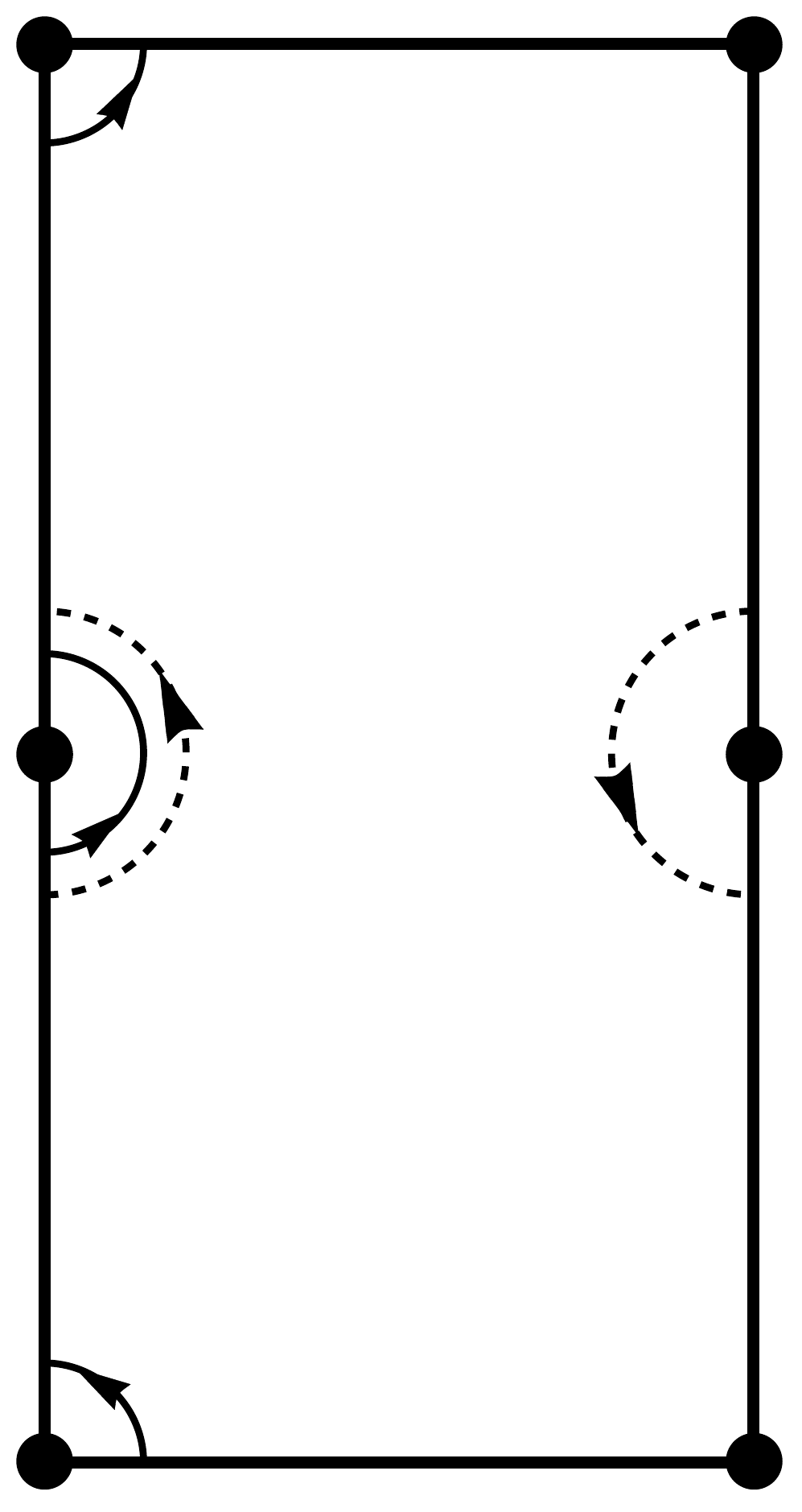}
		\end{overpic}
	}
	\caption{The fundamental domain of $T^2/\mathbb{Z}_2$; the
          black dots denote the orbifold fixed points.  The canonical
          basis of orbifold 1-cycles is presented in
          (a). (b) shows the bulk 1-cycles $\mathcal{T}_{1,2}$, and (c) illustrates the decompositions
$\mathcal{T}_1 \sim \mathcal{C}_3+ \mathcal{C}_4$ and $\mathcal{T}_2
\sim \mathcal{C}_1+ \mathcal{C}_3$. From \cite{Buchmuller:2015eya}.}
	\label{cyclerelation}
\end{figure}
The torus cycles $\mathcal{T}_{1,2}$ can be projected to the
fundamental domain of the orbifold, see Fig.~\ref{projection}, and
then deformed continuously, see Fig.~\ref{projectiondeformation}, 
yielding
$\mathcal{T}_1 \sim \mathcal{C}_3 + \mathcal{C}_4 \sim -
(\mathcal{C}_1 + \mathcal{C}_2)$ and $\mathcal{T}_2 \sim \mathcal{C}_1
+ \mathcal{C}_3$,
where a minus sign indicates a reversed orientation.
The geometry described above has important consequences for gauge
fields on the orbifold, as we shall see in the following sections.

\subsection{Gauge fields on a torus}
\label{sec:WilsonLines}

Consider now a $U(1)$ vector field $A=A_m dy_m$ and a charged complex matter field
$\phi$ on the covering space.
The field theory on the torus is obtained by modding out a
two-dimensional lattice from space-time as well as field space.
Vector and matter fields are required to be invariant under lattice
translations $t = n_1 t_1 + n_2 t_2$ up to a gauge transformation,
\begin{align}\label{periodic}
A_m(t(y)) = A_m(y) - \frac{1}{q} \partial_m \Lambda_t(y)\,, \quad
\phi(t(y)) = e^{i\Lambda_t(y)} \phi(y)\,,
\end{align}
such that the covariant derivative $D_m \phi = (\partial_m + iq A_m)
\phi$ transforms like $\phi$. The periodicity conditions for the matter field
are left invariant under the so-called large gauge transformations,
\begin{align}
\phi(t(y)) \rightarrow  e^{i\Lambda_{(k_1,k_2)}} \phi(y)\,, \quad
\Lambda_{(k_1,k_2)} = 2\pi (k_1 y_1 + k_2 y_2)\,, \quad k_{1,2}
\in \mathbb{Z}\,, 
\end{align}
under which the vector field shifts as
\begin{align}
A_m(y) \rightarrow A_m(y) - \frac{2\pi}{q} k_m \,.
\end{align}
On the covering space constant gauge fields 
\begin{align}
A = \alpha_m dy_m\,, \quad \alpha_m \in [0,2\pi)\,,
\end{align}
are unphysical, they can be removed by gauge transformations. On the
torus, however, they cannot be removed by the remaining large gauge
transformations. Hence, Wilson lines corresponding to 1-cycles 
$\mathcal{T} = n_1 \mathcal{T}_1 + n_2 \mathcal{T}_2$,
\begin{align}
W_{\mathcal{T}} = \exp{\left[- i q \int_{\mathcal{T}} A \right]} = e^{-iq(n_1 \alpha_1
  + n_2 \alpha_2)}\,,
\end{align}
do have a physical meaning and play an important role. 

\begin{figure}[t]
	\begin{center}
	\includegraphics[height = .35 \textheight]{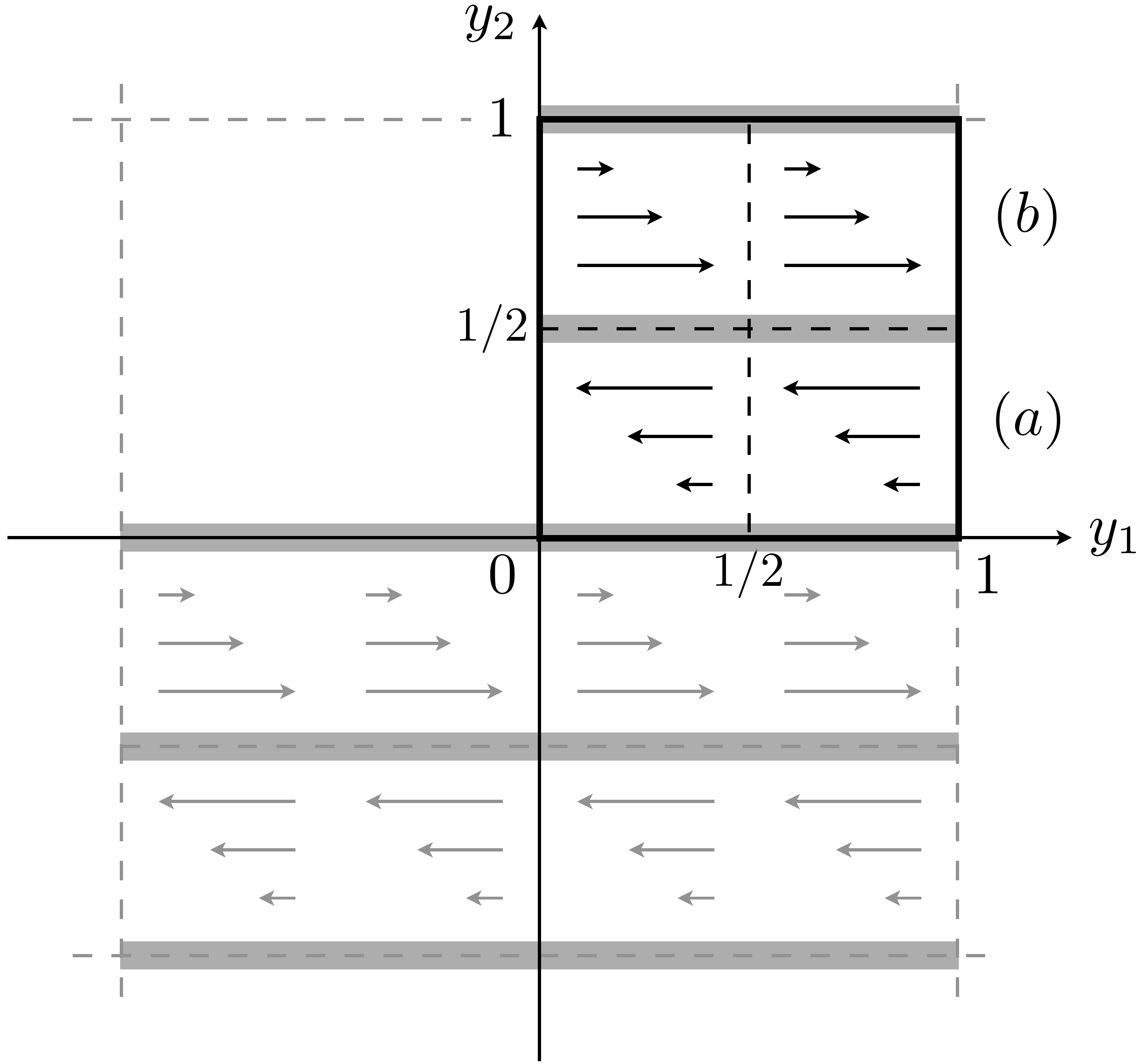} 
	\caption{Three copies of the fundamental domain of the torus
          in the $y_1$$-$$y_2$-plane with a magnetic vector field in
          Landau gauge. The overlap of the two patches of the bundle 
          contains the circles $y_2 = 0$ and $y_2 = 1/2$.}
	\label{torusMF}
	\end{center}
\end{figure}

Since the torus $T^2$ is not simply connected, vector fields and matter
fields are represented by fibre bundles (for a review see, for example
\cite{Eguchi:1980jx}). For the monopole field on a sphere this has been
thoroughy discussed in \cite{Wu:1975es}, and magnetic fields on a torus have
been considered in \cite{Abouelsaood:1986gd}. Two patches covering the
torus are shown in Fig.~\ref{torusMF}\footnote{Here we follow
  \cite{Abouelsaood:1986gd}. Strictly speaking, a patch should not
  contain a non-contractable cycle. However, in Landau gauge, covering
  the torus with more patches would only introduce additional trivial
  transition functions.}, together with a vector field in Landau gauge, 
\begin{align}\label{fluxfield}
A_1 = \left\{
\begin{array}{lll}
-f y_2\ , & 0 \leq y_2 < \tfrac{1}{2} \quad & (a)\\
-f (y_2 - 1) \ , & \tfrac{1}{2} \leq y_2 < 1 \quad & (b)
\end{array}
\right.\,, \quad A_2 = 0\,,
\end{align}
leading to the constant magnetic field
\begin{align}
F = dA = f v\,,
\end{align}
where $v = dy_1\wedge dy_2$. The overlap of the two patches contains the
circles $y_2 = 1/2$ and $y_2 = 0$. The transition function that
relates fields at the same point in the two patches is given by\footnote{We follow the notation and
  conventions of \cite{Wu:1975es}, with the replacement $e \rightarrow -q$.}
\begin{equation}
\begin{split}
\phi_b &= S_{ba} \phi_a\,, \quad S_{ba} = e^{i\Lambda_{ba}}\,, \quad
S_{ba} = S^{-1}_{ab}\,, \\
A_m^b &= A_m^a + \frac{i}{q} S^{-1}_{ba} \partial_m S_{ba}
= A_m^{a} - \frac{1}{q} \partial_m \Lambda_{ba}\,.
\end{split}
\end{equation}
At $y_2 = 1/2$, this implies
\begin{align}
A_1^{b}(y_1,\tfrac{1}{2}) - A_1^{a}(y_1,\tfrac{1}{2}) = 
-\frac{1}{q} \partial_1 \Lambda_{ba} = f\,,
\end{align}
which yields $\Lambda_{ba} = - qfy_1$ and therefore the transition function
\begin{align}\label{transfunc}
S_{ba} = e^{i\Lambda_{ba}} = e^{-iqfy_1}\,.
\end{align}
From the required single-valuedness of the transition function,
\begin{align}
S_{ba}(y_1 + 1) = S_{ba}(y_1)\,,
\end{align}
one obtains the quantization condition for the magnetic flux,
\begin{align}\label{fluxquant}
qf = 2\pi M\,, \quad M \in \mathbb{Z} \,.
\end{align}
At $y_2 = 0$, the vector field in the patches $(a)$ and $(b)$ is the same,
$A^a_1 (y_1,0) =  A^b_1 (y_1,0) = 0$. The transition function at $y_2 = 0$ is therefore
trivial, $S_{ab} = 1$.

Starting at $y_2 = 1/2$ in patch $(b)$ and going around the torus in
$y_2$-direction via patch $(b)$ and patch $(a)$ until $y_2 = 1/2$ in patch $(a)$,
the vector field changes from $A_1 = f/2$ to $A_1 = -f/2$. 
This necessitates a non-trivial transition
function $S_{ba}(y_1)$, which in the literature on magnetized tori 
is usually treated as twisted boundary condition,
\begin{align}
\phi(y_1,y_2 + 1) = S_{ba}^{-1}(y_1) \phi(y_1,y_2) = e^{iqfy_1} \phi(y_1,y_2)\,,
\end{align}
i.e. the twist factor corresponds to the transition function
$S_{ab}$ on the torus.

Constant vector fields $A_m = \alpha_m$ can be chosen to be the
same in both patches. Hence, the transition functions are trivial and
all values $\alpha_m \in [0,2\pi)$ are allowed. On the covering space
a constant vector field can be removed by a gauge transformation.
Writing $A_m(y) = \alpha_m + A'_m(y)$, one has
\begin{equation}\label{redefine}
\begin{split}
A_m(y) &\rightarrow A'_m(y)  = A_m(y) - \frac{1}{q} \partial_m \Lambda_{(\alpha)} (y) \,, \\
\phi(y)  &\rightarrow \phi'(y) = e^{i\Lambda_{(\alpha)}(y)}\phi(y) \,, \quad
%\end{equation}
%where 
%\begin{equation}
\Lambda_{(\alpha)}(y) = q(\alpha_1 y_1 + \alpha_2 y_2)\,.
\end{split}
\end{equation}
For a translation $t$ by a lattice vector $\lambda$,
the boundary condition \eqref{periodic} changes to
\begin{align}
\phi(y+\lambda) = e^{i(q(\alpha_1 y_1 + \alpha_2 y_2) + \Lambda_t)}\phi(y) \,.
\end{align}
This means that the effect of a constant background field can be represented by the
term $\exp{(iq(\alpha_1 y_1 + \alpha_2 y_2))}$ in a twisted boundary condition.

\subsection{Regular gauge fields on orbifolds}
\label{subsec:fields_regular}

For vector fields on the covering space which are odd under
reflections up to a gauge transformation,
\begin{align}\label{reflective}
A_m(p(y)) = -A_m(y) - \frac{1}{q} \partial_m \Lambda_p(y)\,, \quad
\phi(p(y)) = e^{i\Lambda_p(y)} \phi(y)\,,\quad p(y) = -y\,,
\end{align}
%with $p(y) = -y$, 
one can mod out a $\mathbb{Z}_2$ symmetry, which leads to a field theory on the orbifold $T^2/\mathbb{Z}_2$
(see, for example \cite{Hebecker:2001jb}).
The vector field \eqref{fluxfield} is odd under
reflections. Hence, the projection to the orbifold does not require an
additional gauge transformation and we have $\Lambda_p = 0$.

It is instructive to compare Wilson lines in the bulk with Wilson
lines around fixed points. For the line integral between two points we define
\begin{align}\label{WLQP}
W_{QP} = \exp{\left[-iq\int_P^Q A\right]}\,.
\end{align}
The bulk Wilson line  shown in Fig.~\ref{cWLflux}a is then given by
(see, for example \cite{Wu:1975es}),
\begin{align}
W_{AFEDCBA} = W^{(a)}_{AF} S_{ab}(F) W^{(b)}_{FE} W^{(b)}_{ED}
W^{(b)}_{DC} S_{ba}(C) W^{(a)}_{CB} W^{(a)}_{BA} \,.
\end{align}
Here the superscripts denote the relevant patches.
Using Eqs.~\eqref{transfunc} and \eqref{WLQP} one obtains
\begin{align}
W_{AFEDCBA} = e^{iqf(y_{1F} - y_{1C} + \epsilon)} e^{-iqf(y_{2E} -
  y_{2A})\epsilon} = e^{-iq\Delta F}\,,
\end{align}
where $\Delta F = f\epsilon\delta$, with $\epsilon = \overline{AB} = \overline{ED}$
and $\delta/2 = \overline{AF} = \overline{FE} = \overline{CD} =
\overline{BC}$. 
This is the expected result that the line integral is given by the
enclosed flux according to Stokes' theorem. To obtain this result it
is crucial to take the transition function $S_{ba}$ and $S_{ab}$ into account.
\begin{figure}[t]
	\begin{center}
	\includegraphics[height = .3 \textheight]{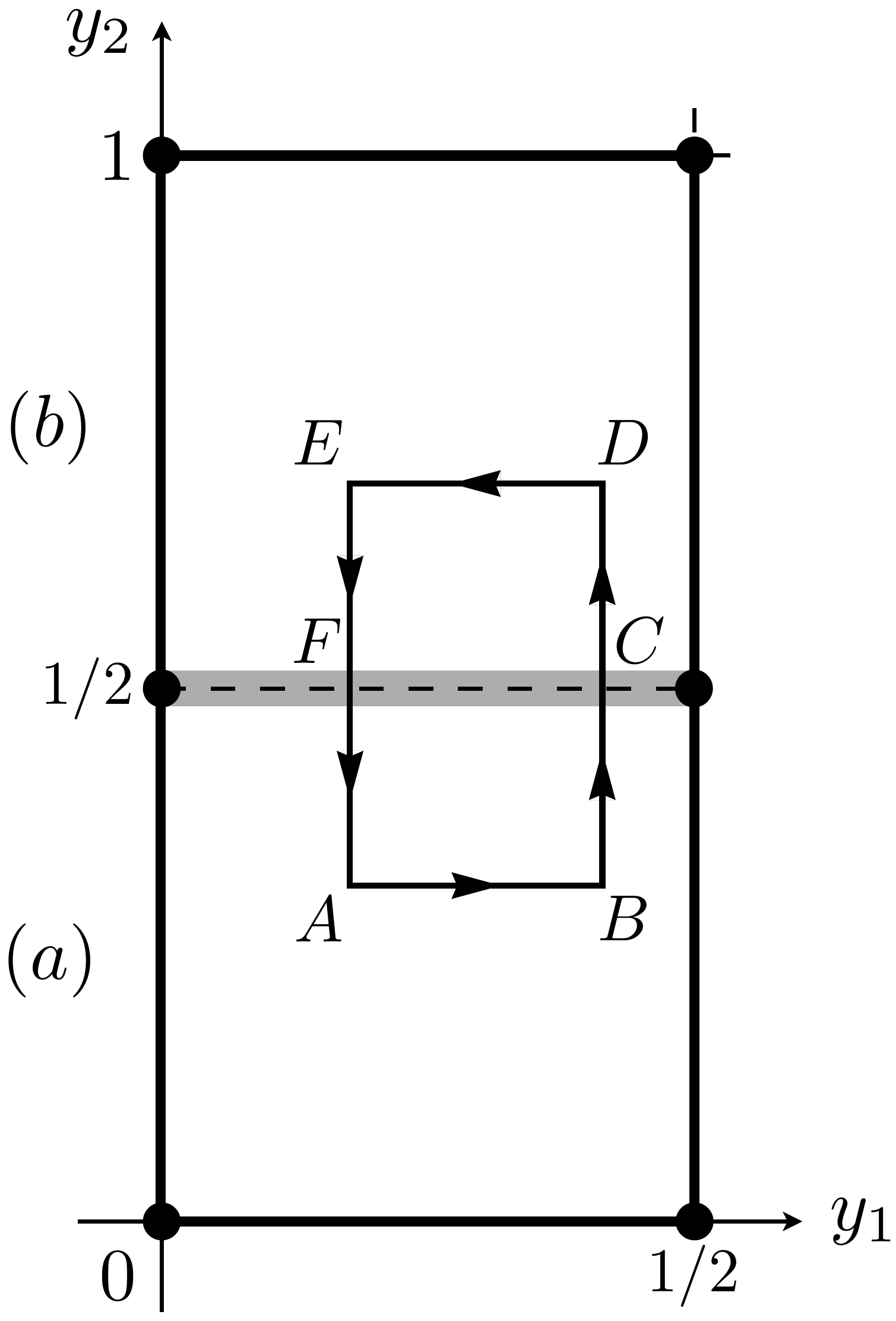} \hspace{1cm}
        \includegraphics[height = .3 \textheight]{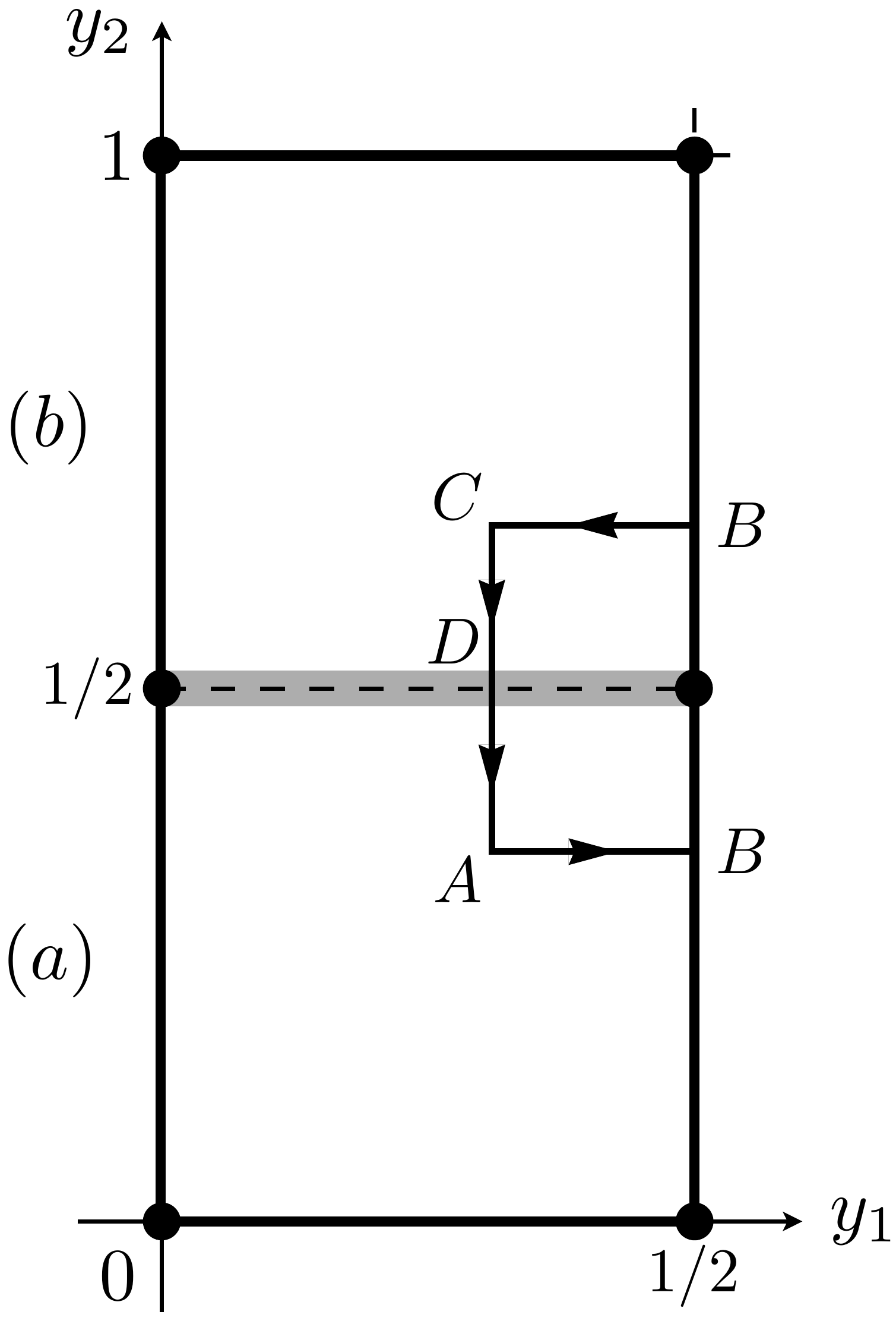} 
	\caption{a) (left) Wilson line integral in the bulk; b)
          (right) Wilson line integral around the fixed point at $\zeta_4$.} 
	\label{cWLflux}
	\end{center}
\end{figure}
The Wilson line integral around the fixed point shown in Fig.~\ref{cWLflux}b
can be calculated in the same way,
\begin{align}
W_{ADCBA} = W^{(a)}_{AD} S_{ab}(D) W^{(b)}_{DC} W^{(b)}_{CB}
S_{ba}(B) W^{(a)}_{BA} \,.
\end{align}
At the boundary $y_1 = 1/2$ between the patches $(a)$ and $(b)$ the
vector field does not change. Hence, the transition function is
trivial, $S_{ba}(B) = 1$. 
Using Eqs.~\eqref{transfunc} and \eqref{WLQP} one now finds
\begin{align}\label{cWL4}
W_{ADCBA} = e^{iqf/2} e^{-iq\Delta F}\,,
\end{align}
where again $\Delta F = f\epsilon\delta$, with $\epsilon = \overline{AB}$
and $\delta/2 = \overline{AD} = \overline{CD}$. The result differs
from the naive expectation by a factor which does not vanish in the
limit where the enclosed area goes to zero,
\begin{align}\label{W4}
W_4 = \lim_{\epsilon,\delta \rightarrow 0} W_{ADCBA} =
e^{iqf/2}\,.
\end{align}
The flux quantization \eqref{fluxquant} then implies
$W_{4} = \pm 1$. One easily verifies that the line integrals
around the other fixed points are $W_{i} = 1$, $i = 1,2,3$.
The flux quantization condition can therefore be expressed as
\begin{align}\label{trueflux}
e^{iqf/2} = W_{1} W_{2} W_{3} W_{4} \equiv W = \pm 1\,.
\end{align}
Hence, for $W = 1$,
the bulk flux $F = \int_{T^2/\mathbb{Z}_2} dA$
satisfies $qF = qf/2 = 2\pi M$, $M \in \mathbb{Z}$, whereas
%for a non-trivial product, 
for $W = -1$ one has $qF = \pi + 2\pi M$.

This is an interesting result. Originally, the magnetic flux
density on the orbifold was assumed to be twice as large as on
the torus, $qf \in 4\pi \mathbb{Z}$,
since the area of the orbifold is half the area of the torus.
The bulk flux on the orbifold is then quantized as the one on the torus, $qF = qf/2 \in 2\pi \mathbb{Z}$
\cite{Bachas:1995ik,Braun:2006se}. This immediately follows from Stokes' theorem if one assumes
that the surface integral receives no contribution from the fixed
points. On the contrary, it has been argued that a flux density $qf \in 2\pi \mathbb{Z}$
is consistent also on the orbifold since it allows normalizable wave
functions \cite{Abe:2008fi,Abe:2013bca}. Eq.~\eqref{trueflux} shows how this is indeed possible due
to the effect of non-trivial Wilson line integrals around the fixed points.
Such Wilson line integrals can be interpreted as localized
flux \cite{Buchmuller:2015eya}. In general, one has at each fixed point
\begin{align}
W_{i} = e^{-iq F_i}\,, \quad F_i = \frac{\pi}{q}(\delta_{(W_i,-1)} + 2k_i)\,,
\quad k_i \in \mathbb{Z}\,.
\end{align}
From Eq.~\eqref{trueflux} one then obtains
\begin{align}
q(F + \sum_i F_i) = \pi (\delta_{(W,-1)} + 2M + \sum_i
(\delta_{(W_i,-1)} + 2k_i))\,.
\end{align}
Since $(\delta_{(W,-1)} + \sum_i \delta_{(W_i,-1)}) \in 2\mathbb{Z}$, 
the total flux of bulk and fixed points always satisfies the torus quantization
condition
\begin{align}
q(F + \sum_i F_i) \in 2\pi \mathbb{Z}\,.
\end{align} 
The bulk flux alone, however, can be odd, $qF \in \pi \mathbb{Z}$.

For completeness, let us point out that the above discussion is
independent of the choice of the fundamental domain of the orbifold.
\begin{figure}[t]
	\begin{center}
	\includegraphics[height = .25 \textheight]{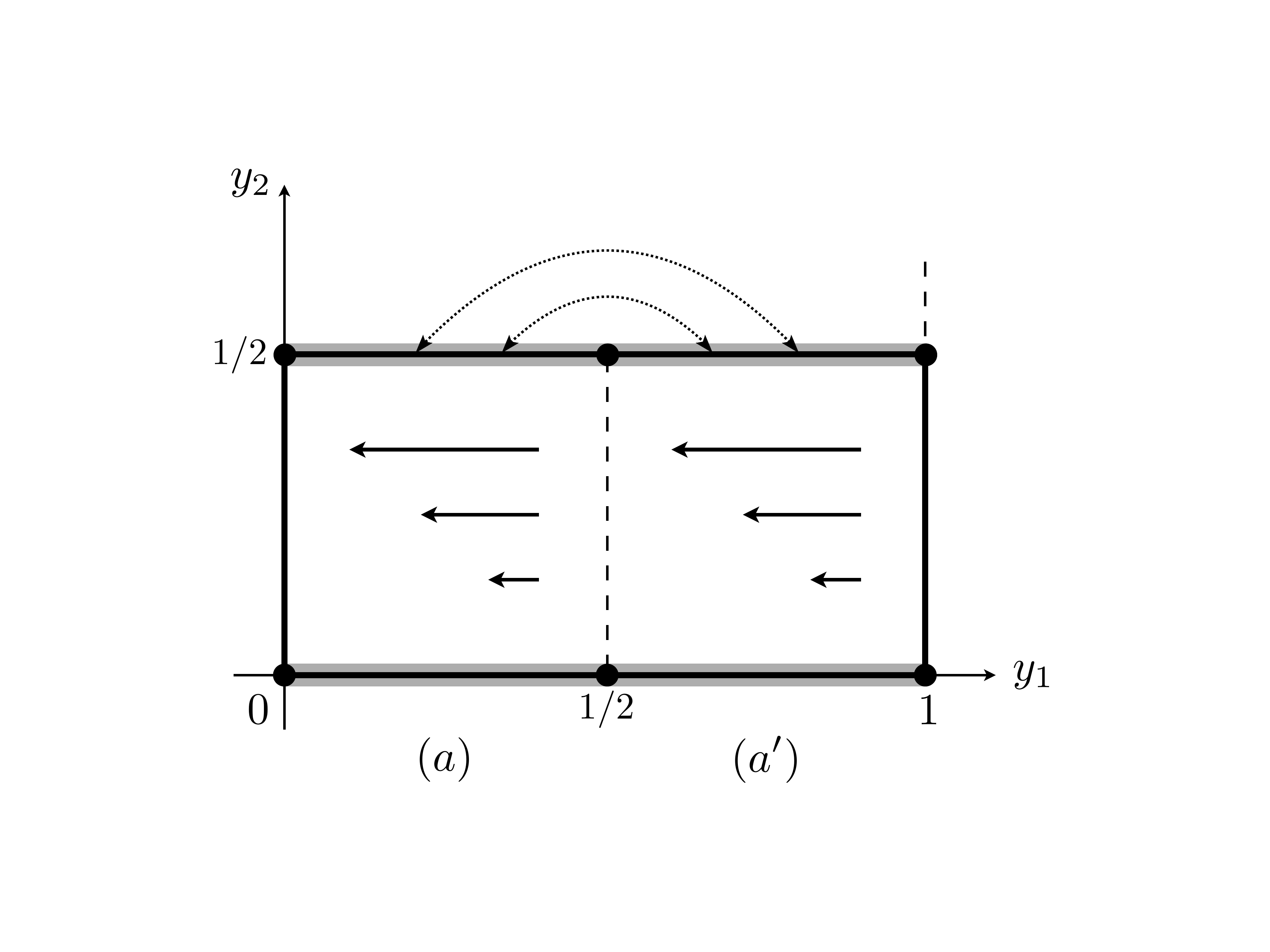} 
	\caption{Possible choice of fundamental domain on the magnetized
          orbifold $T^2/\mathbb{Z}_2$. At the boundary $y_2 = 1/2$ points connected by
          dotted lines are identified.}
	\label{fundom2}
	\end{center}
\end{figure}
Fig.~\ref{fundom2} shows the second choice for a fundamental domain of
the orbifold with a field configuration projected from the torus,
\begin{align}\label{fluxdom2}
A_1 = 
-f y_2\,, \quad  0 \leq y_1 < 1\,,\quad 0 \leq y_2 < \tfrac{1}{2}\quad
(a), (a')\,.
\end{align}
We introduce patches $(a)$ and $(a')$, separated by the
boundaries $y_1 = 0$ and $y_1 = 1/2$. They
 correspond to the patches $(a)$ and $(b)$ in Fig.~\ref{torusMF}.  
The transition functions at these boundaries are obviously
trivial. On the orbifold the $y_1$-intervals $[0,1/2]$ and $[1/2,1]$
at the boundaries are identified. At $y_2 = 0$ and $y_2 =
1/2$ one has
\begin{align}
y_1 \sim s(y_1) = 1 - y_1\,, \quad y_1 \in [0,1/2]\,.
\end{align}
Since the identification involves a reflection (see
Fig.~\ref{fundom2}), the corresponding transition functions are
obtained from
\begin{align}\label{reflective}
A^{a'}_1(s(y)) = -A^{a}_1(y) - \frac{1}{q} \partial_1 \Lambda_{a'a}(y)\,.
\end{align}
At $y_2 = 0$, the vector field vanishes and the transition function is
therefore trivial. At $y_2 = 1/2$, one has $A^{a'}_1(s(y)) +
A^{a}_1(y) = -f$, which yields $\Lambda_{a'a} = -qfy_1$, and therefore
the transition function
\begin{align}\label{Saa}
S_{a'a} = e^{-iqfy_1}\,.
\end{align}
This transition function is identical to the transition function
$S_{ba}$ given in Eq.~\eqref{transfunc}. This has to be the case, as
the patches $(a')$ and $(b)$ both describe the ``back'' of the
``orbifold pillow'', as a comparison of Figs.~\ref{torusMF} and
\ref{fundom2} shows.

Let us now consider ``Wilson lines'', i.e.~constant vector fields,
on the orbifold. Fig.~\ref{constcov} shows a field configuration on
the torus, which is odd under reflection so that it can be projected on the orbifold, 
\begin{align}\label{constfield}
A_1 = \left\{
\begin{array}{lll}
\alpha_1\ , & 0 \leq y_2 < \tfrac{1}{2} \quad & (a)\\
-\alpha_1 \ , & \tfrac{1}{2} \leq y_2 < 1 \quad & (b)
\end{array}
\right.\,, \quad A_2 = 0\,.
\end{align}
Analogously to the discussion below Eq.~\eqref{fluxfield},
\begin{figure}[t]
	\begin{center}
	\includegraphics[height = .35 \textheight]{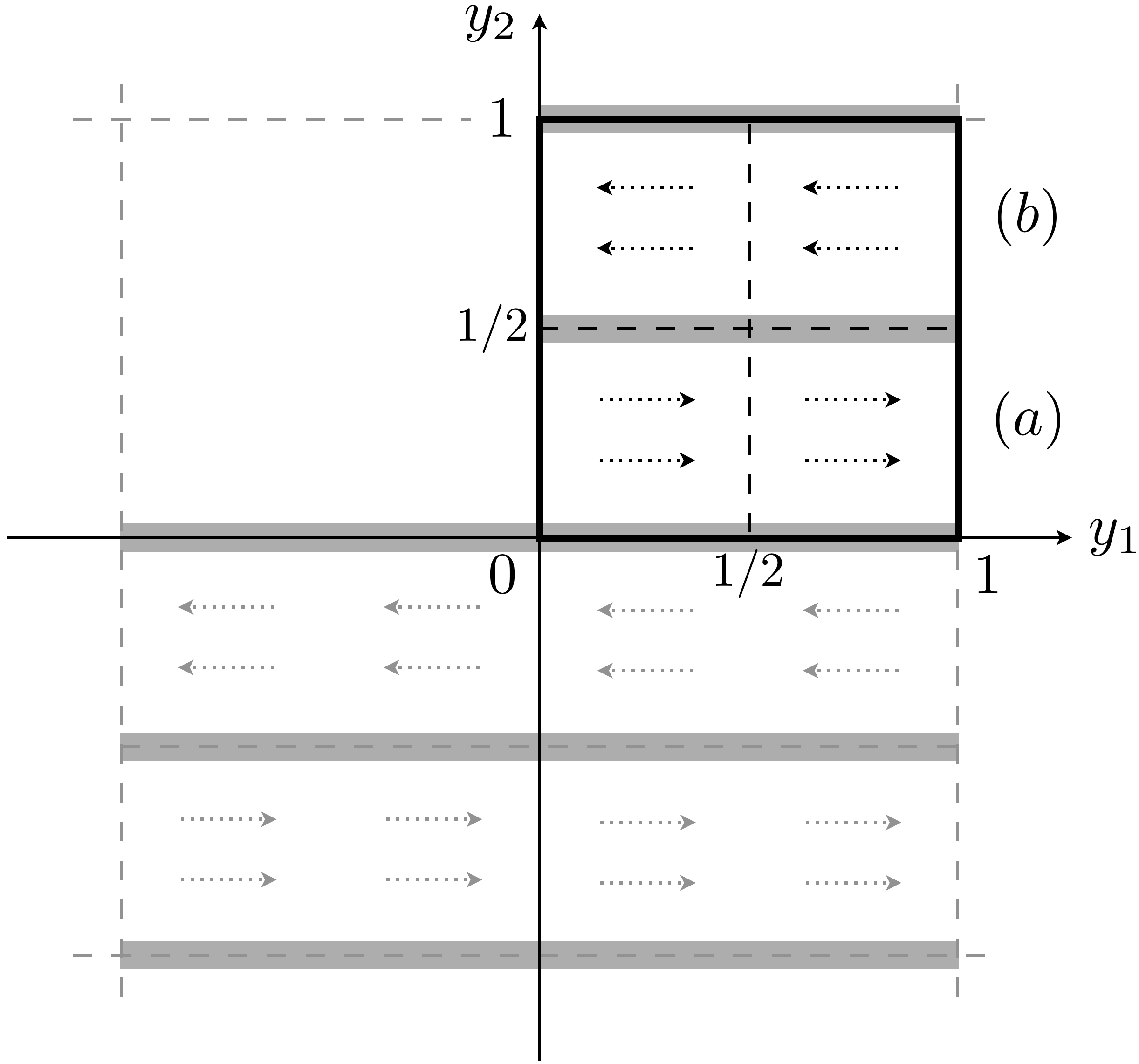} 
	\caption{Three copies of the fundamental domain of the torus
          in the $y_1$$-$$y_2$-plane with constant vector field.
          Two patches of the bundle are
          seperated by the circles $y_2 = 0$ and $y_2 = 1/2$.}
	\label{constcov}
	\end{center}
\end{figure}
the transition function at $y_2 = 1/2$ is determined by
\begin{align}
A_1^{b}(y_1,\tfrac{1}{2}) - A_1^{a}(y_1,\tfrac{1}{2}) = 
-\frac{1}{q} \partial_1 \Lambda_{ba} = -2\alpha_1 \,,
\end{align}
from which one obtains
%which yields $\Lambda_{ba} = - qfy_1$ and therefore the transition function
\begin{align}\label{trafuconst}
S_{ba}(y_1,\tfrac{1}{2}) = e^{i\Lambda_{ba}(y_1,y_2=1/2)} = e^{2iq\alpha_1 y_1}\,.
\end{align}
The required single-valuedness of the transition function,
$
S_{ba}(y_1 + 1) = S_{ba}(y_1)\,,
$ implies
\begin{align}\label{quantization}
\alpha_1 = \frac{\pi k_1}{q}\,, \quad k_1 \in \mathbb{Z} \,.
\end{align}
This is the well-known fact that Wilson lines on
orbifolds are discrete. The same holds for $\alpha_2$. In the transition from $(b)$
to $(a)$ the vector field changes by $2\pi k_1/q$. Hence, 
$S_{ab}(y_1,0) = S^{-1}_{ab}(y_1,1/2) = S_{ba}(y_1,1/2)$.

\begin{figure}[t]
	\begin{center}
	\includegraphics[height = .3 \textheight]{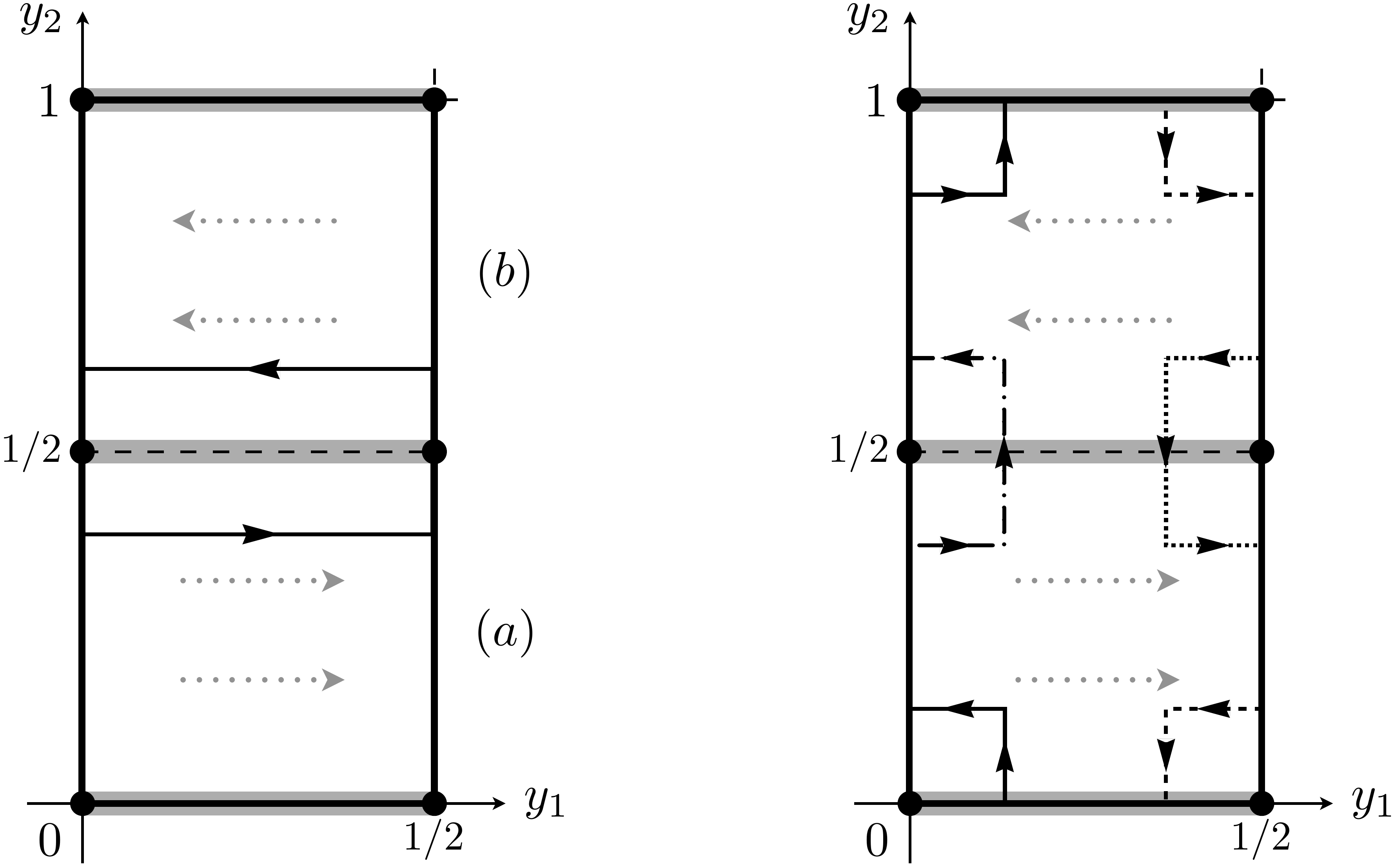} 
	\caption{Wilson line integrals on orbifold with constant
          vector field. Left: bulk Wilson line $W_{\mathcal{T}_1}$;
          right: fixed-point Wilson lines $W_i$.}
	\label{constTCi}
	\end{center}
\end{figure}

It is now straightforward to compute bulk Wilson lines and 
Wilson lines around fixed points (see Fig.~\ref{constTCi}). For 
$k_1$ odd, one finds
\begin{equation}\label{Wialpha1}
\begin{split}
W_1 &= W_3 = 1\,, \quad W_2 = W_4 = -1\,, \\
W_{\mathcal{T}_1} &= W_3 W_4 = W_1^{-1} W_2^{-1} = -1\,,  \\
W_{\mathcal{T}_2} &= W_1 W_3 = W_2^{-1} W_4^{-1} = 1\,.
\end{split}
\end{equation}
Note that the projections of the torus Wilson lines
$W_{\mathcal{T}_{1,2}}$ now factorize into products of fixed-point
Wilson line integrals. 
For a constant vector fields $A^a = \alpha_2 dy_2$ and $A^b = -\alpha_2
dy_2$ on the patches $(a)$ and $(b)$, respectively (see Fig.\ref{constcov}), one obtains the 
transition functions $S_{ba} = e^{2iq\alpha_2 y_2}$ at $y_2 \approx 1/2$ and
$S_{ab} = e^{-2iq\alpha_2 y_2}$ at $y_2 \approx 0$.
The transition functions at $y_1 \approx 1/2$ and $y_1 \approx 0$ are trivial.
The Wilson line integrals now read ($k_2$ odd)
\begin{equation}\label{Wialpha2}
\begin{split}
W_1 &= W_2 = 1\,, \quad W_3 = W_4 = -1\,, \\
W_{\mathcal{T}_1} &= W_3 W_4 = W_1^{-1} W_2^{-1} = 1\,,  \\
W_{\mathcal{T}_2} &= W_1 W_3 = W_2^{-1} W_4^{-1} = -1\,.
\end{split}
\end{equation}
For $k_1$ and $k_2$ odd, the Wilson line
integrals $W_i$ are given by the product of the factors given in
Eqs.~\eqref{Wialpha1} and \eqref{Wialpha2}.

\begin{figure}[t]
	\begin{center}
	\includegraphics[height = .3 \textheight]{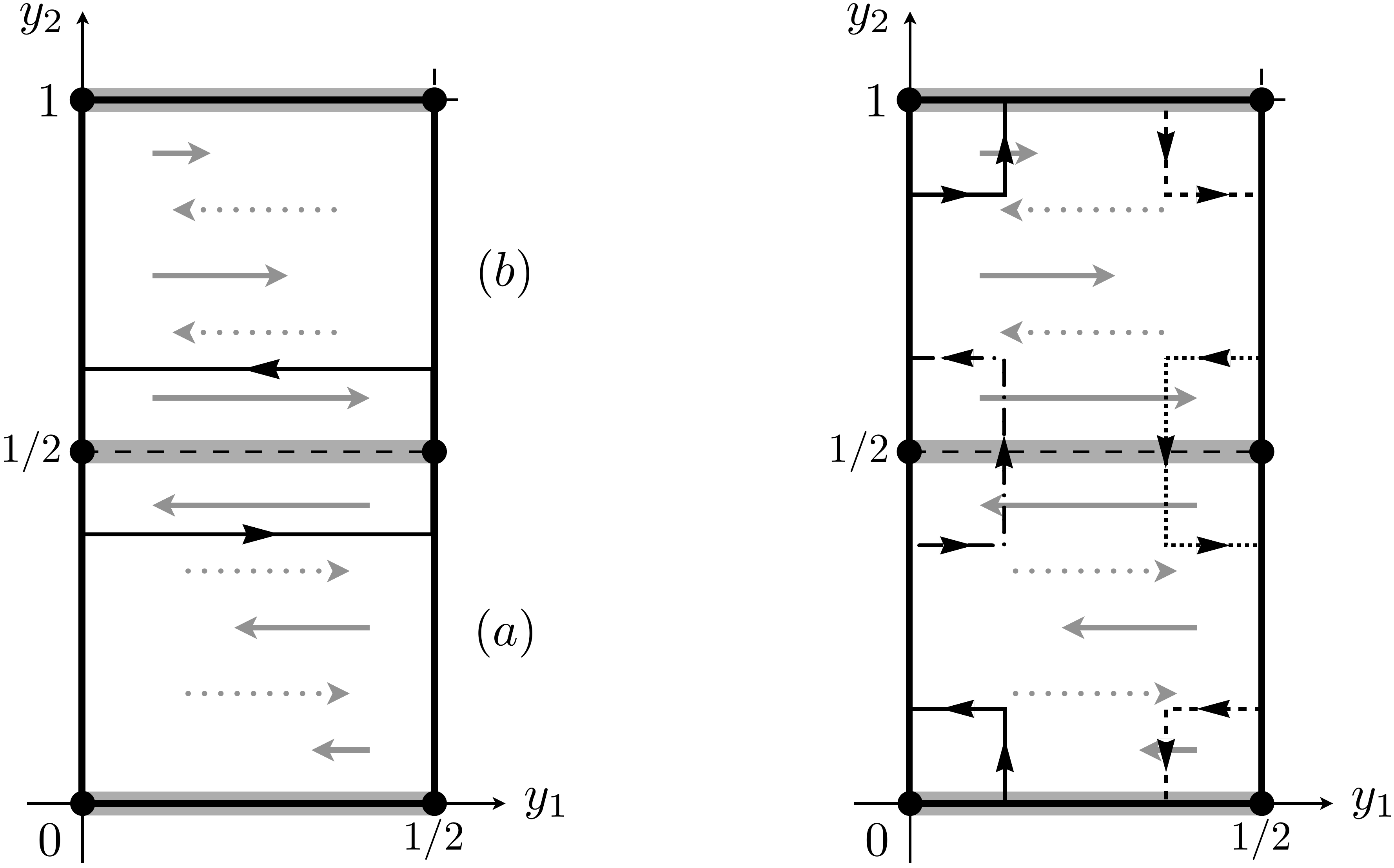} 
	\caption{Wilson line integrals on orbifold with magnetic flux
            and constant
          vector field. Left: bulk Wilson line $W_{\mathcal{T}_1}$;
          right: fixed-point Wilson lines $W_i$.}
	\label{fluxTCi}
	\end{center}
\end{figure}

Our particular interest concerns the Wilson line integrals in the case
of non-vanishing magnetic flux together with constant background
fields, see Fig.~\ref{fluxTCi}. The transition functions
can be read off from Eqs.~\eqref{transfunc} and \eqref{trafuconst},
\begin{equation}\label{totaltf}
\begin{array}{ll}
y_2 \approx \tfrac{1}{2}: \quad & S_{ba} = e^{i(2\pi(k_1y_1 +k_2 y_2) - qfy_1)} \,, \\
y_2 \approx 0: \quad & S_{ba} = e^{2i\pi k_1 y_1} \,,
\end{array}
\end{equation}
where $qf = 2\pi M$, $M \in \mathbb{Z}$. Knowing the transition
functions one easily determines the fixed-point Wilson line integrals
$W_i$ for different values of $M$, $k_1$ and $k_2$. The results are
summarized in Table~\ref{tab:WL}.

\begin{table}
\begin{center}
$\begin{array}{| c | c | c c c c |}\hline
M & (k_1,k_2) & W_1 & W_2 & W_3 & W_4\\\hline\hline
\text{even} & (0,0) & + & + & + & + \\
& (1,0) & + & - & + & - \\
& (0,1) & + & + & - & - \\
& (1,1) & + & - & - & + \\\hline
\text{odd} & (0,0) & + & + & + & - \\
& (1,0) & + & - & + & + \\
& (0,1) & + & + & - & + \\
& (1,1) & + & - & - & - \\ 
\hline
\end{array}$
\caption{Wilson line integrals around the orbifold fixed points
  $\zeta_i$, $i=1,\ldots,4$, for different transition functions
  determined by $k_1$, $k_2$ and $M$.}
\label{tab:WL} 
\end{center}
\end{table}
 
In the case of even bulk flux the number of negative
Wilson lines $W_i$ is even. Which Wilson lines are negative depends
on the values of $k_1$ and $k_2$. This pattern has already been
discussed in \cite{Buchmuller:2015eya}. For odd bulk flux
\cite{Abe:2008fi,Abe:2013bca}, an odd number of Wilson lines has to be
negative. Again it depends on the values of $k_1$ and $k_2$, which
Wilson lines are negative.  In this way the standard quantization
condition, which holds on the torus, is restored for the total flux,
the sum of bulk and localized fluxes.

\subsection{Singular gauge fields}
\label{subsec:fields_singular}

So far we have considered regular gauge fields defined on the orbifold by
means of transition functions, which are related to twisted boundary
conditions on the covering space. It is remarkable that due to the
singular fixed points, magnetic fields on orbifolds can also be
described by means of singular gauge fields without the need to
introduce transition functions. This possibility has previously been
discussed in connection with localized Fayet-Iliopoulos terms \cite{Lee:2003mc}.

Consider the Green's function of the bosonic string on 
a torus \cite{polchinski1998string}
with the singularity located at the position of one of the orbifold fixed points,
\begin{align}\label{torusGreen}
G(z-\zeta,\tau) = \frac{c}{2} \ln{|\vartheta_1(z-\zeta,\tau)|^2} -
\frac{c\pi}{\tau_2}(\text{Im}(z-\zeta))^2\,.
\end{align}
Here $z = y_1 + \tau y_2$, $\zeta = \rho + \tau \eta$, and
$\vartheta_1$ is the Jacobi theta-function listed in Appendix~A. The Green's
function satisfies the differential equation
\begin{align}
\dc\db G(z-\zeta,\tau) = \pi c~\delta^2(z-\zeta) - \frac{c\pi}{2\tau_2}\,,
\end{align}
where we have used\footnote{For simplicity, we use the same symbol
  $\zeta$ for $\rho + \tau \eta$ and $(\rho,\eta)$.}
\begin{align}\label{Dcomplex}
\dc = \partial_z = \frac{i}{2\tau_2}(\bar{\tau}\partial_1
- \partial_2)\,, \;\;
\db = \partial_{\bar{z}} = -\frac{i}{2\tau_2}(\tau\partial_1
- \partial_2)\,, \quad
\delta^2(z-\zeta) = \frac{1}{2\tau_2} \delta^2(y - \zeta)\,.
\end{align}
Defining the vector field \cite{Lee:2003mc}
\begin{align}\label{complexvector}
A = A_z dz +  A_{\bar{z}} d\bar{z}\,, \quad  A_z = i\dc G\,,\;
A_{\bar{z}} = -i\db G\,, 
\end{align}
one obtains 
\begin{align}
F = dA = F_{z\bar{z}} dz\wedge d\bar{z} = \tfrac{1}{2} F_{mn}
dy_m\wedge dy_n\,, \quad F_{z\bar{z}} = \dc A_{\bar{z}} - \db A_z = -2i
\dc\db G\,,
\end{align}
and therefore
\begin{align}\label{Fsing}
F_{mn} &= -2i \epsilon_{mn} \tau_2 F_{z\bar{z}} = -4 \epsilon_{mn}
\tau_2 \dc\db G \nonumber\\
& = \epsilon_{mn} (-2\pi c~\delta^2(y - \zeta) + 2\pi c)\,.
\end{align}
Clearly, the vector field $A$ describes a constant bulk
flux density $c$, which is related to a flux density of opposite
sign localized at an orbifold fixed point. From the discussion of flux
quantization in the previous section we know $qc \in \mathbb{Z}$.

In the vicinity of the orbifold fixed point, $z \simeq \zeta$, the vector field is
singular,
\begin{align}
\vartheta_1(z-\zeta,\tau) \propto (z-\zeta)\,, \quad \db G(z-\zeta,\tau) \simeq
\frac{c}{2} \frac{z-\zeta}{|z-\zeta|^2}\,,
\end{align}
and with
\begin{align}\label{Acomplex}
A_z = \frac{i}{2\tau_2}(\bar{\tau}A_1 - A_2)\,, \quad
A_{\bar{z}} = -\frac{i}{2\tau_2}(\tau A_1 - A_2)\,, 
\end{align}
one obtains
\begin{align}\label{local}
A_m = c\tau_2 \epsilon_{mn} \frac{(y-\zeta)_n}{|y-\zeta|^2}\,.
\end{align}
This is precisely the vortex field introduced in \cite{Buchmuller:2015eya}
to account for localized flux.

The Green's function $G(z,\tau)$ is even and invariant under lattice
translations,
\begin{align}\label{Gsymmetries}
G(z,\tau) = G(-z,\tau)\,, \quad G(z,\tau) = G(z+1,\tau)\,, \quad G(z,\tau) = G(z+\tau,\tau)\,.
\end{align}
Since $2\zeta_i$ is a lattice vector for all fixed points,
$\zeta_1 = (0,0)$, $\zeta_2 = (1/2,0)$, $\zeta_3 = (0,1/2)$ and
$\zeta_4 = (1/2,1/2)$, 
$G(z-\zeta_i)$ is invariant under reflections
at the origin, $G(z-\zeta_i,\tau) = G(-z-\zeta_i,\tau)$. Hence, the
vector field $A_z(z-\zeta;c)$ is invariant under lattice translations and odd
under reflection at the origin. It can therefore be projected to the
orbifold. Because of the invariance under lattice translations it is not necessary to intoduce patches and
transition functions for the singular vector field.

Using Eq.~\eqref{local} one obtains for the Wilson line integral
around the 1-cycle $\mathcal{C}_i$ for the vector field $A_m(y-\zeta_i;c_i)$,
\begin{align}
W_i = \exp{\left(- i q \oint_{\mathcal{C}_i} A \right)} = e^{-i q \pi c_i  } \,,
\label{sec:wilsonlines:eq:Wilson_line_phase}
\end{align}
which implies
$W_i = \pm 1$ for $c_i = k_i/q$, $k_i \in \mathbb{Z}$. Here we have
taken into account that the orbifold fixed point $\zeta_i$ has
a deficit angle $\pi$.

\section{Wave functions}
\label{sec:Wavefunctions}

We now turn to 6d Weyl fermions in a background $U(1)$ gauge field,
\begin{align}\label{fermionR}
\mathcal{L}_f = i\bar{\Psi}(x) \Gamma^a e_a^M D_M \Psi(x) \,, \quad \Gamma^7
\Psi = -\Psi\,.
\end{align}
Here $\Gamma^0, \dots, \Gamma^6$ are gamma-matrices in six
dimensions (see Appendix~B), $e_a^M$ is the inverse vielbein, $\Gamma^7 = \Gamma^0\cdot
\ldots \cdot\Gamma^6$ and
$D_M = \partial_M + i q A_M$ is the gauge covariant derivative. 
For the torus metric \eqref{torusmetric},  the inverse
zweibein ($a=5,6$; $m=1,2$) is given by\footnote{Lower and upper
  indices label rows and columns, respectively.}
\begin{align}
\left(e_a^m\right) = \frac{1}{\sqrt{\tau_2}} 
\left(\begin{array}{cc} \tau_2 & 0 \\ -\tau_1 &
    1 \end{array}\right)\,,
\end{align}
where we have used the definitions
\begin{align}
e_m^a\delta_{ab} e_n^b = g_{mn}\,, \quad e_a^m e_m^b = \delta_{ab}\,.
\end{align}
The 6d Weyl fermion $\Psi$ contains two 4d Weyl fermions of opposite
chirality. For gamma-matrices in Weyl representation, one has
\begin{align}
\Psi = \left(\begin{array}{c} \psi_L \\
    \psi_R \end{array}\right)\,,\quad \gamma_5 \psi_L = - \psi_L\,,
\quad \gamma_5 \psi_R = \psi_R\,.
\end{align}
On the orbifold, we impose chiral boundary conditions,
\begin{align}
\psi_L(x^\mu,y_m) = \psi_L(x^\mu,-y_m) \,, \quad \psi_R(x^\mu,y_m) =
-\psi_R(x^\mu,-y_m) \,,
\label{chiral_bc}
\end{align}
which correspond to one possible embedding of the orbifold twist into
the $SU(2)_R$ symmetry of the 6d theory. $\Gamma^a e_a^m D_m$ is the
mass operator of the fermion fields in the effective 4d theory, and from 
the 6d Dirac equation
%theory satisfy the equations\footnote{In the following we drop the
%  dependence on the 4d space-time coordinates for simplicity.}
\begin{align}
 \Gamma^a e_a^M D_M \Psi(x,y) = 0\,
\end{align}
one obtains a coupled system of equations for the two 4d Weyl fermions
$\psi_L$ and $\psi_R$,
\begin{equation}\label{diffwf}
\begin{split}
i\gamma^\mu \partial_\mu \psi_L &= -\frac{i}{\sqrt{\tau_2}} (\tau D_1 -
D_2) \psi_R = 2\sqrt{\tau_2}(\db + iqA_{\bar{z}}) \psi_R\,,\\
i\gamma^\mu \partial_\mu \psi_R &=  -\frac{i}{\sqrt{\tau_2}} (\bar{\tau}D_1 -
D_2) \psi_L = -2\sqrt{\tau_2}(\dc + iqA_{z}) \psi_L\,. 
\end{split}
\end{equation}
Here we have assumed a background gauge field in the compact
dimensions, which contains a constant part and magnetic flux part,
$A_m = \alpha_m + A_m^\text{flux}$. The constant part can be removed
by a field redefinition, see Eq.~\eqref{redefine}. This changes the
boundary conditions to\footnote{In the following we drop the
  dependence on the 4d space-time coordinates for simplicity.}
\begin{align}\label{twistedpsi}
\psi_{L,R}(y+\lambda) = e^{iq(\alpha_1 y_1 + \alpha_2
  y_2)}S_{ba}^{-1}(y)\psi_{L,R}(y)\,,
\end{align}
where $S_{ba}$ is the transition function of the vector bundle.

For comparison, a complex scalar $\Phi$ with charge $q$ satisfies the
equation of motion
\begin{align}\label{diffscalar}
(\eta^{\mu\nu} \partial_\mu \partial_\nu + (g_2)^{mn} D_m D_n)\Phi = 0\,.
\end{align}
Eqs.~\eqref{diffwf} and \eqref{diffscalar}
are a convenient starting point to construct mass spectra and wavefunctions.

\subsection{Twisted wave functions}
\label{sec:fields_twisted}

For the magnetic vector field in Landau gauge, $A_1 = - fy_2$,
Eqs.~\eqref{diffwf} can be rewritten as 
\begin{align}\label{diffwf2}
i\gamma^\mu \partial_\mu \psi_L = - i \sqrt{2qf} a^\dagger \psi_R\,, \quad
i\gamma^\mu \partial_\mu \psi_R = i \sqrt{2qf} a~\psi_L\,,
\end{align}
where we have introduced the differential operators ($qf >0$)
\begin{align}
a = - (2qf\tau_2)^{-1/2} (\bar{\tau}(\partial_1 -iqfy_2) - \partial_2)\,, \quad
a^\dagger = (2qf\tau_2)^{-1/2} (\tau(\partial_1 -iqfy_2) - \partial_2)\,.
\end{align}
The operators $a$ and $a^\dagger$ satisfy the commutation relation
$[a,a^\dagger]=1$, and they can
therefore be interpreted as annihilation and creation
operators.\footnote{For the symmetric gauge, $A_1 = -fy_2/2, A_2 =
  fy_1/2$, and $\tau = i$, one obtains, in the conventions of
  \cite{Buchmuller:2018eog}:
$a = i(\partial_z + qf\bar{z})/\sqrt{2qf}, a^\dagger = i(\partial_{\bar{z}} - qfz)/\sqrt{2qf}$.}
An orthonormal set of mode functions is given by $(qf = 2\pi M$, $M\in \mathbb{N}$)
\begin{equation}\label{modes}
\begin{split}
\xi_{n,j} &= \frac{i^n}{\sqrt{n!}}\left(a^\dagger\right)^n \xi_j\,,
\quad a~\xi_j = 0\,, \\ 
a~\xi_{n,j} &= i\sqrt{n}~\xi_{n-1,j}\,, \quad 
a^\dagger \xi_{n,j} = -i\sqrt{n+1}~\xi_{n+1,j}\,,
\end{split}
\end{equation}
where $j = 1,\dots,M$ labels the degeneracy of the ground state, with
the corresponding mode functions $\xi_j$.
Expanding the chiral fermions $\psi_L$ and $\psi_R$ in terms of the
mode functions $\xi_{n,j}$, 
\begin{align}
\psi_L(x,y) = \sum_{n,j} \psi_{Ln,j}(x) \xi_{n,j}(y)\,, \quad
\psi_R(x,y) = \sum_{n,j} \psi_{Rn,j}(x) \xi_{n,j}(y)\,, 
\end{align}
one obtains from Eqs.~\eqref{diffwf2} and \eqref{modes}
\begin{equation}
\begin{split}
i\gamma^\mu \partial_\mu \psi_{L 0,j}  &= 0 \,,\quad
i\gamma^\mu \partial_\mu \psi_{L n+1,j} = -\sqrt{2qf(n+1)}\ 
\psi_{R n,j}\,, n\geq 0\,, \\
 i\gamma^\mu \partial_\mu \psi_{R n,j} &= -\sqrt{2qf(n+1)} \psi_{L
   n+1,j} \,, n\geq 0\,.
\end{split}
\end{equation}
The fermions $\psi_{L 0,j}$ are the $M$ expected zero-modes, and the pair
of chiral fermions $(\psi_{L n+1,j}, \psi_{R n,j})$ form 4d Dirac fermions with masses
$m_{n,j} = \sqrt{2qf(n+1)}$.

Correspondingly, for the complex scalar $\Phi$ one finds
(cf.~\eqref{diffscalar})
\begin{align}\label{diffscalar2}
(\eta^{\mu\nu} \partial_\mu \partial_\nu - 2qf(a^\dagger a +
\tfrac{1}{2}))\Phi = 0 \,.
\end{align}
After a mode expansion, $\Phi = \sum_{n,j} \Phi_{n,j}\xi_{n,j}$, one
obtains the scalar mass spectrum $M^2_{n,j} = 2qf(n+1)$, $n\geq 0$.

We are particularly interested in the fermionic zero-modes $\psi_{L
  0,j}$. Their mode functions are determined by the equation
\begin{align}\label{Rzeromode}
a~\xi_{j} \propto (\bar{\tau} D_1 - D_2) \xi_{j} = 0\,,
\end{align}
where $D_{1,2}$ are now the covariant derivatives in the flux
background. From Eq.~\eqref{twistedpsi} and the transition functions 
\eqref{totaltf}
%As discussed in Section~\ref{}, all mode functions have to satisfy 
one obtains the twisted boundary conditions ($qf = 2\pi M, M \in \mathbb{N}$)
\begin{equation}\label{twistedbc}
\begin{split}
\xi_{n,j}(y+\lambda_1) &= e^{-\pi i k_1} \xi_{n,j}(y)\,, \\
\xi_{n,j}(y+\lambda_2) &= e^{- \pi i (k_2 - 2M y_1) } \xi_{n,j}(y)\,.
\end{split}
\end{equation}
Clearly, the mode functions $\xi_{n,j}$ depend on the number of flux
quanta $M$ as well as $k_1$ and $k_2$, which determine the boundary conditions.

For comparison with the untwisted wave functions, which we will
construct in the following section, we briefly recall the derivation
of the zero-mode functions, following \cite{Cremades:2004wa}. The
boundary conditions \eqref{twistedbc} are satisfied by functions
$\xi$ which can be written as
\begin{equation}\label{ansatz}
\xi(y) = e^{-i\pi (k_1 y_1 + k_2 y_2)} \sum_n f_n(y_2) e^{2
  \pi i n y_1}\,,
\end{equation}
where the coefficient functions $f_n(y_2)$, $n\in \mathbb{Z}$,
fulfill the recurrence relation
\begin{equation}
	f_n(y_2 + 1) = f_{n - M}(y_2)\,.
	\label{recurrence}
\end{equation}
The zero-mode equation \eqref{Rzeromode}, $a~\xi = 0$, yields first-order
differential equations for the functions $f_n(y_2)$, whose solutions
are given by
\begin{align}\label{fn}
f_n(y_2) = c_n e^{-i\pi \bar{\tau} My_2^2 - i\pi (\bar{\tau}k_1 -
  k_2)y_2 + 2\pi i \bar{\tau} n y_2}\,.
\end{align}
Because of the recurrence relation \eqref{recurrence}, only $M$ of the
constants $c_n$ are independent. Writing $n = lM+j$, with $l \in
\mathbb{Z}$ and $j = 0,\ldots,M-1$, one has
\begin{align}\label{constant}
c_n = \mathcal{N}_j e^{-\frac{i\pi\bar{\tau}}{M}(lM+j)^2 +
  \frac{i\pi}{M}(\bar{\tau}k_1 - k_2)(lM+j)}\,,
\end{align}
where the $\mathcal{N}_j$ are normalization constants.
Hence, there are indeed $M$ independent zero-mode functions $\xi_j$ on the torus. 
Combining Eqs.~\eqref{ansatz}, \eqref{fn} and \eqref{constant}, and
replacing the sum over $n$ by a double sum over $j$ and $l$, one  
obtains the wave functions $\xi_j$ as infinite sums over $l$, which can be
conveniently expressed in terms of Jacobi theta-functions
(see Appendix B) \cite{Cremades:2004wa},
\begin{align}
\xi_j(y) &= \mathcal{N}_j e^{-i\pi M\bar{\tau} y_2^2 - i\pi
  k_1\bar{z}} \sum_l e^{-i\pi M\bar{\tau}(l+j/M)^2 
+ 2\pi i(l+j/M)(M\bar{z} + (\bar{\tau} k_1 - k_2)/2)} \nonumber \\
&= \mathcal{N}_j e^{-i\pi M\bar{\tau} y_2^2 - i\pi k_1\bar{z}}\ 
\vartheta\left[\begin{array}{l} j/M\\(k_1\bar{\tau}
    -k_2)/2\end{array}\right](M\bar{z},-M\bar{\tau})\,,\;\;\; \bar{z}=y_1 + \bar{\tau}y_2\,.
\end{align}
Since $\tau_2 = -\mathrm{Im}\bar{\tau} > 0$, these zero-mode functions
are normalizable for $M > 0$.

Under reflection the zero-mode $\xi_j$ turns into another zero mode
$\xi_{j'}$. One finds the relation
\begin{align}\label{reflection}
\xi_j(-y) = e^{\frac{2\pi i}{M}\left(j'-k_1/2\right)k_2}
\frac{\mathcal{N}_j}{\mathcal{N}_{j'}} \xi_{j'} (y)\ ,\quad j' = mM +k_1-j\ ,\quad m\in \mathbb{Z}\,,
\end{align}
with $m$ chosen such that $j,j'=1,\ldots,
M-1$. From $\xi_j$ and $\xi_{j'}$ one can form even and odd linear
combinations,
\begin{equation}
\begin{split}
\xi^\eta_j(y) &\propto \xi_j(y) + \eta\xi_j(-y)\,, \quad \eta = +,-\,,\\
\xi^\eta_j(y) &= \eta\xi^\eta_j(-y)\,.
\end{split}
\end{equation}
For $j\neq j'$ one obtains one even and one odd zero-mode. In the case
$j=j'$ the zero-mode is either even or odd. From
Eq.~\eqref{reflection} one easily derives the relations between $j$
and $j'$ for given $M$, $k_1$ and $k_2$ (see Table~\ref{even_odd}).
For $k_1=0$ the $j=0$ mode is even; moreover, for $M$ even the $j=M/2$
mode is even ($k_2 =0$) or odd ($k_2 =1$). For $k_1=1$ and $M$ odd the
$(M+1)/2$ mode is even ($k_2=0$) or odd ($k_2=1$). This leads to the
numbers of even and odd zero-modes listed in Table~\ref{even_odd}.
These results have previously been obtained in \cite{Abe:2013bca}.

\begin{table}[t]
\begin{center}
$\begin{array}{|c | c | c | c c | c c|}\hline
 & j & j' & \begin{array}{c} k_2=0\\ \eta = +\end{array}
 &\begin{array}{c} k_2=0\\ \eta = -\end{array} & \begin{array}{c}
   k_2=1\\ \eta = +\end{array} & \begin{array}{c} k_2=1\\ \eta = -\end{array}\\\hline\hline
\begin{array}{c} k_1=0\\ M\ \text{even}\end{array} & 0,1,\ldots,
\frac{M}{2} & 0, \text{$M$$-$$1$}, \ldots,\frac{M}{2}& 
\frac{M}{2}+1& \frac{M}{2}-1 & \frac{M}{2} & \frac{M}{2}\\ \hline
\begin{array}{c} k_1=0\\ M\ \text{odd}\end{array} & 0,1,\ldots, \frac{M-1}{2} & 0,\text{$M$$-$$1$},\ldots,\frac{M+1}{2}& 
\frac{M+1}{2} & \frac{M-1}{2} & \frac{M+1}{2} & \frac{M-1}{2}\\ \hline
\begin{array}{c} k_1=1\\ M\ \text{even}\end{array} & 0,1,2,\ldots,
\frac{M}{2} & 1,0,\text{$M$$-$$1$},\ldots,\frac{M}{2}+1 & 
\frac{M}{2} & \frac{M}{2} & \frac{M}{2} & \frac{M}{2}\\ \hline
\begin{array}{c} k_1=1\\ M\ \text{odd}\end{array} & 0,1,2\ldots, \frac{M+1}{2} & 1,0,\text{$M$$-$$1$},\ldots,\frac{M+1}{2}& 
\frac{M+1}{2} & \frac{M-1}{2} & \frac{M-1}{2} & \frac{M+1}{2}\\ \hline
\end{array}$
\end{center}
\caption{Zero-modes $j$ and $j'$, which are related by reflection,
  with the choice $j' \geq j$. The last four columns give the number of even and odd linear
  combinations that depend on the values of $M$, $k_1$ and
  $k_2$. Their sum is always $M$.}
\label{even_odd}
\end{table}

The field theory on the orbifold is defined by the chiral boundary
conditions \eqref{chiral_bc} which require that 4d left-handed
fermions are linear combinations of even mode functions. For the
zero-modes these are\footnote{Note, that these zero-mode functions are
  the complex conjugate of the wave functions given in \cite{Buchmuller:2015eya}.}
\begin{equation}\label{Orbimfs}
\begin{split}
\xi^+_j(y) &= \mathcal{N}_j e^{-i\pi M\bar{\tau} y_2^2}\bigg(e^{- i\pi k_1\bar{z}}\ 
\vartheta\left[\begin{array}{l} j/M\\(k_1\bar{\tau}
    -k_2)/2\end{array}\right](M\bar{z},-M\bar{\tau}) \\
&\hspace{2.8cm} + e^{ i\pi k_1\bar{z}}\ 
\vartheta\left[\begin{array}{l} j/M\\(k_1\bar{\tau}
    -k_2)/2\end{array}\right](-M\bar{z},-M\bar{\tau})\bigg) \\
&= \mathcal{N}_j e^{-i\pi M\bar{\tau} y_2^2}\sum_l e^{-i\pi M\bar{\tau}(l+j/M)^2 
+ i\pi (\bar{\tau} k_1 - k_2)(l+j/M)} \cos{\left[2\pi \left(lM+j-\frac{k_1}{2}\right) \bar{z}\right]}\,,
\end{split}
\end{equation}
where $\mathcal{N}_j$ are adjusted normalization constants on the orbifold.

Let us now consider some examples. For $M=0$, one has the standard
orbifold result without flux, a single constant mode function for $k_1
= k_2 =0$, and there are no non-vanishing mode functions otherwise. For $M=1$,
one obtains a single zero-mode $j=0$ for each pair $(k_1,k_2)$. For
$k_1 = k_2 =1$, the zero-mode is odd, otherwise it is even. Using
properties of the Jacobi theta-functions (see Appendix~A) one has
\begin{equation}\label{RM1}
\begin{split}
\xi^+_0(y) &= \mathcal{N}_0 e^{-i\pi \bar{\tau} y_2^2} e^{- i\pi k_1\bar{z}}\ 
\vartheta\left[\begin{array}{l} 0\\(k_1\bar{\tau}
    -k_2)/2\end{array}\right](\bar{z},-\bar{\tau})   \\
&= \mathcal{N}_0 e^{-i\pi \bar{\tau} y_2^2} e^{- i\pi
    k_1\bar{z}}\ \vartheta (\bar{z} + (k_1\bar{\tau} -k_2)/2, -\bar{\tau})\,.
\end{split}
\end{equation}
The three even zero-mode functions are shown in Figure~\ref{fig:M1}. For
each $(k_1,k_2)$ combination the zero-mode function vanishes at one
fixed point.
\begin{figure}[t]
	\begin{center}
	\subfloat[\(k_1 = 0, k_2 = 0\)]{\includegraphics[height = .25 \textheight]{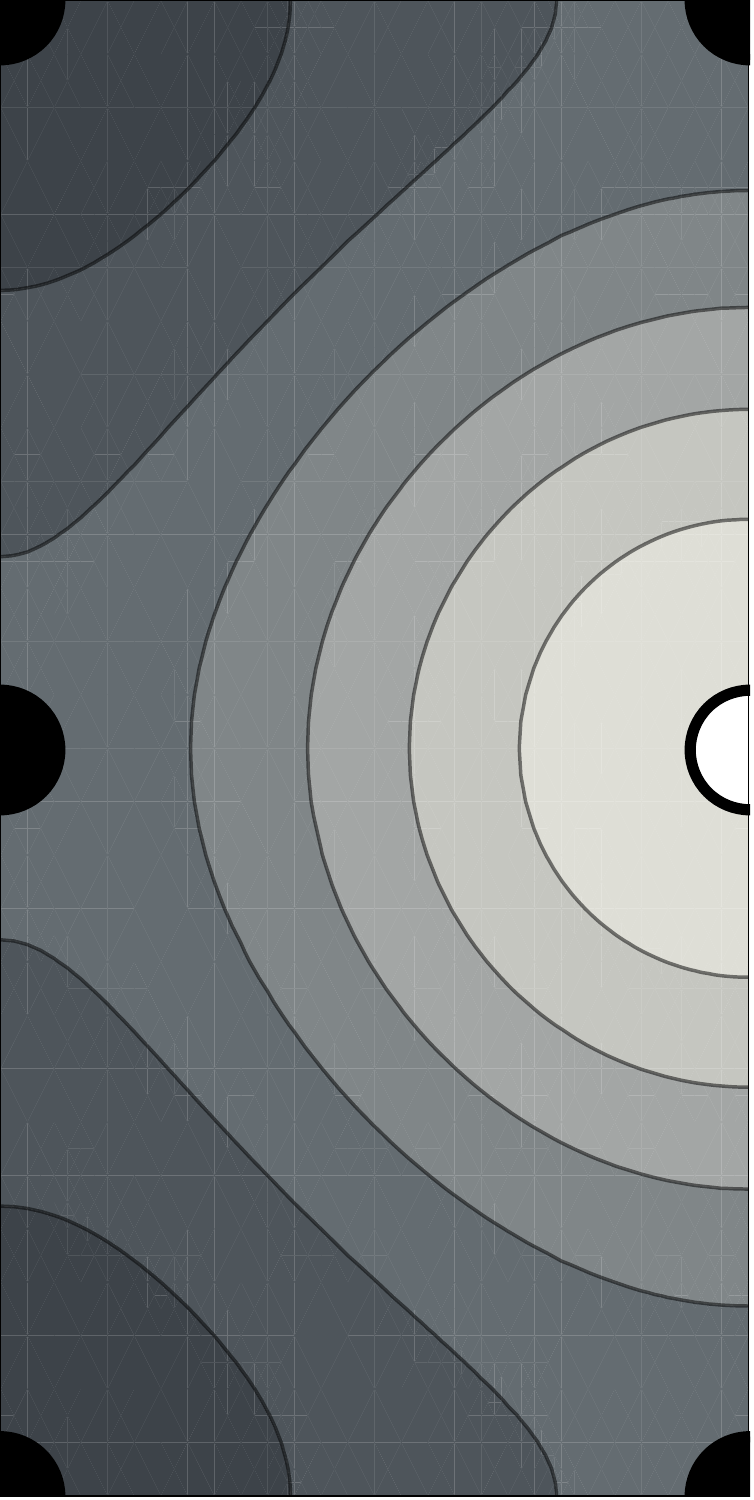} } \hspace*{1cm}
	\subfloat[\(k_1 = 1, k_2 = 0\)]{\includegraphics[height = .25 \textheight]{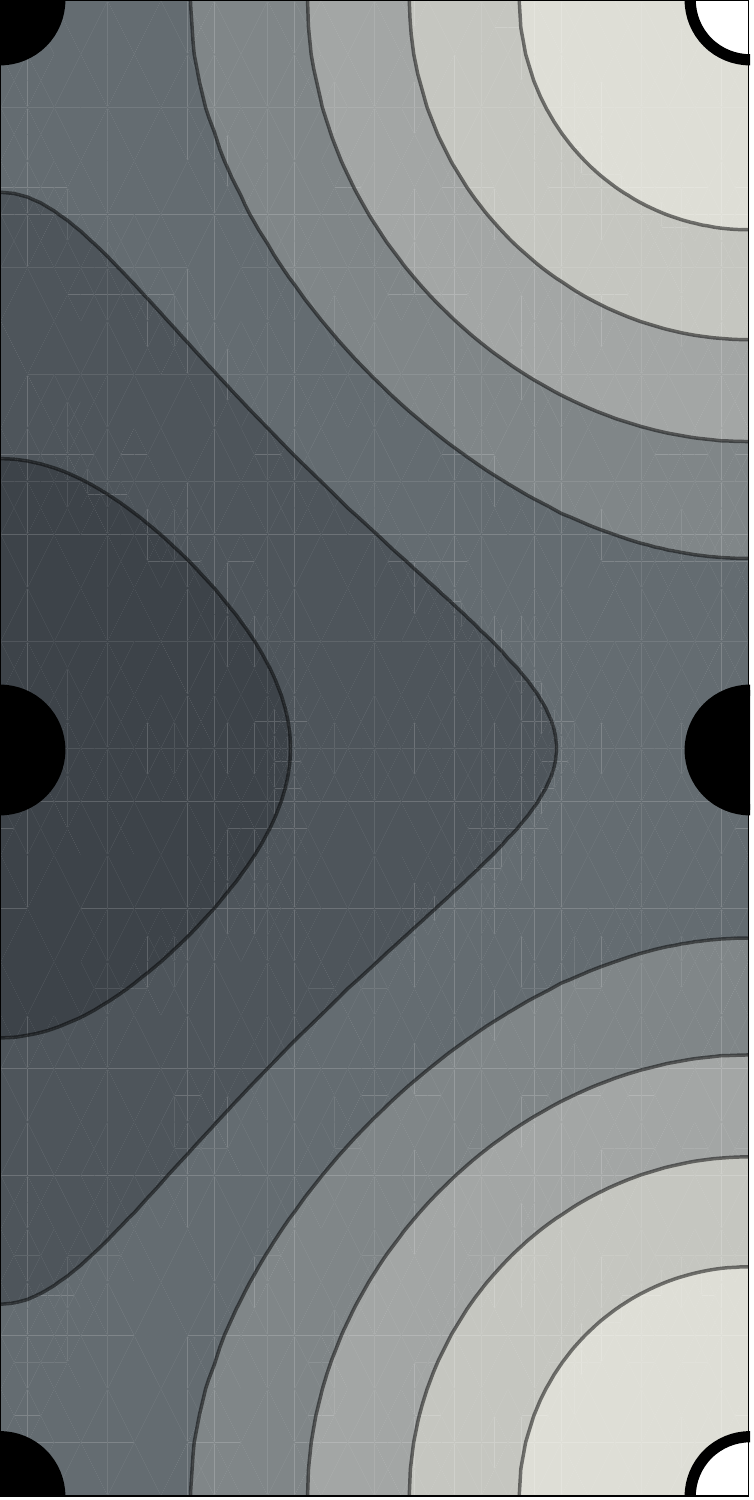} } \hspace*{1cm}
	\subfloat[\(k_1 = 0, k_2 = 1\)]{\includegraphics[height = .25 \textheight]{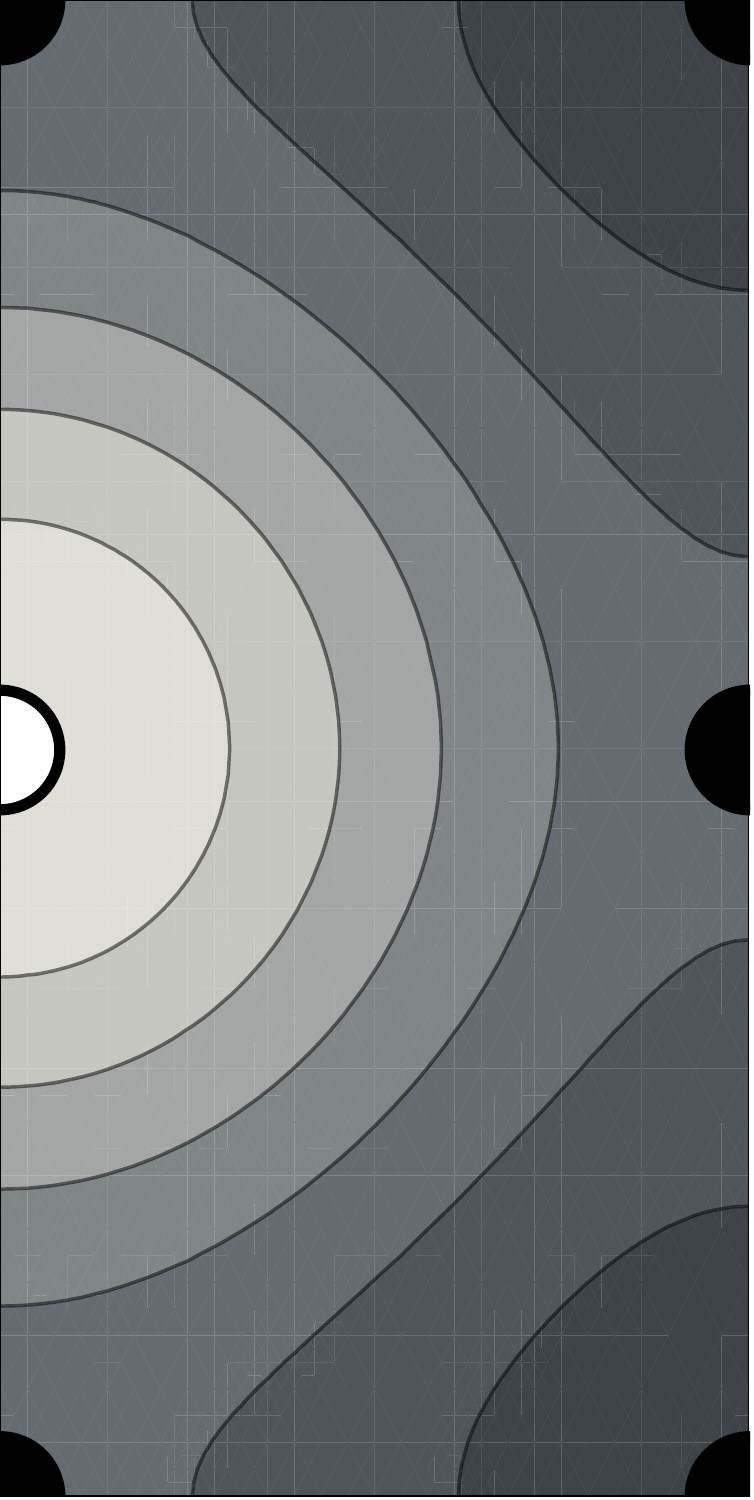} } \hspace*{0.1cm}	\subfloat{\includegraphics[height = .25 \textheight]{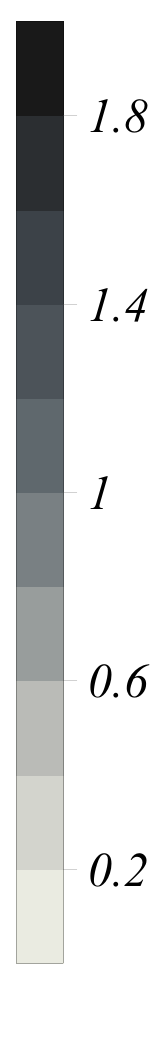}}
	\caption{Absolute square of even zero-mode functions for $M=1$ and different
          values of $k_1$ and $k_2$. The white circles indicate where
          the mode functions vanish.}
	\label{fig:M1}
	\end{center}
\end{figure}

For $M=2$ there are two zero-modes, $j=0,1$. In the case $k_1= k_2 =0$
both of them are even. Their mode functions read
\begin{equation}\label{RzeroM2_1}
\begin{split}
\xi^+_j(y) &= \mathcal{N}_j e^{-2\pi i \bar{\tau} y_2^2} \ 
\vartheta\left[\begin{array}{l} j/2\\ 0 \end{array}\right](2\bar{z},-2\bar{\tau})\\   
&= \mathcal{N}_j e^{-2\pi i \bar{\tau} y_2^2} e^{-i\pi \bar{\tau} j^2/2
  +2\pi i j \bar{z}}\ \vartheta (2\bar{z} -j\bar{\tau},
-2\bar{\tau})\,,\quad j=0,1\,.
\end{split}
\end{equation}
The two mode functions are depicted in Fig.~\ref{fig:M2}(a,b). Note
that the functions are non-zero at all fixed points.
In the remaining cases there is always one even and one odd mode
functions. For the even mode functions one obtains from
Eq.~\eqref{Orbimfs}, after some manipulations,
\begin{align}\label{RzeroM2_2}
%\begin{split}
\xi^+_0(y) &= \mathcal{N}_0 e^{-2\pi i \bar{\tau} y_2^2} 
\left( e^{-i\pi \bar{z}}\ \vartheta (2\bar{z} + \bar{\tau}/2,
-2\bar{\tau}) + e^{i\pi \bar{z}}\ \vartheta (2\bar{z} -
\bar{\tau}/2,-2\bar{\tau}) \right)\,, \nonumber\\
&\hspace*{9cm} k_1=1, k_2=0\,,\nonumber\\
\xi^+_0(y) &= \mathcal{N}_0 e^{-2\pi i \bar{\tau} y_2^2} 
\ \vartheta (2\bar{z} - 1/2,-2\bar{\tau})\,, \, k_1=0, k_2=1\,,\\
\xi^+_0(y) &= \mathcal{N}_0 e^{-2\pi i \bar{\tau} y_2^2} 
\left( e^{-i\pi \bar{z}}\ \vartheta (2\bar{z} + \bar{\tau}/1-1/2,
-2\bar{\tau}) + e^{i\pi \bar{z}}\ \vartheta (2\bar{z} -
\bar{\tau}/2+1/2,-2\bar{\tau}) \right)\,, \nonumber\\ 
&\hspace*{9cm} k_1=1, k_2=1\,. \nonumber
%\end{split}
\end{align}
The even functions are shown in Fig.~\ref{fig:M2}(c,d,e).
They all vanish at two fixed points.
\begin{figure}[t]
	\begin{center}
	\subfloat[\(k_{1,2}=0, j = 0\)]{\includegraphics[height = .25 \textheight]{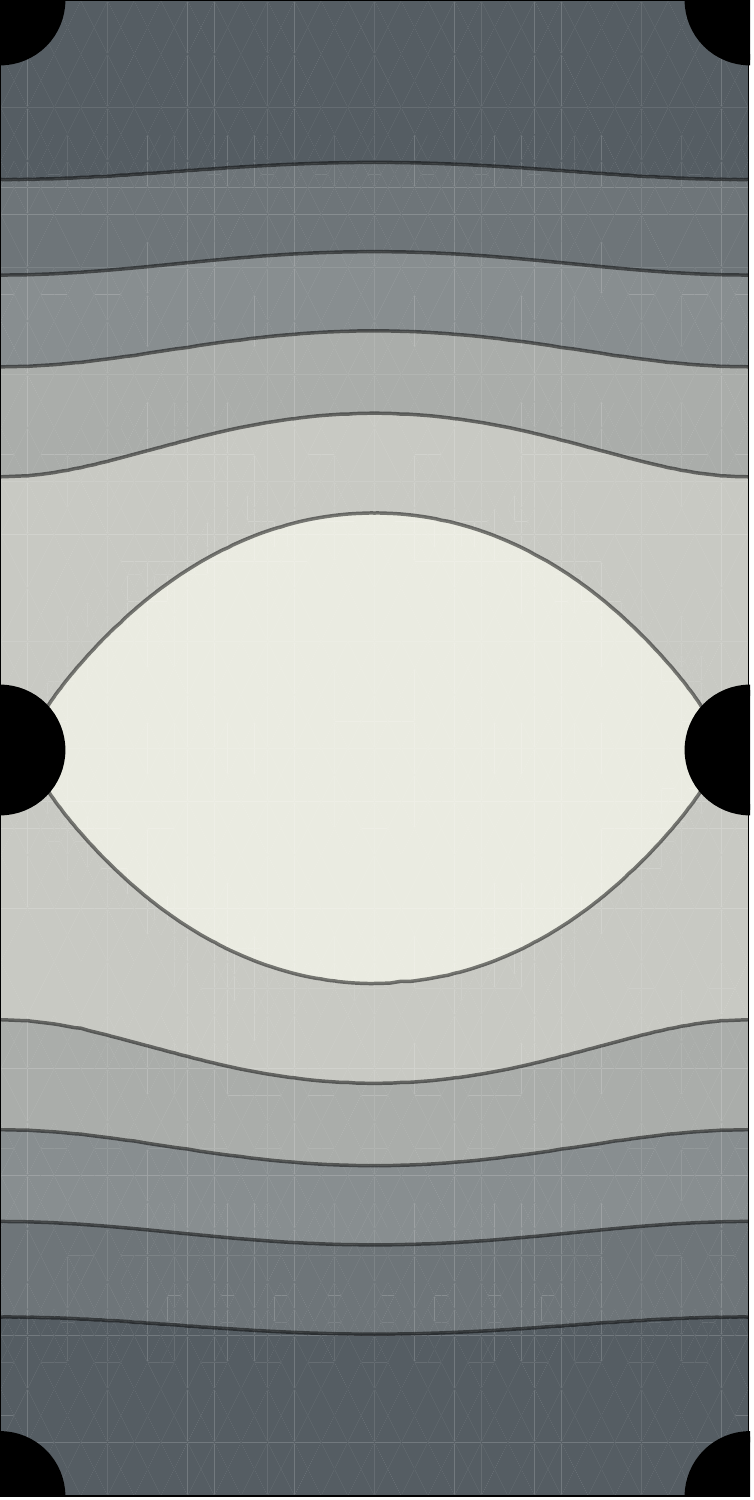} } \hspace*{1.0cm}
	\subfloat[\(k_{1,2}=0, j = 1\)]{\includegraphics[height = .25 \textheight]{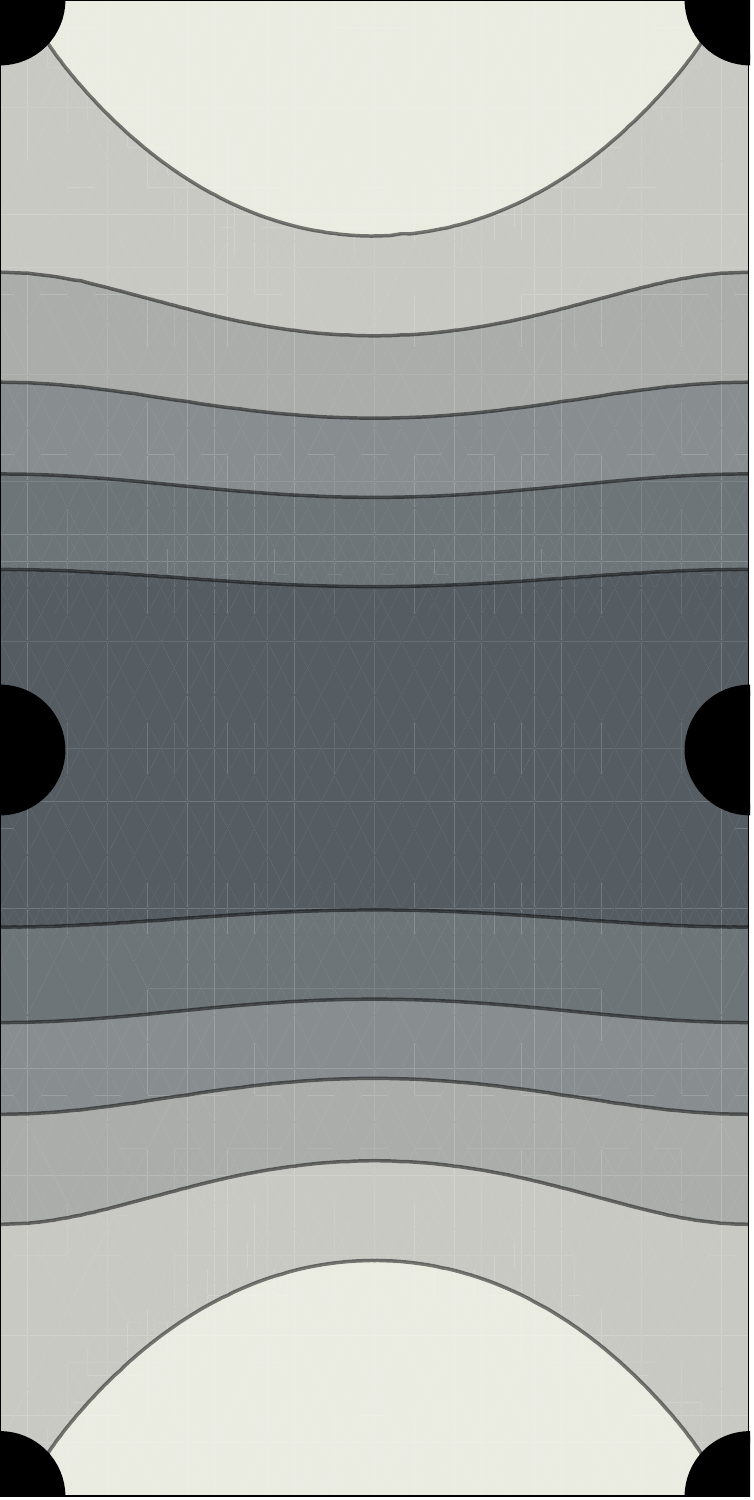} } \hspace*{0.1cm}\\
	\subfloat[\(k_1 = 1, k_2 = 0\)]{\includegraphics[height = .25 \textheight]{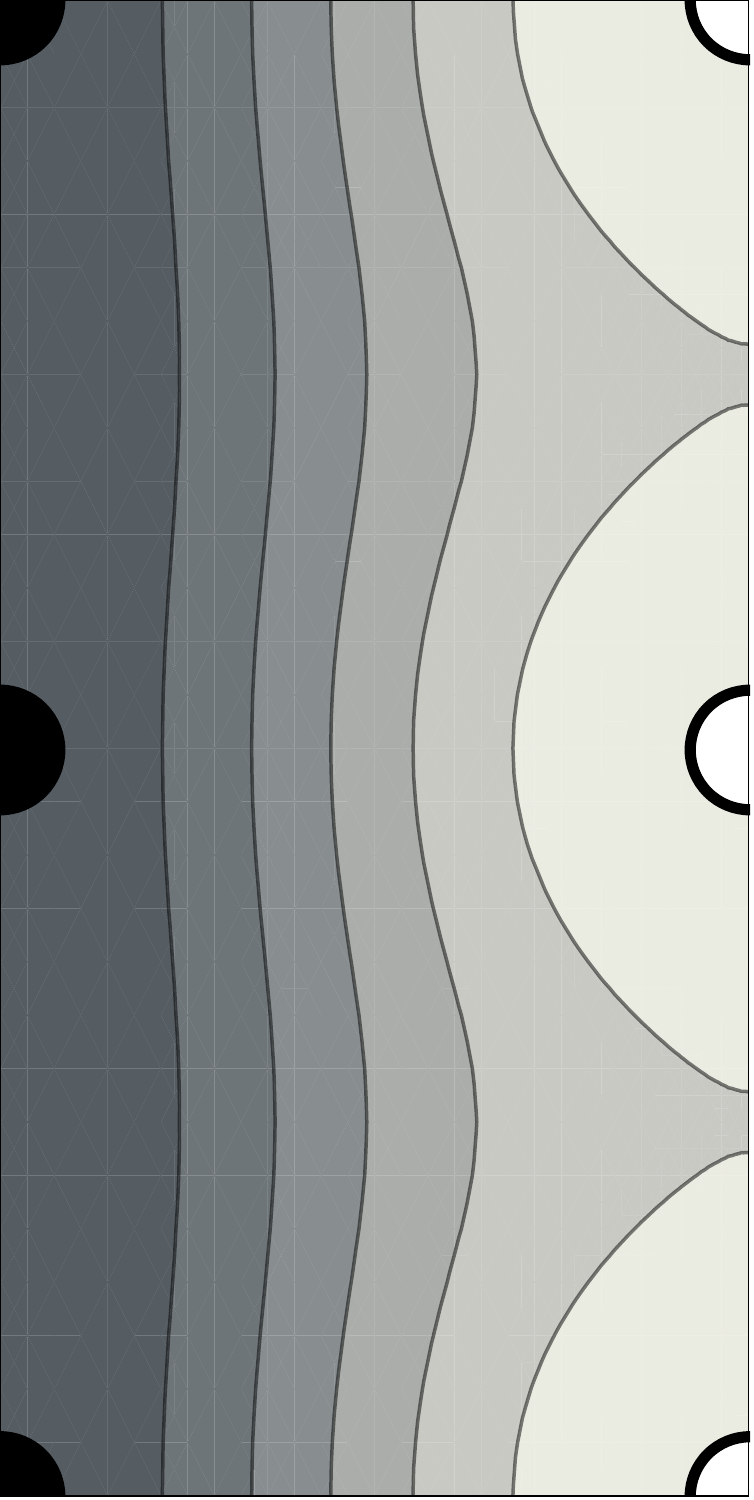} } \hspace*{1.0cm}
	\subfloat[\(k_1 = 0, k_2 = 1\)]{\includegraphics[height = .25 \textheight]{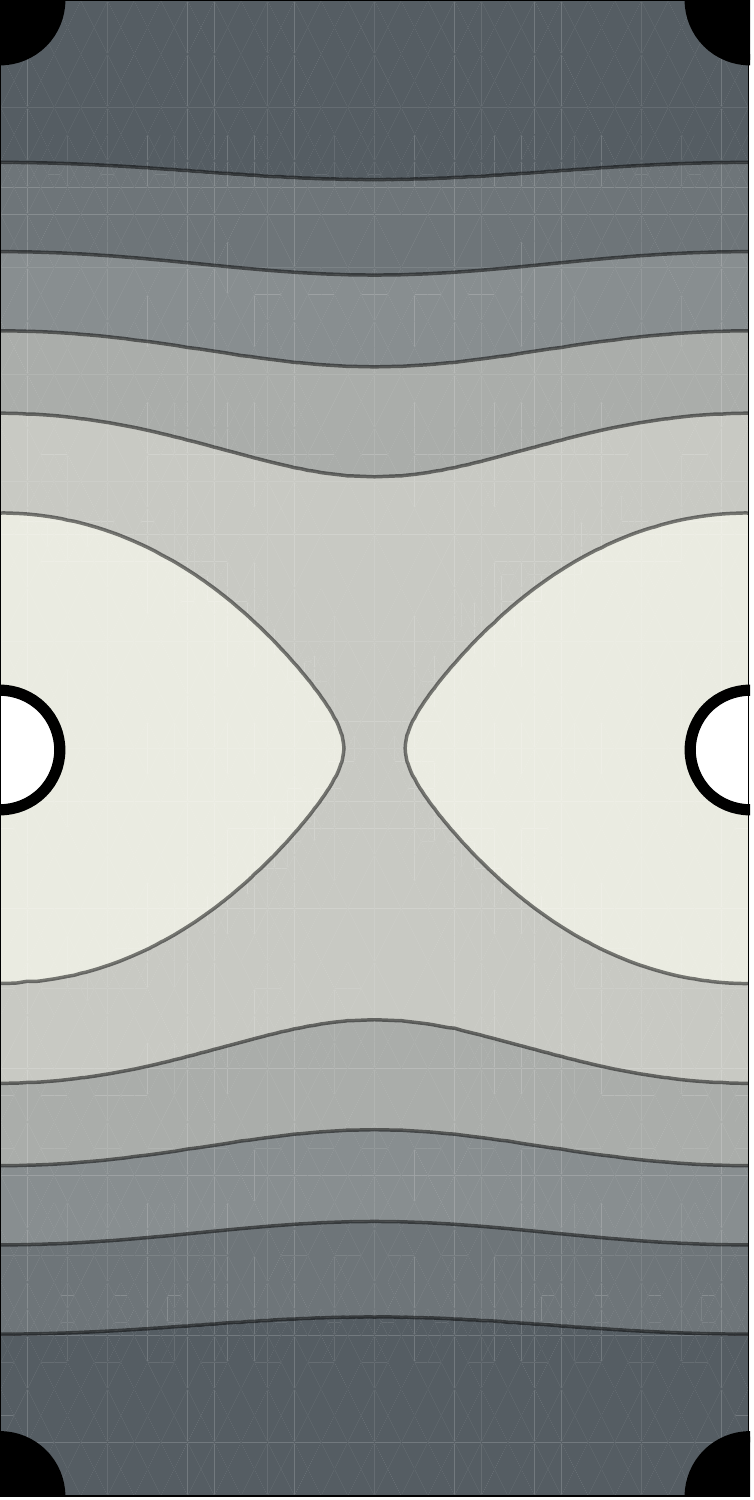} } \hspace*{1.0cm}
	\subfloat[\(k_1 = 1, k_2 = 1\)]{\includegraphics[height = .25 \textheight]{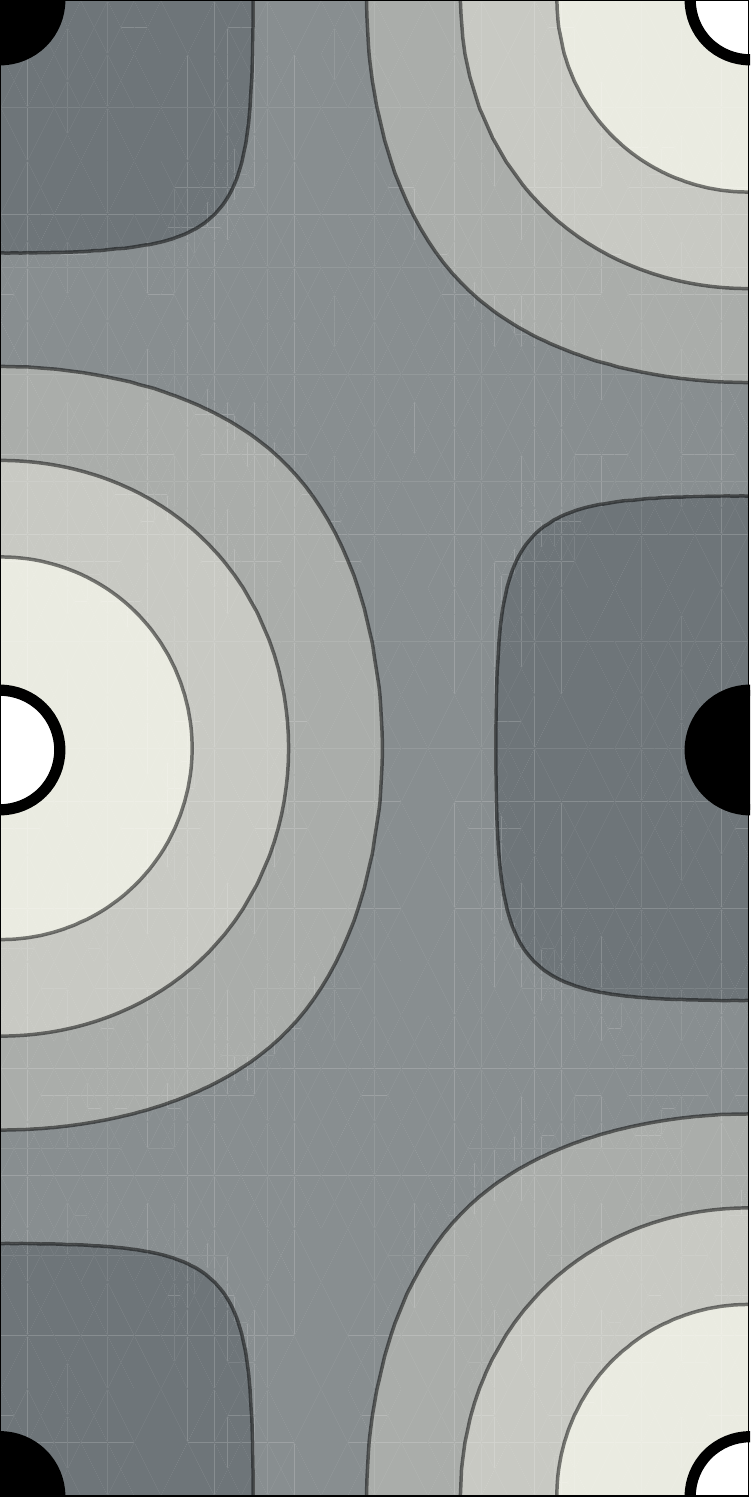} } \hspace*{0.1cm}
	\subfloat{\includegraphics[height = .25 \textheight]{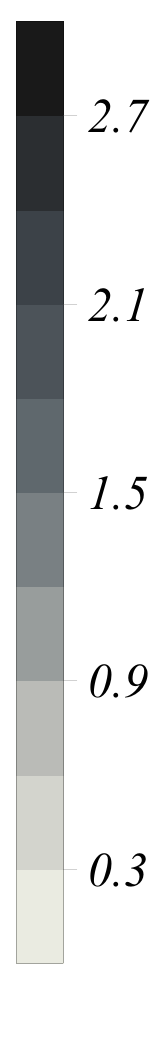}}
\caption{Absolute square of even zero-mode functions for $M=2$ and different
          values of $k_1$ and $k_2$. The white circles indicate where
          the mode functions vanish.}	
	\label{fig:M2}
	\end{center}
\end{figure}

It is straightforward to continue the discussion to larger values of
$M$. For $M=3$ one obtains a pattern similar to the one for $M=1$.
For $(k_1,k_2)=(0,0), (1,0), (0,1)$ one finds
two zero-modes, $j=0$ and $j=1$ (see Fig.~\ref{fig:M3}(a,b,c)). The zeros of the mode
functions appear at the same fixed points as for $M=1$. An interesting
new aspect is that now an even mode function also exists for
$(k_1,k_2)=(1,1)$ with zeros at three fixed points (see
Fig.~\ref{fig:M3}(d)).
\begin{figure}[t]
	\begin{center}
	\subfloat[\(k_1 = 0, k_2 = 0\)]{\includegraphics[height = .25 \textheight]{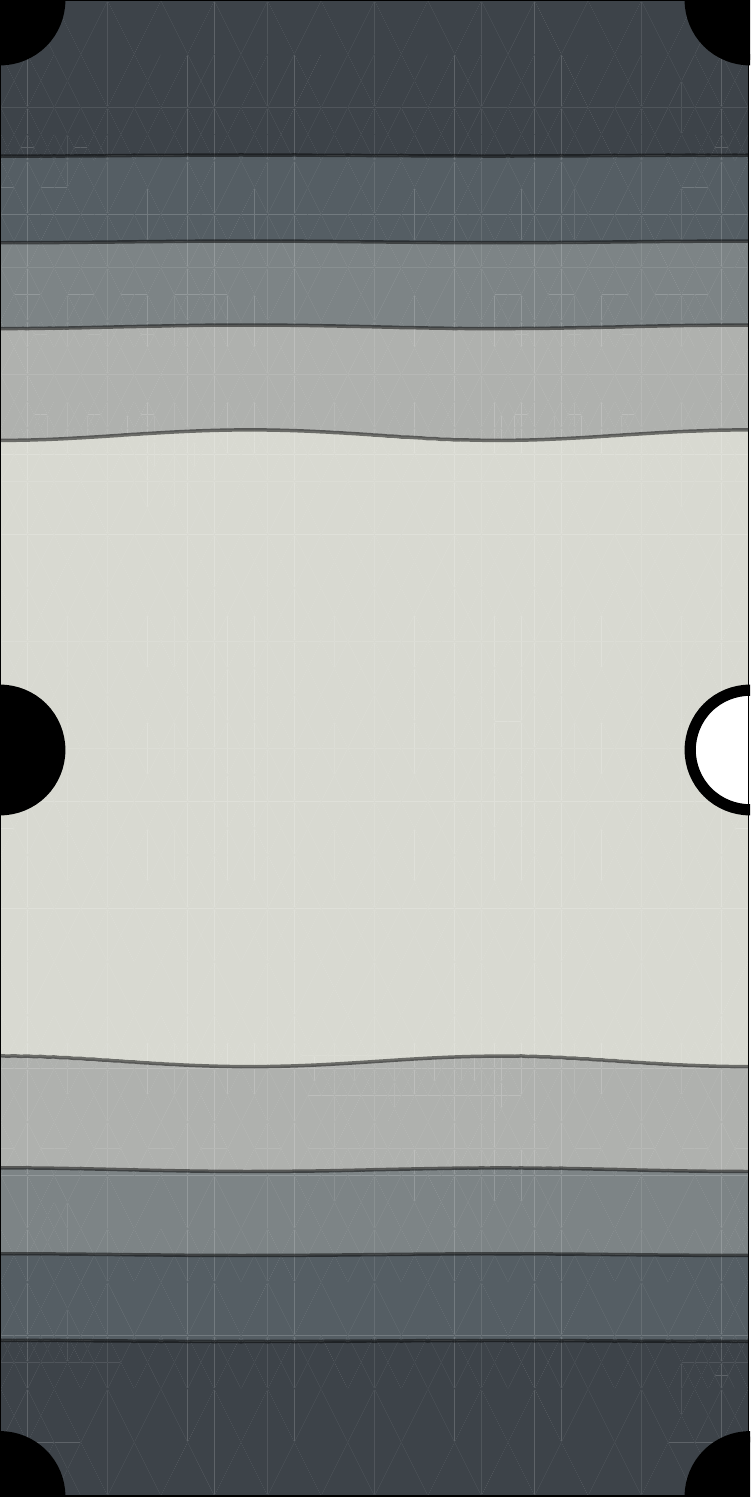} \hspace*{0.1cm} \includegraphics[height = .25 \textheight]{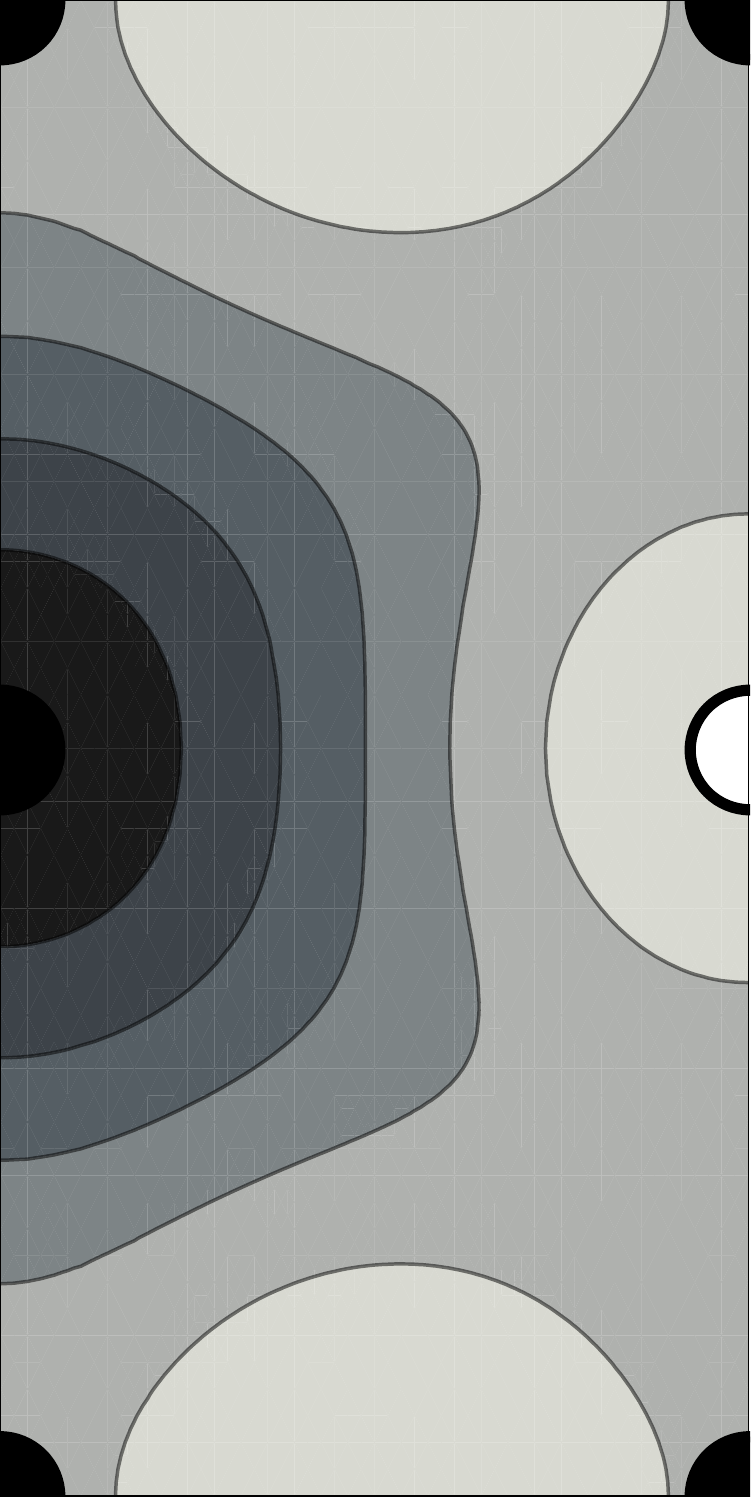} } \hspace*{1.0cm}
	\subfloat[\(k_1 = 1, k_2 = 0\)]{\includegraphics[height = .25 \textheight]{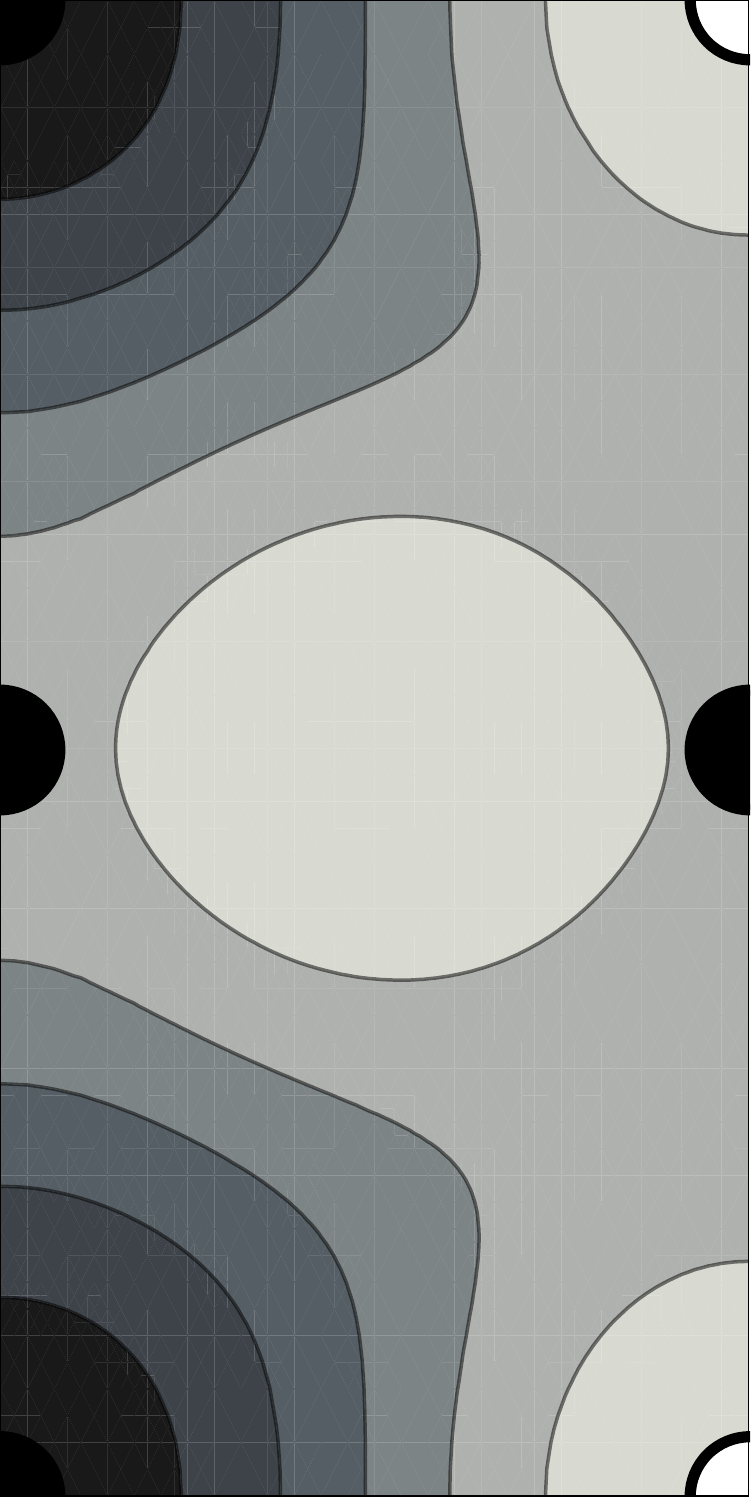} \hspace*{0.1cm} \includegraphics[height = .25 \textheight]{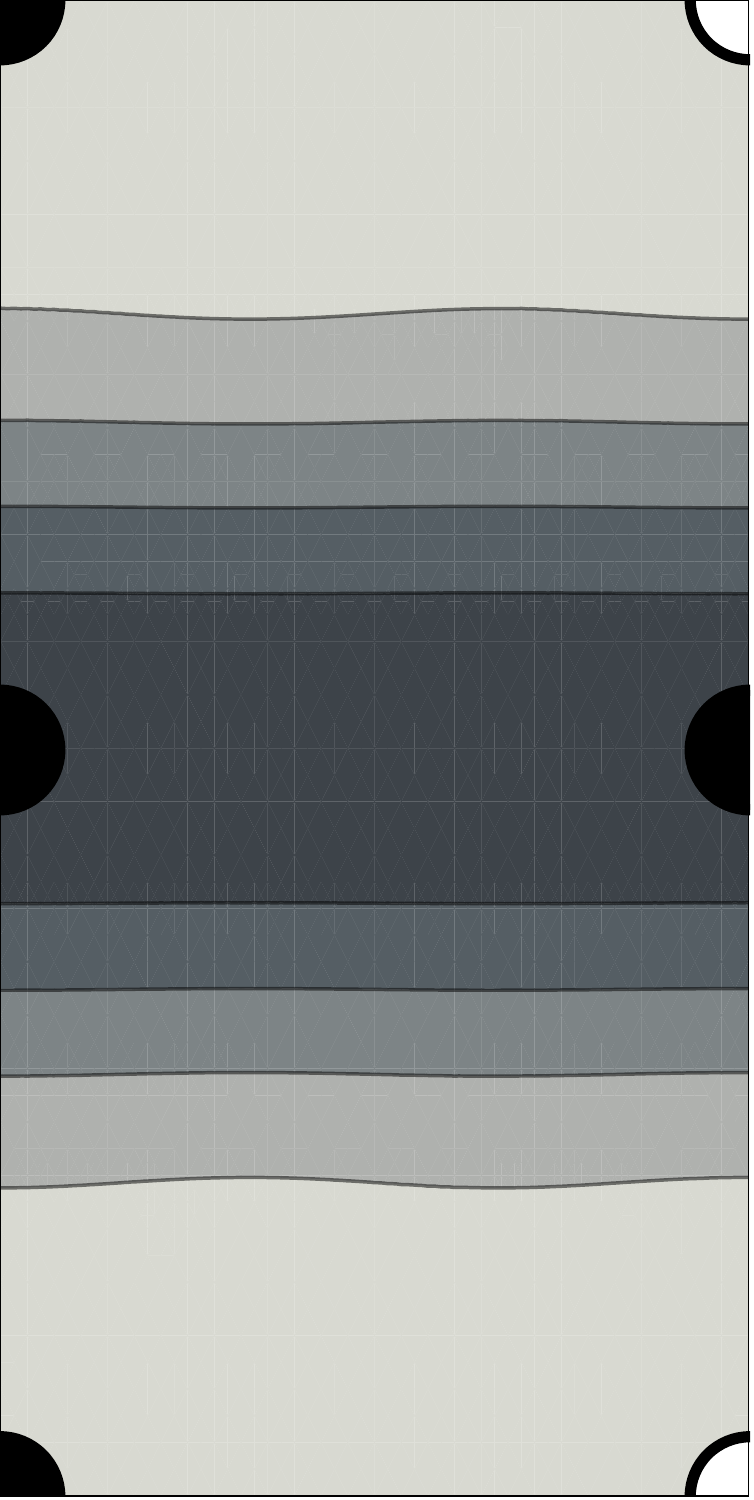} } \hspace*{0.1cm}
\\
	\subfloat[\(k_1 = 0, k_2 = 1\)]{\includegraphics[height = .25 \textheight]{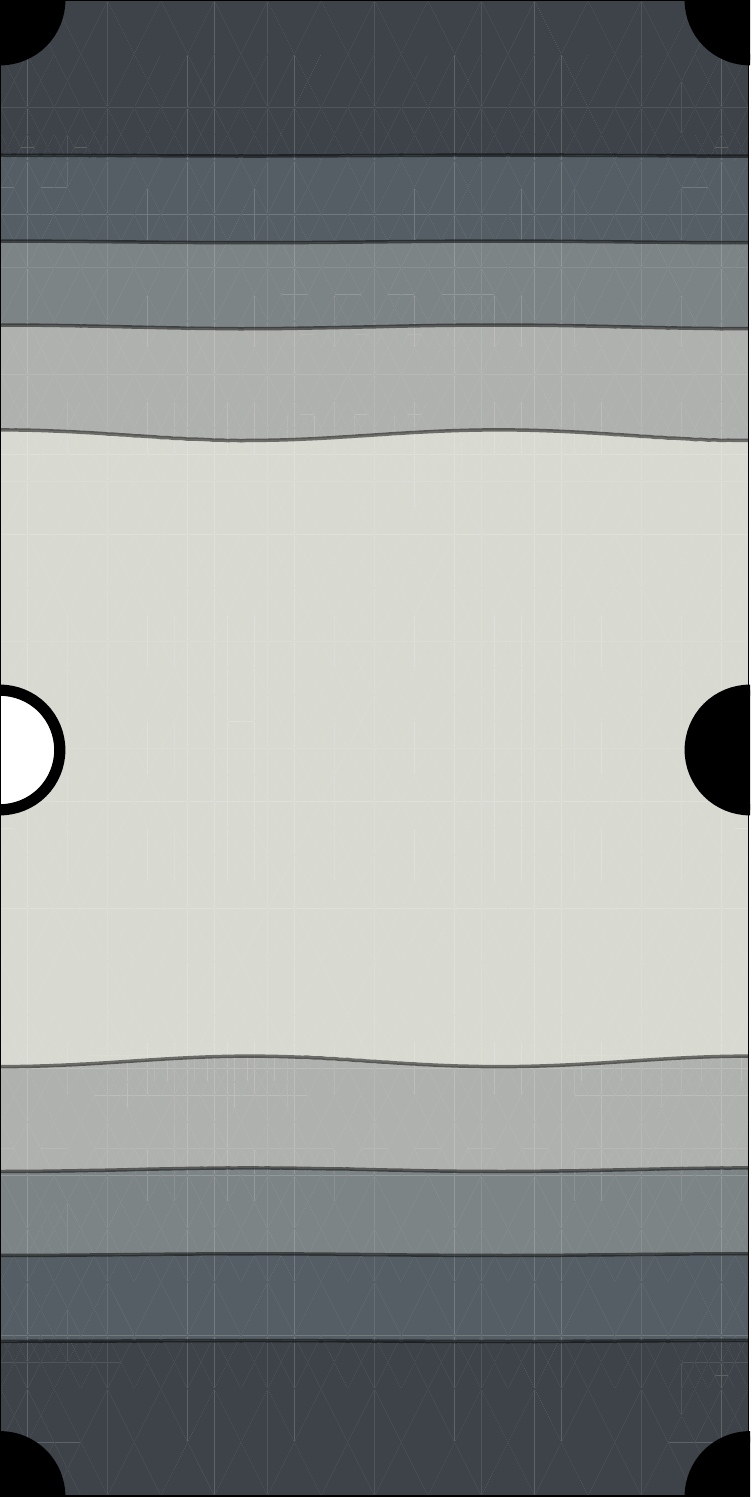} \hspace*{0.1cm} \includegraphics[height = .25 \textheight]{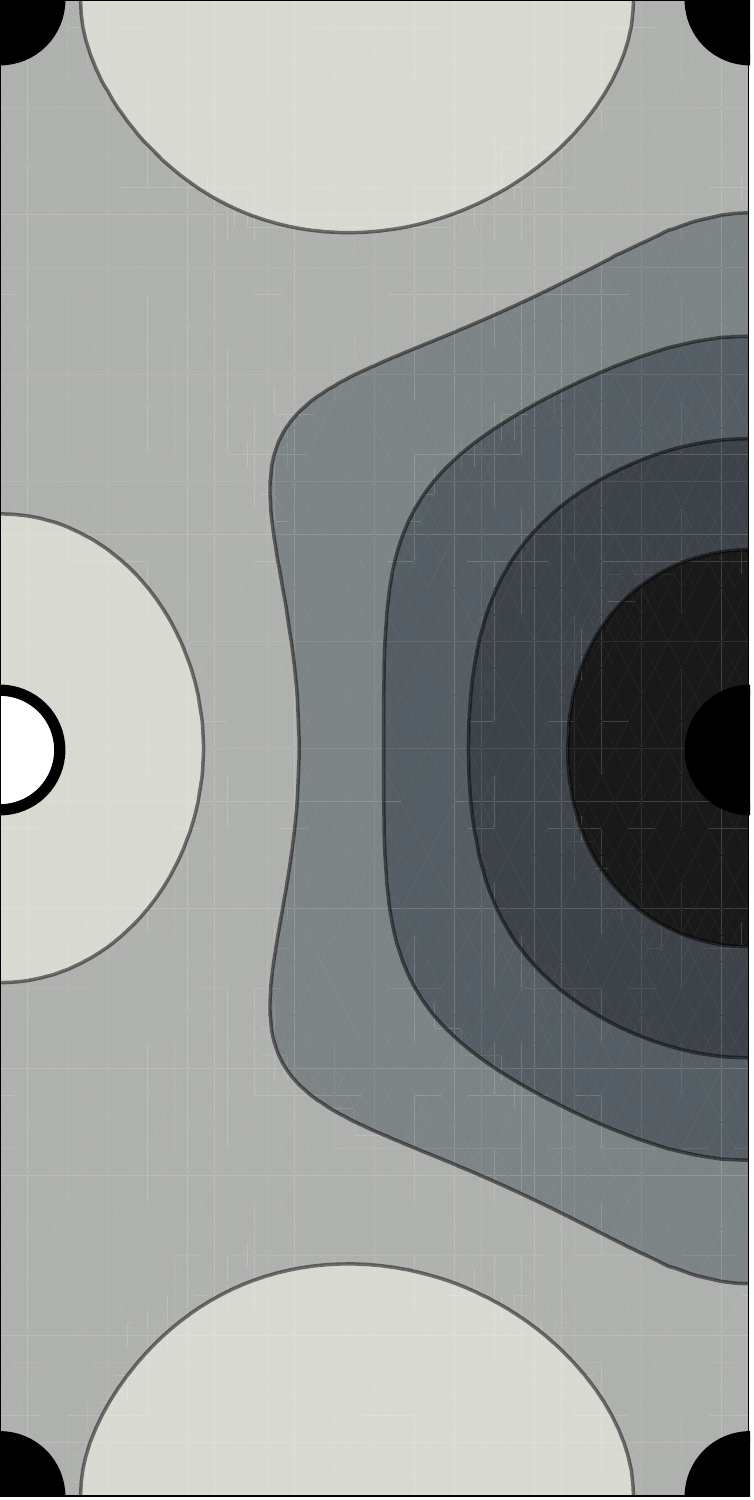} } \hspace*{1.0cm}
	\subfloat[\(k_1 = 1, k_2 = 1\)]{\includegraphics[height = .25 \textheight]{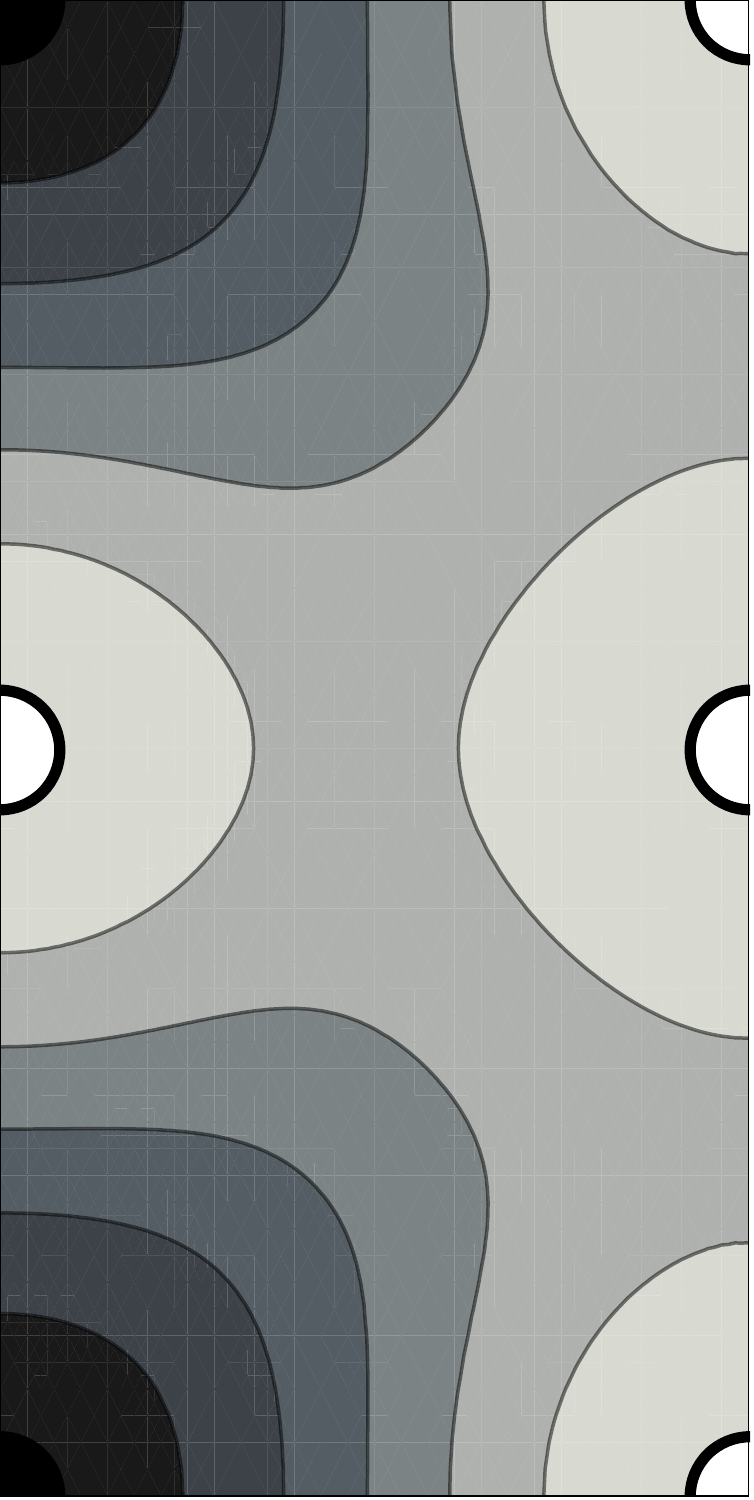} } \hspace*{0.1cm}
	\subfloat{\includegraphics[height = .25 \textheight]{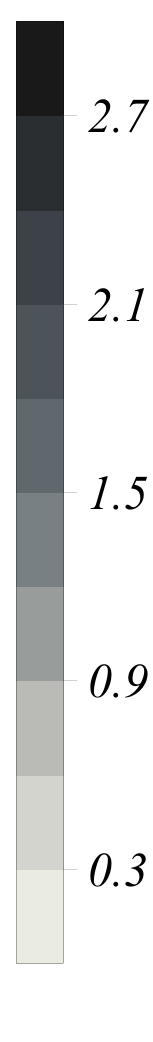}}
\caption{Absolute squre of even zero-mode functions for $M=3$ and different
          values of $k_1$ and $k_2$. The white circles indicate where
          the mode functions vanish.}	
	\label{fig:M3}	
	\end{center}
\end{figure}

Finally, the mode functions for $M=4$ are shown in
Fig.~\ref{fig:M4}. The pattern completely agrees with the case
$M=2$, especially with respect to the distribution of zeros at fixed
points. The only difference is that for each combination of $k_1,k_2$
the number of zero modes has increased by one. The mode functions
agree with those obtained in \cite{Buchmuller:2015eya}.
\begin{figure}
	\begin{center}
%	\subfloat[No flux \label{sec:wavefunctions:fig:wavefunctions_comparison_m0}]{
%		\includegraphics[height = .25 \textheight]{./wavefunctions_m0.pdf}}\hspace*{1.0cm}
	\subfloat[\(k_{1,2}=0, j = 0\) \label{sec:wavefunctions:fig:wavefunctions_comparison_m2_0}]{
		\includegraphics[height = .25 \textheight]{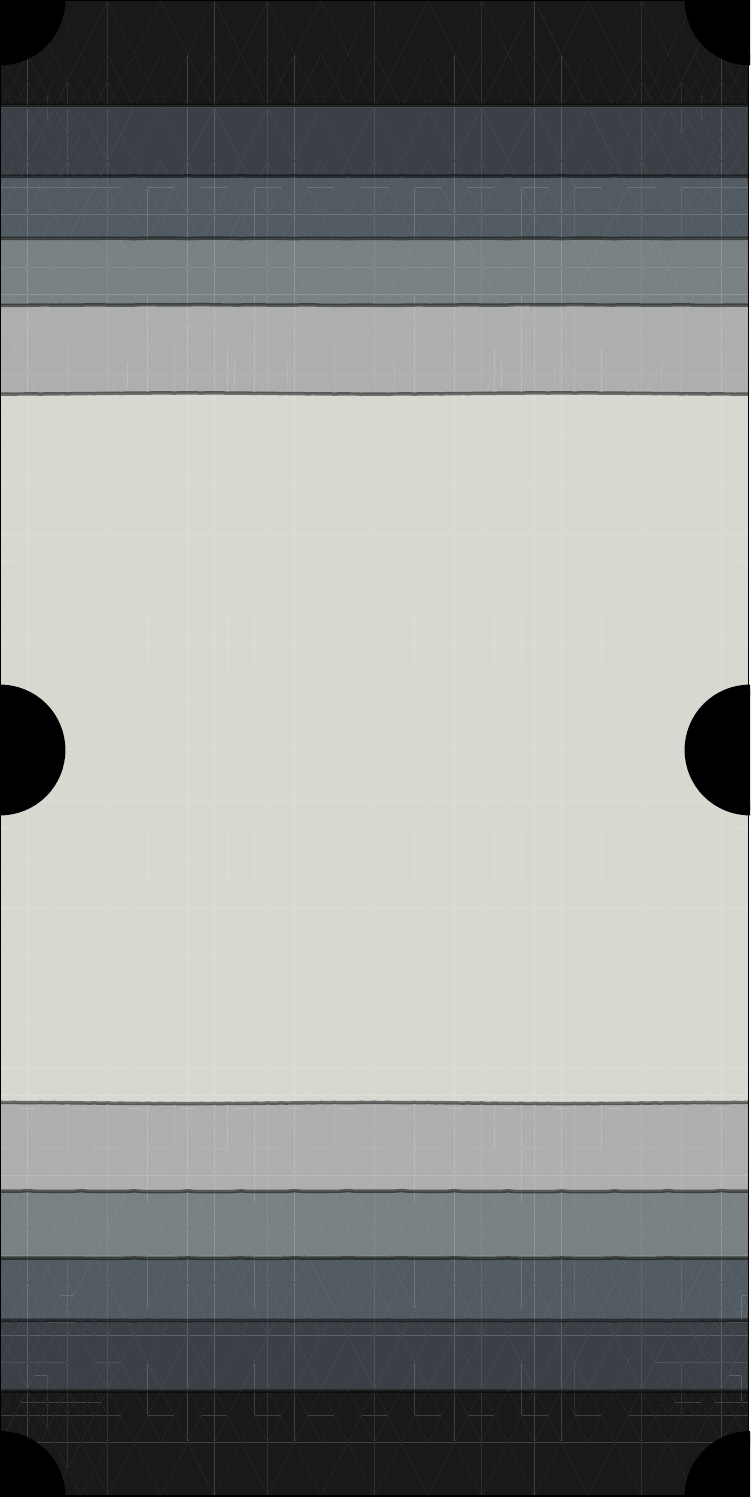}}\hspace*{0.1cm}
	\subfloat[\(k_{1,2}=0, j = 1\) \label{sec:wavefunctions:fig:wavefunctions_comparison_m2_1}]{
		\includegraphics[height = .25 \textheight]{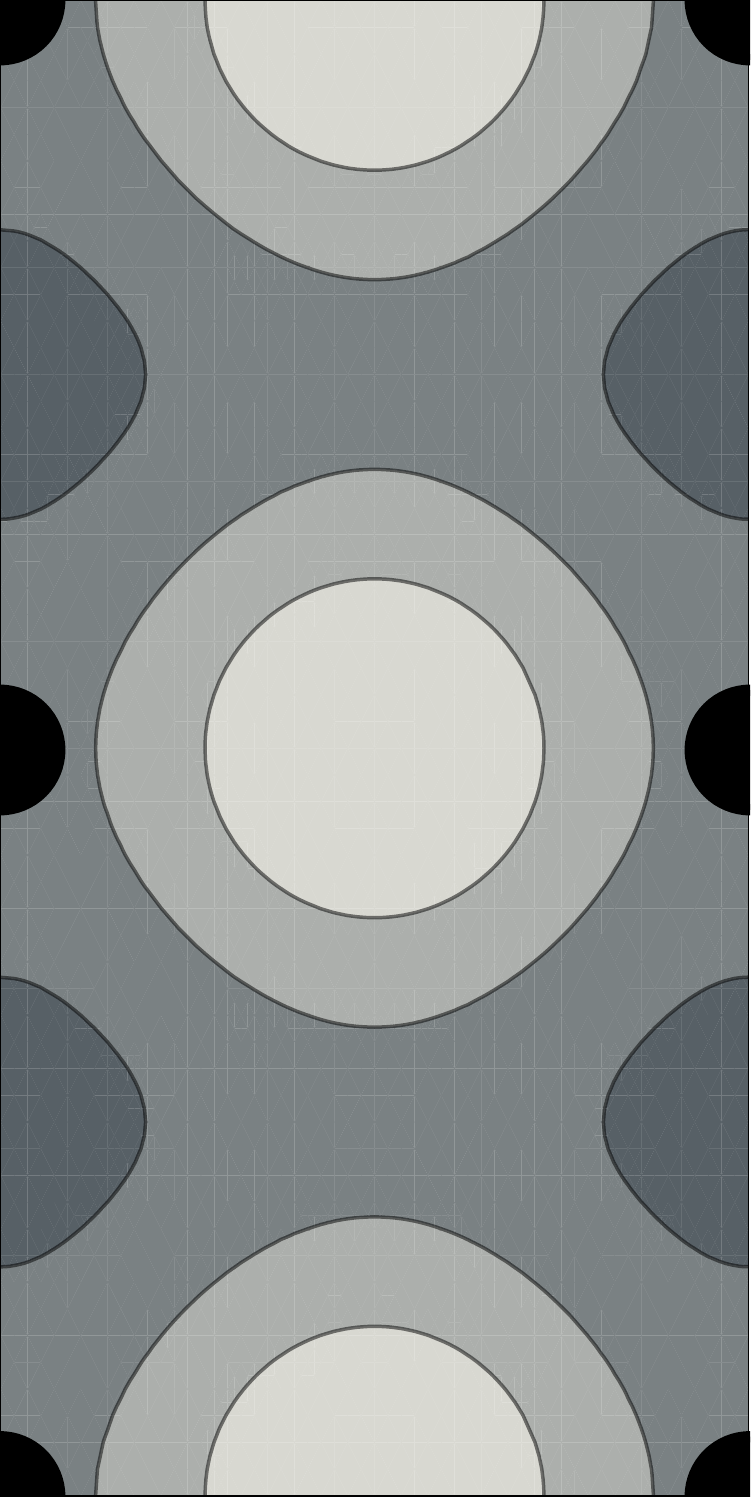}}\hspace*{0.1cm}
	\subfloat[\(k_{1,2} = 0, j = 2\) \label{sec:wavefunctions:fig:wavefunctions_comparison_m2_2}]{
		\includegraphics[height =
                .25\textheight]{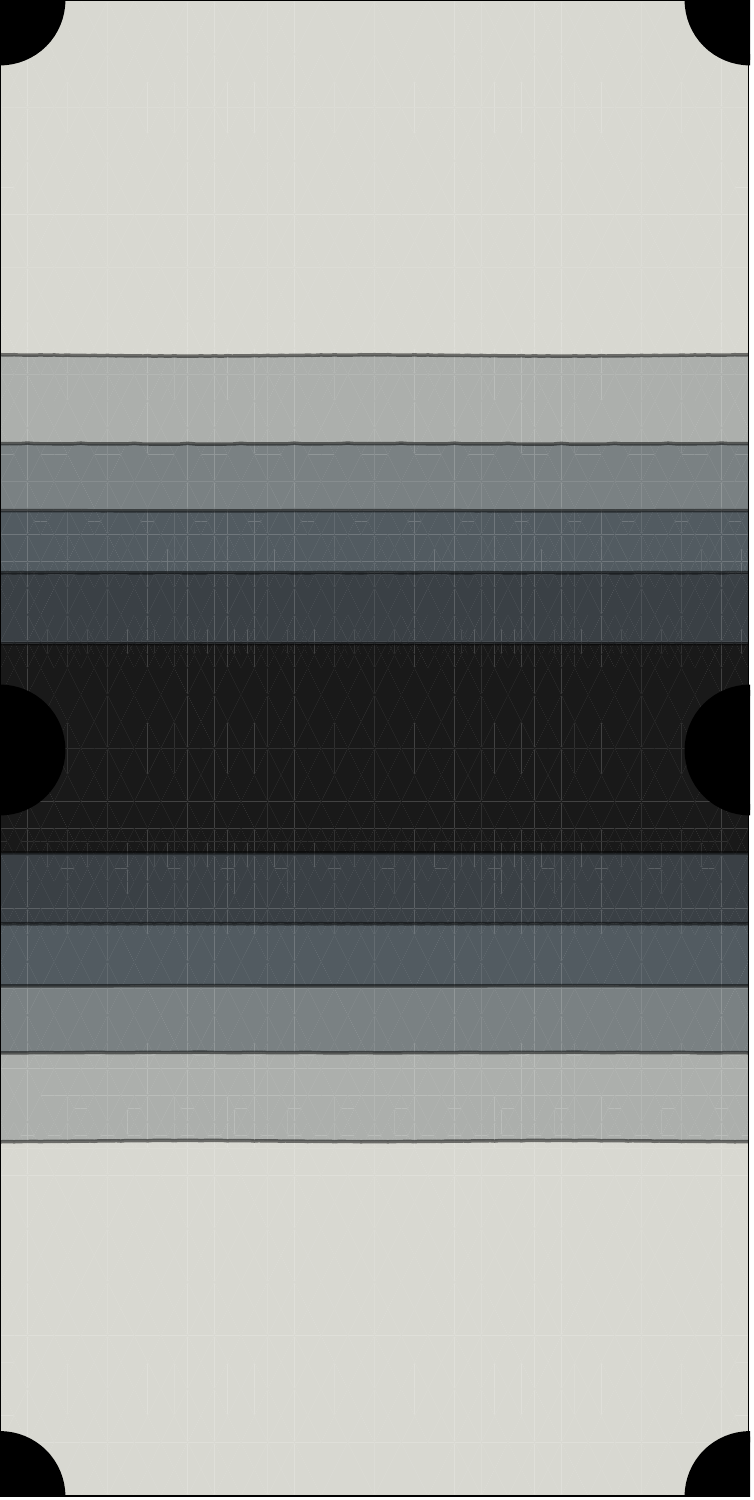}}\\
%\hspace*{1.0cm}
	\subfloat[\(k_1 = 1, k_2 = 0\)]{\includegraphics[height = .25\textheight]{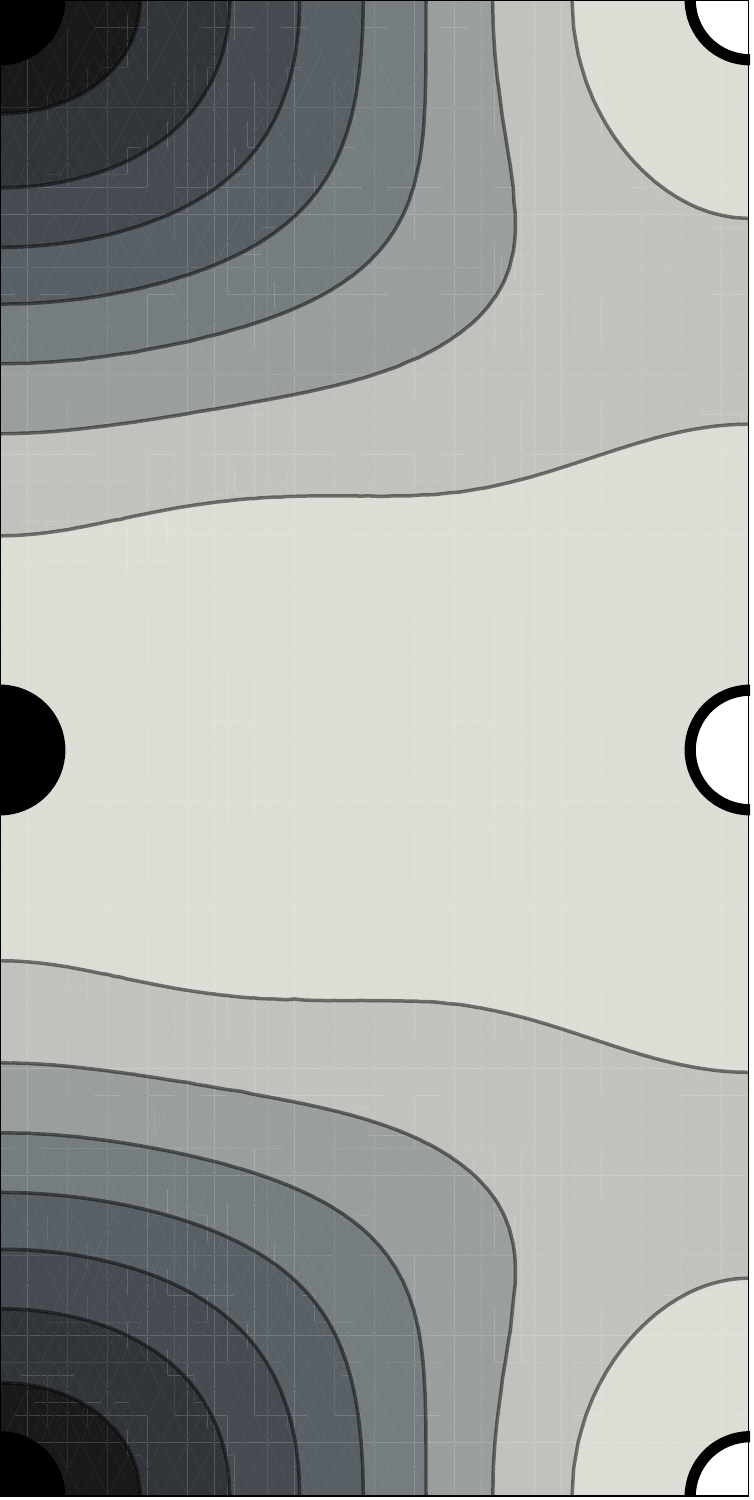} 
         \hspace*{0.1cm} \includegraphics[height = .25 \textheight]{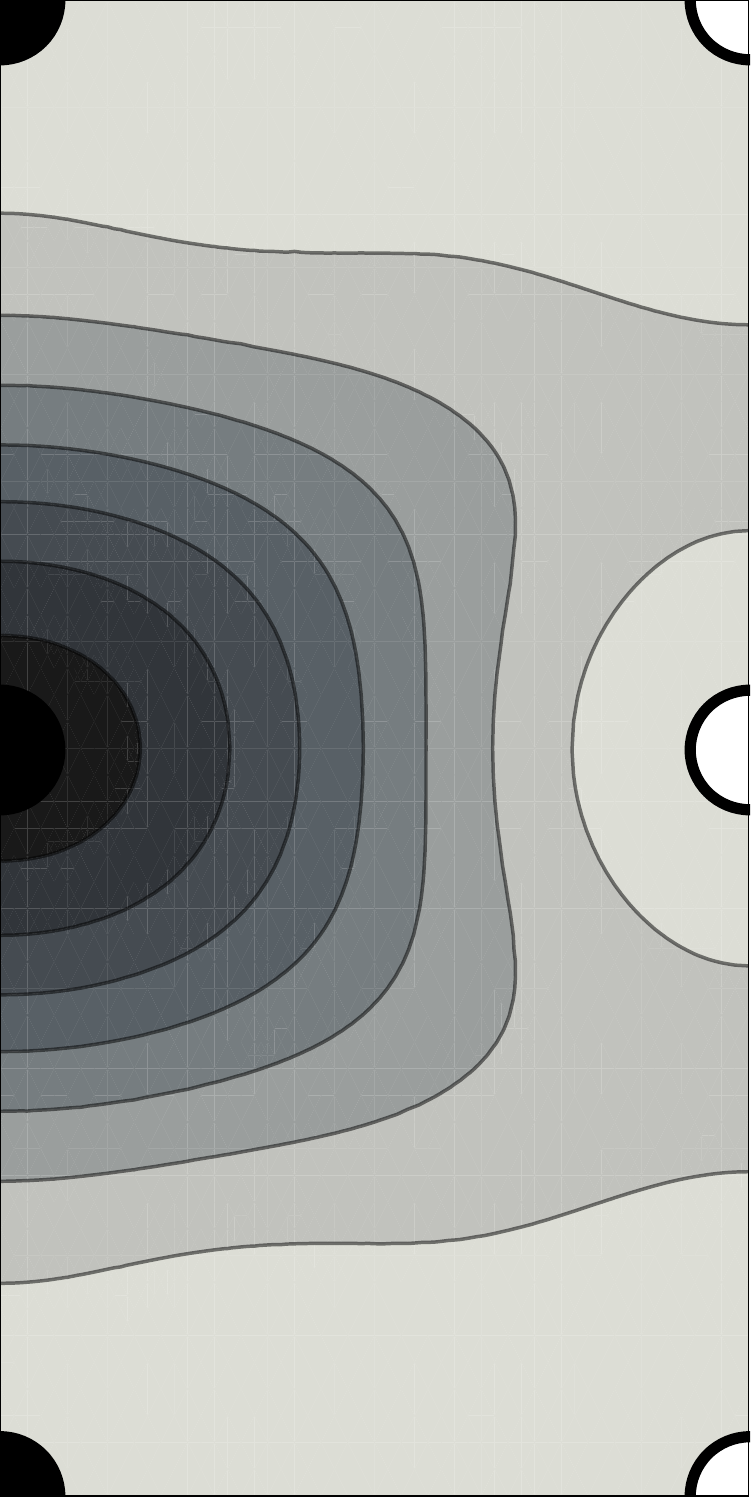} } \hspace*{1.0cm}
	\subfloat[\(k_1 = 0, k_2 = 1\)]{\includegraphics[height = .25\textheight]{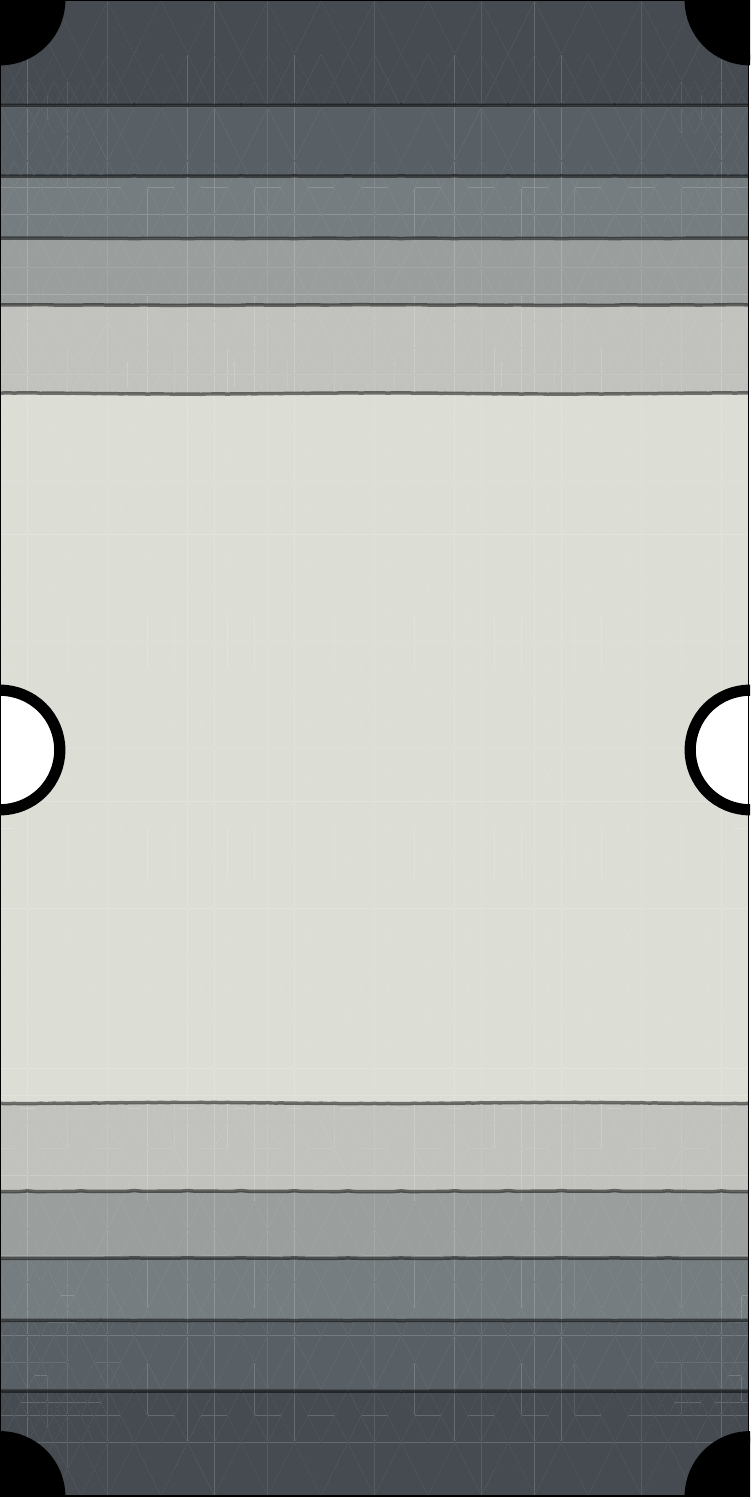} 
        \hspace*{0.1cm} \includegraphics[height =
        .25\textheight]{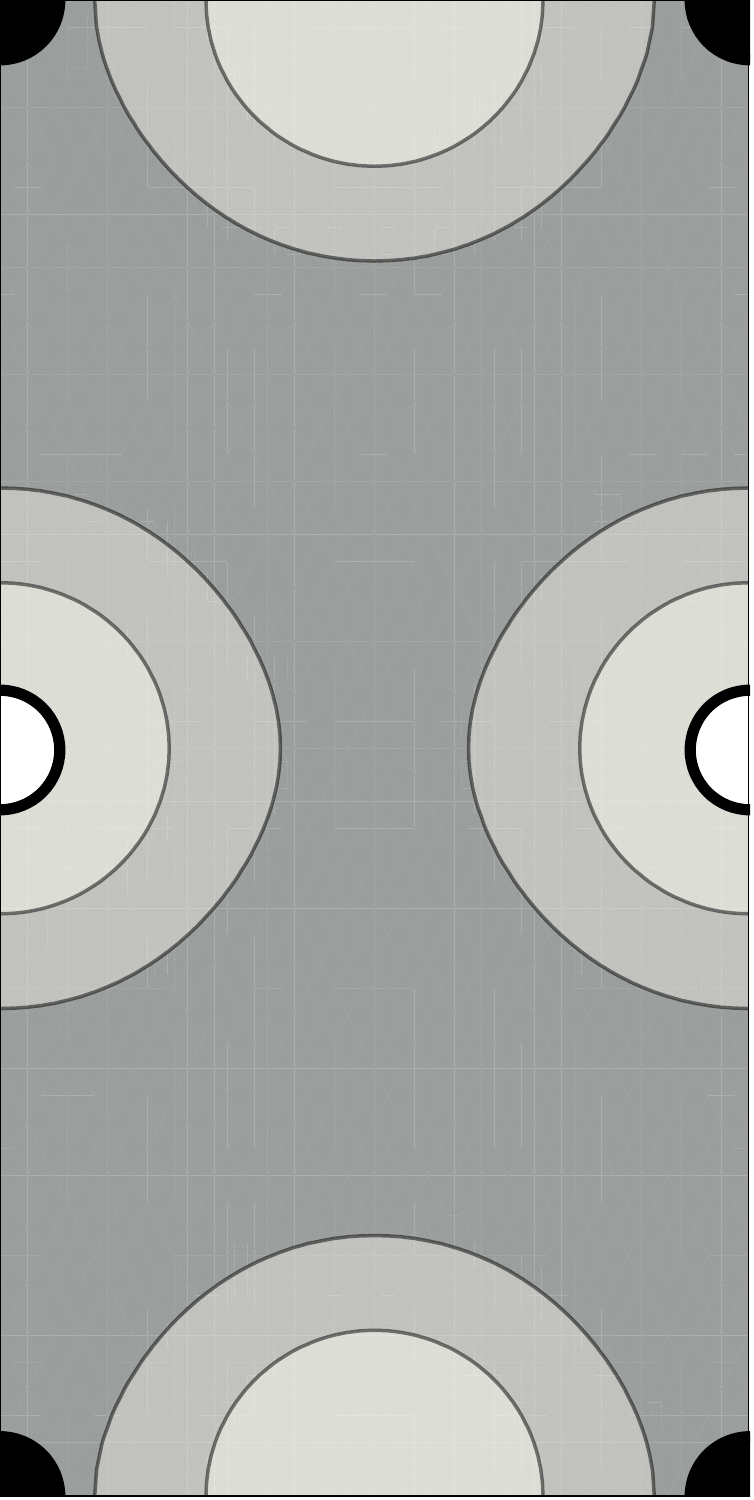} } 
\\
	\subfloat[\(k_1 = 1, k_2 = 1\)]{\includegraphics[height = .25 \textheight]{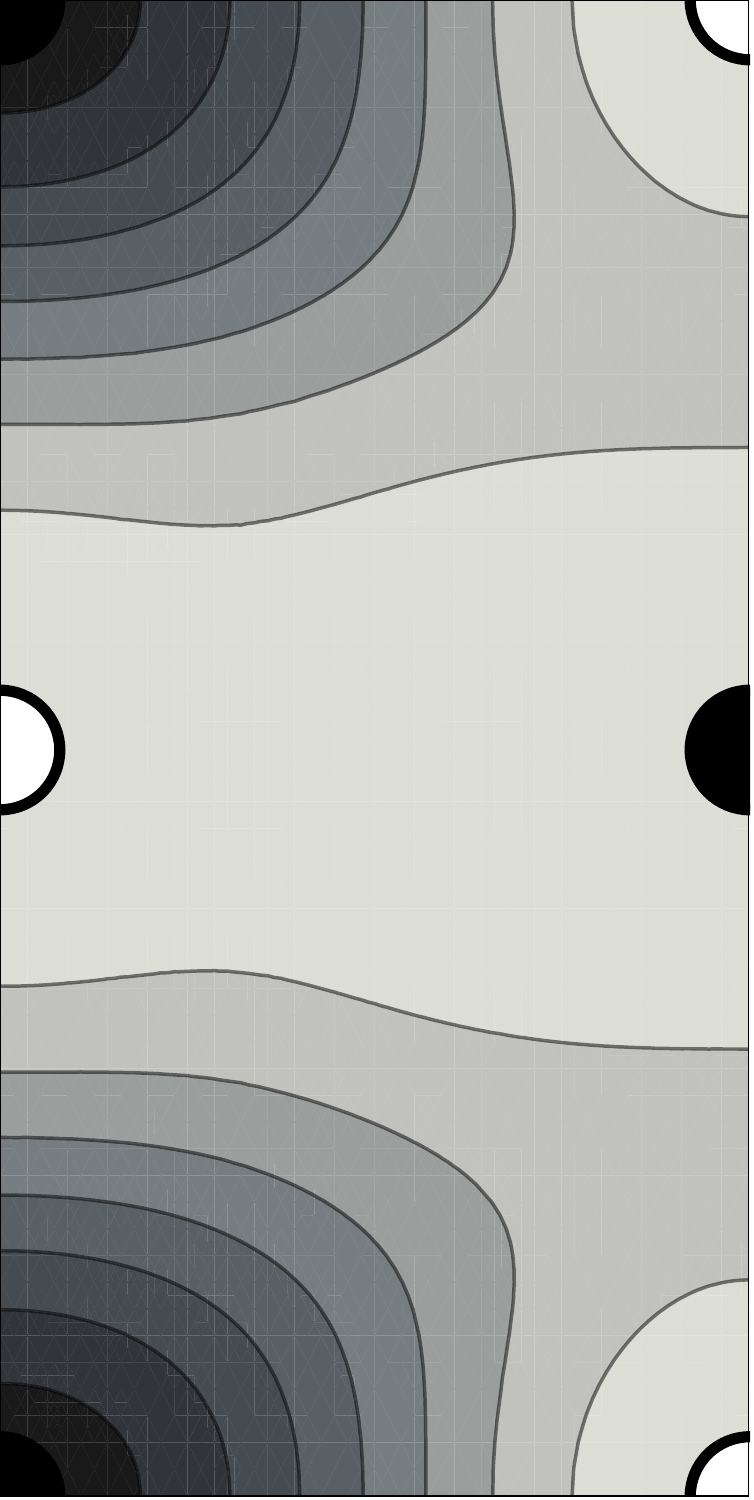} \hspace*{0.1cm} \includegraphics[height = .25 \textheight]{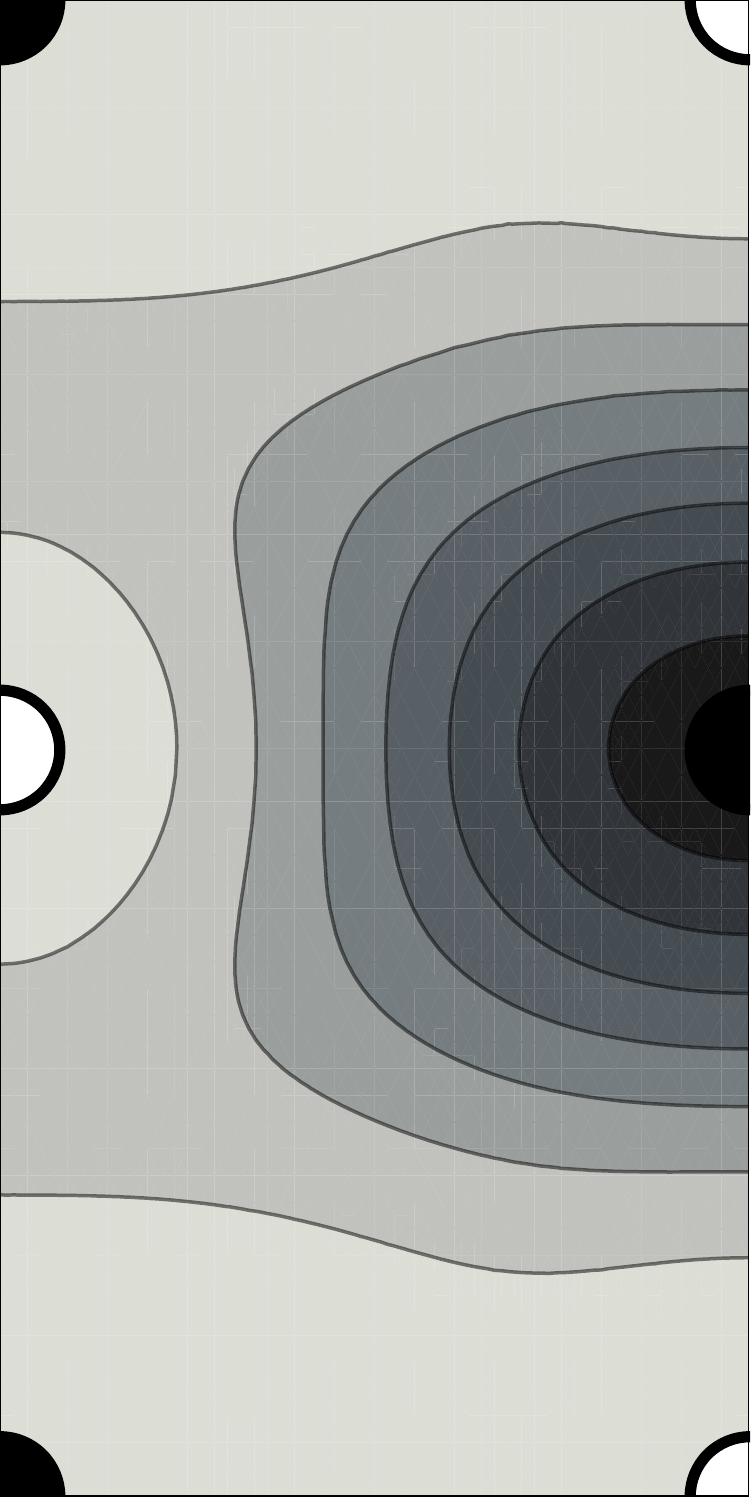} } \hspace*{0.1cm}
	\subfloat{\includegraphics[height = .25
          \textheight]{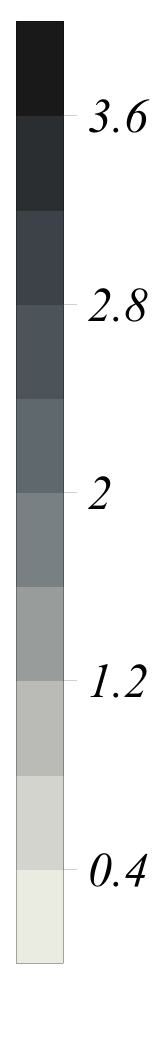}}
\caption{Absolute square of even zero-mode functions for $M=4$ and different
          values of $k_1$ and $k_2$. The white circles indicate where
          the mode functions vanish.}	
	\label{fig:M4}	
	\end{center}
\end{figure}

\subsection{Untwisted wave functions}
\label{subsec:fields_untwisted}

In \cite{Buchmuller:2015eya} it has been pointed out that the zeros of
the zero-mode functions are related to negative Wilson line integrals
around the orbifold fixed points, which might be interpreted in terms
of localized flux, envoking Stokes' theorem. Such localized flux has
previously been discussed in connection with fixed point anomalies \cite{Scrucca:2004jn,vonGersdorff:2006nt}
and also with respect to localized Fayet-Iliopoulos terms
\cite{Lee:2003mc}. In the latter case a consistent description of the
localized flux is obtained by means of the Green's function on a torus,
whose gradient yields a singular vector field, 
as discussed in Section~\ref{subsec:fields_singular}.

From the torus Green's function one obtains a localized flux together with a
bulk flux (see \eqref{Fsing}),
\begin{align}
F_{12} = -2\pi c~\delta^2(y - \zeta) + 2\pi c\,.
\end{align}
Hence, in our convention for the bulk flux, $qf = qF_{12} = 2\pi M >
0$, one has $c = M/q$. The torus Green's function $G(z-\zeta,\tau)$ has been used to
describe zero-modes of a charged bulk field \cite{Lee:2003mc}.
Indeed, for a vector field $A_z =i\dc G$, the wave function
\begin{align}\label{WFsing}
\chi_\zeta(z) \propto e^{qG(z-\zeta,\tau)}
\end{align}
solves the field equation
\begin{align}\label{Szeromode}
(\dc + iq A_z) \chi_\zeta(z) = 0\,.
\end{align}
Using Eqs.~\eqref{Dcomplex} and \eqref{Acomplex} it is apparent that
Eq.~\eqref{Szeromode} is nothing but the field equation
\eqref{Rzeromode}, with the regular vector field of
Section~\ref{subsec:fields_regular} replaced by the singular vector
field $A_z =i\dc G$. Using \eqref{torusGreen}, the zero-mode function \eqref{WFsing} can
be written as
\begin{align}\label{WFzeta}
\chi_\zeta(z) \propto |\vartheta_1(z-\zeta,\tau)|^M e^{-\frac{\pi M}{\tau_2}(\text{Im}(z-\zeta))^2}\,,
\end{align}
where $\zeta$ denotes the Green's function's singularity.
The asympotic behaviour of the wave function close to 
the singularity is given by
\begin{align}\label{zerosM1}
\chi_\zeta(z) \underset{z\rightarrow \zeta}\propto |z-\zeta|^M\,.
\end{align}
Clearly, the wave function is normalizable for $M > 0$.

Wave functions for bulk flux $M=1$ can be obtained from Green's
functions \mbox{$G(z-\zeta_i)$}, where ${\zeta_i}$ are the four
orbifold fixed points. From Eq.~\eqref{WFzeta}, and using relations among
theta-functions listed in Appendix~A,
one obtains for the fixed points $\zeta_2$,
$\zeta_3$ and $\zeta_4$:
\begin{equation}\label{SM1}
%\begin{split}
\chi_\zeta(z) = \left\{\begin{array}{l} |e^{-i\pi\bar{\tau}y_2^2
        -i\pi\bar{z}} \vartheta(\bar{z}+\bar{\tau}/2,-\bar{\tau})|\,,\quad
\zeta_2 = 1/2\,\, (k_1=1, k_2=0)\\
|e^{-i\pi\bar{\tau}y_2^2} \vartheta(\bar{z}-1/2,-\bar{\tau})|\,,\quad
\zeta_3 = \tau/2\,\, (k_1=0, k_2=1)\\
|e^{-i\pi\bar{\tau}y_2^2}
\vartheta(\bar{z}+\bar{\tau}/2,-\bar{\tau})|\,,\quad
\zeta_4 = (1+\tau)/2\,\,(k_1=0, k_2=0)\,.
\end{array}\right.
\end{equation}
These wave functions are precisely the modulus of the $M=1$ zero-mode
functions given in Eq.~\eqref{RM1} (see Fig.~\ref{fig:M1}). This
connection is not unexpected since both sets of functions are zero
modes for the same bulk flux, $M=1$, just for different choices of the
vector field. By construction, the zeros are obvious for the untwisted
wave functions, see Eq.~\eqref{zerosM1}. The three fixed points
$\zeta_2$, $\zeta_3$ and $\zeta_4$ correspond to three pairs
$(k_1,k_2)$, which are given in brackets in Eq.~\eqref{SM1}. The wave
function $\chi_{\zeta_1}$ corresponds to the odd zero mode among
the untwisted wave functions.

At first sight the appearance of the modulus in Eq.~\eqref{SM1} is
surprizing. As a consequence, the untwisted wave functions transform
trivially under translations by the lattice vectors $\lambda_1$ and
$\lambda_2$. This, however, is expected. Like the Green's function,
the vector field is invariant under translations by lattice
vectors. Therefore, no non-trivial transition functions occur, and the
boundary conditions are trivial. The situation is different for the
regular vector field, where the non-trivial transition functions lead
to twisted boundary conditions, which can be satisfied by appropriate
phase factors.

The case $M=2$ can be treated in a similar way. Two bulk flux quanta
can be obtained by localizing one quantum at two fixed points. The
vector field is now the sum $A_z =i (\dc G(z-\zeta_1,\tau) + \dc
G(z-\zeta_2,\tau))$, and the wave function is the product
\begin{align}\label{productWF}
\chi_{\zeta_1,\zeta_2} &= \chi_{\zeta_1}(z) \chi_{\zeta_2} (z) \propto
e^{q(G(z-\zeta_1,\tau) + G(z-\zeta_2,\tau))} \nonumber\\
& \propto|\vartheta_1(z-\zeta_1,\tau) \vartheta_1(z-\zeta_2,\tau) | 
e^{-\frac{\pi}{\tau_2}((\text{Im}(z-\zeta_1))^2 +(\text{Im}(z-\zeta_2))^2)}\,.
\end{align}
Using Eq.~\eqref{addition} the product of theta-functions can
be witten as
\begin{align}\label{product_theta}
\vartheta_1(z-\zeta_1,\tau)\ &\vartheta_1(z-\zeta_2,\tau) \nonumber\\
&=\vartheta\!\begin{bmatrix} 1/2 \\ 0 \end{bmatrix}\! (z-\zeta_1+1/2, \tau)\
\vartheta\!\begin{bmatrix} 1/2 \\ 0 \end{bmatrix}\! (z-\zeta_2+1/2, \tau) \nonumber\\
&=\sum_{m=0}^{1}\vartheta\!\begin{bmatrix} (m+1)/2 \\ 0 \end{bmatrix}\!
(2z-\zeta_1-\zeta_2 +1, 2\tau)\
\vartheta\!\begin{bmatrix} m/2 \\ 0 \end{bmatrix}\! (\zeta_1-\zeta_2, 2\tau)\,.
\end{align}
The second theta-function in the last line does not depend on
$z$. Hence, the product of $\vartheta_1$-functions can be expressed as
a linear combination of $z$-dependent theta-functions. A look at
Fig.~\ref{fig:M2} suggests to consider the combination of fixed points
$(\zeta_2,\zeta_4)$, $(\zeta_3,\zeta_4)$ and
$(\zeta_2,\zeta_3)$. After some algebra, one then obtains from
Eqs.~\eqref{productWF} and \eqref{product_theta}:
 \begin{align}\label{twozeta}
%\begin{split}
\chi_{\zeta,\zeta'} \propto 
e^{-2\pi \tau_2 y_2^2}\left\{ 
\begin{array}{l}
|e^{-i\pi \bar{z}}\ \vartheta (2\bar{z} + \bar{\tau}/2,
-2\bar{\tau}) + e^{i\pi \bar{z}}\ \vartheta (2\bar{z} -
\bar{\tau}/2,-2\bar{\tau})|\,,\\
\hspace*{1.5cm} 
\zeta_2, \zeta_4\quad (k_1=1, k_2=0)\\
|\vartheta (2\bar{z} - 1/2,-2\bar{\tau})|\,, 
\quad\zeta_3, \zeta_4\quad (k_1=0, k_2=1)\\
|e^{-i\pi \bar{z}}\ \vartheta (2\bar{z} + (\bar{\tau}-1)/2,
-2\bar{\tau}) + e^{i\pi \bar{z}}\ \vartheta (2\bar{z} -
(\bar{\tau}-1)/2,-2\bar{\tau})|\,,\\
\hspace{1.5cm}
\zeta_2,\zeta_3\quad (k_1=1, k_2=1)\,. 
\end{array}\right.
%\end{split}
\end{align}
These wave functions are again the modulus of the $M=2$ twisted
zero-mode functions listed in Eq.~\eqref{RzeroM2_2}. There are three
more untwisted zero-mode functions corresponding to the pairs of fixed
points $(\zeta_1,\zeta_2)$, $(\zeta_1,\zeta_3)$ and
$(\zeta_1,\zeta_4)$. They correspond to the three odd twisted
zero modes that are not shown in Fig.~\ref{fig:M2}.

\begin{figure}[t]
	\begin{center}
	\subfloat[\(\chi^2_{\zeta_1}\)]{\includegraphics[height = .25 \textheight]{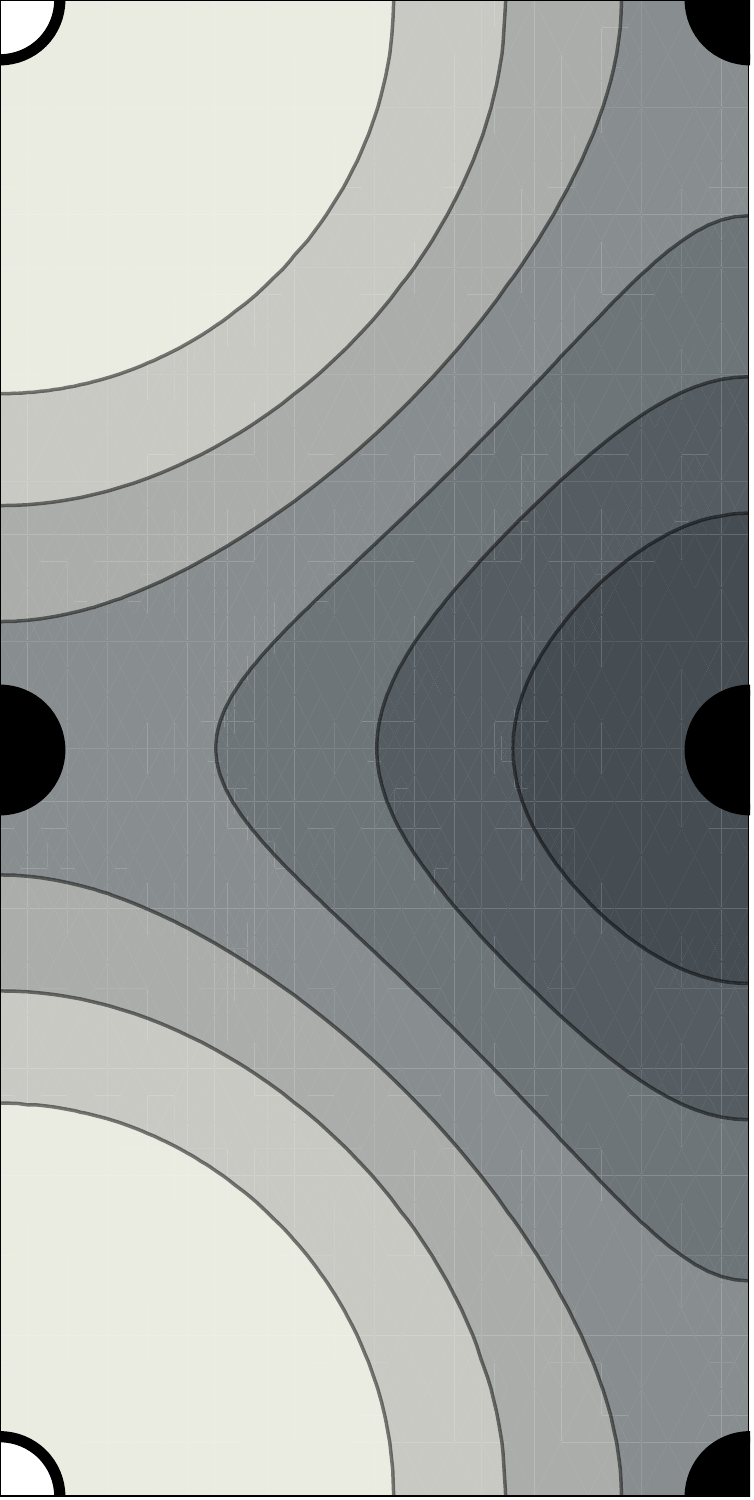} } \hspace*{1.0cm}
	\subfloat[\(\chi^2_{\zeta_2}\)]{\includegraphics[height = .25 \textheight]{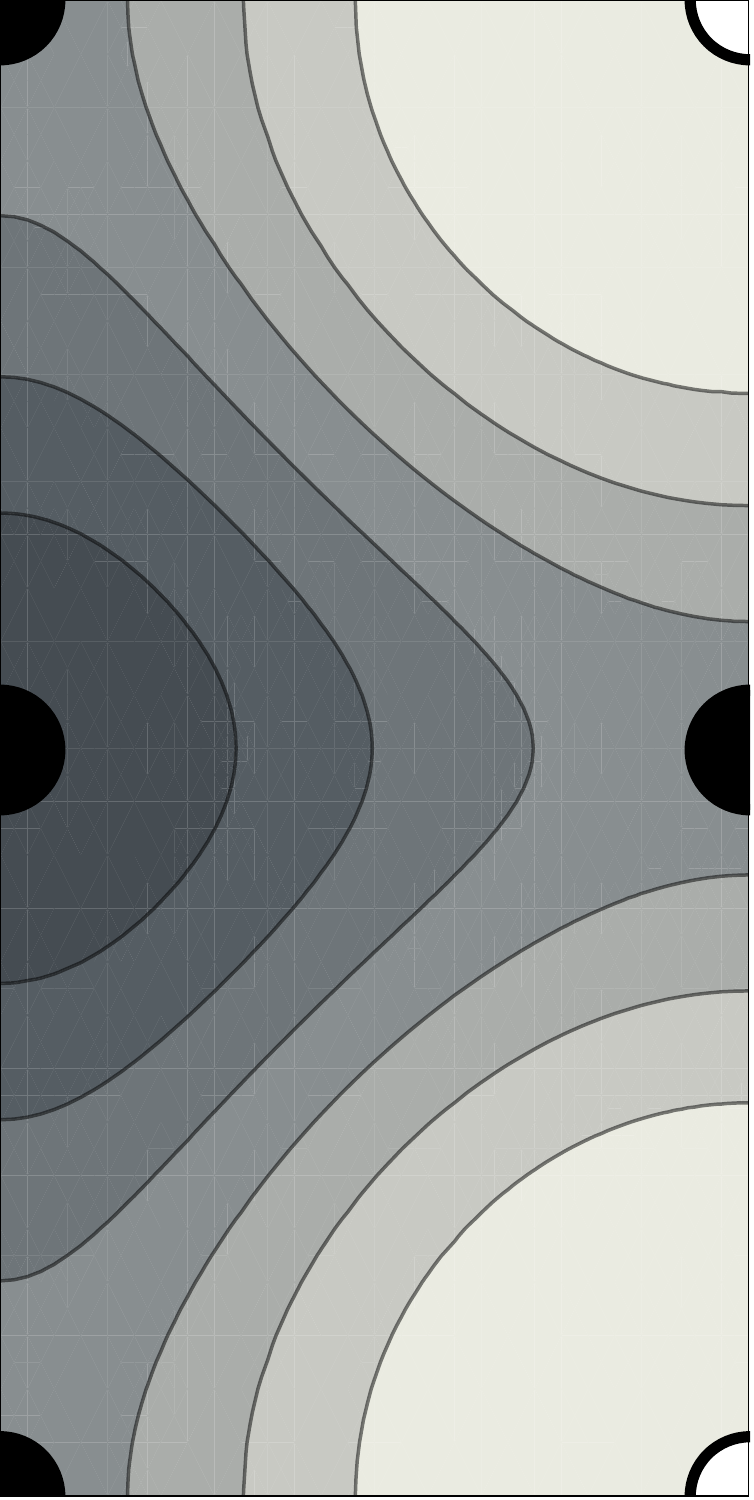} } \hspace*{0.1cm}
	\subfloat{\includegraphics[height = .25 \textheight]{./wavefunctions_m2_legend.pdf}}
	\caption{Untwisted zero-mode functions with two flux quanta at
          a fixed point.}
\label{fig:M2_untwisted}
\end{center}
\end{figure}
Further $M=2$ untwisted zero-mode functions are obtained by localizing
two flux quanta at one fixed point. The corresponding mode functions
read
\begin{align}\label{productWF2}
\chi^2_{\zeta_i} &  \propto e^{qG(z-\zeta_i,\tau)}\, \nonumber\\
& \propto|\vartheta_1(z-\zeta_i,\tau)^2|
e^{-\frac{2\pi}{\tau_2}(\text{Im}(z-\zeta_i))^2}\,, \quad i=1,\ldots,4\,.
\end{align}
Using Eq.~\eqref{product_theta} the product of $\vartheta_1$-functions
can be written as
\begin{align}
\vartheta_1(z-\zeta_i,\tau)^2 =
 \vartheta\!\begin{bmatrix} 0 \\ 0 \end{bmatrix}\! (0, 2\tau)\
&\vartheta\!\begin{bmatrix} 1/2 \\ 0 \end{bmatrix}\! (2z-2\zeta_i+1,
\tau) 
\nonumber\\
&+\vartheta\!\begin{bmatrix} 1/2 \\ 0 \end{bmatrix}\! (0, 2\tau)\
\vartheta\!\begin{bmatrix} 0 \\ 0 \end{bmatrix}\! (2z-2\zeta_i, \tau) \,.
\end{align}
The wave functions $\chi^2_{\zeta_i}$ can
easily be expressed in terms of two $k_1=k_2=0$ mode functions
$\xi^+_j$, see Eq.~\eqref{RzeroM2_1}. For $\chi^2_{\zeta_1}$ and $\chi^2_{\zeta_2}$, for instance, one finds
\begin{equation}\label{M2_untwisted}
\chi^2_{\zeta_{1}} = |a \xi^+_0 - b \xi^+_1|\,,\quad
\chi^2_{\zeta_{2}} = |a \xi^+_0 + b \xi^+_1|\,,
\end{equation}
where $a = \vartheta_1(-1/2,-2\bar{\tau})/\mathcal{N}_0$ and 
 $b = \vartheta(0,-2\bar{\tau})/\mathcal{N}_1$. Analogous expressions
 can be given for $\chi^2_{\zeta_3}$ and $\chi^2_{\zeta_4}$. The wave functions
 $\chi^2_{\zeta_1}$ and $\chi^2_{\zeta_2}$ are shown in
 Fig.~\ref{fig:M2_untwisted}. By construction, they vanish at
 $z=\zeta_1$ and $z=\zeta_2$, respectively.

The construction of untwisted zero-mode functions described above can
be extended to larger values of $M$ in a straightforward way. These
wave functions have characteristic zeros which may be interesting in
physical applications. However, the combinatorics becomes more
involved because of the many possibilities to distribute flux quanta
over the fixed points, and the extension to larger $M$ goes beyond the
scope of this paper.

\subsection{Singular gauge transformations}

The regular and singular gauge fields discussed in Section~\ref{sec:orbifoldGT} yield
the same bulk flux, and the corresponding zero-mode functions
constructed in the previous subsections are closely related. As we
shall now show, these two descriptons of a gauge theory on orbifolds
can indeed be directly mapped into each other by means of a singular
gauge transformation. 
 
The regular gauge field in Section~\ref{subsec:fields_regular} is given by Eq.~\eqref{fluxdom2},
\begin{align}\label{regular}
A^{(r)} = A^{(r)}_1 dy_1 + A^{(r)}_2 dy_2\,,\quad A^{(r)}_1 = -2\pi M
y_2\,, \quad A^{(r)}_2 = 0\,.
\end{align}
Alternatively, the singular gauge field 
\begin{align}
A^{(s)} &= A^{(s)}_z dz + A^{(s)}_{\bar{z}} d\bar{z} = A^{(s)}_1 dy_1 +
A^{(s)}_2 dy_2
\end{align}
is defined by means of the 
Green's function $G(z-\zeta,\tau)$, where $\zeta$ denotes one of the 
orbifold singularities (see \eqref{torusGreen}, \eqref{complexvector}).
With $c=M/q$ and $\zeta = \rho + \tau \eta$, complex and real
components of the singular vector field are given by
\begin{align}
qA^{(s)}_z &= iq \dc G = iM \dc\left(\frac{1}{2} \ln{|\vartheta_1(z-\zeta,\tau)|^2} -
\frac{\pi}{\tau_2}(\text{Im}(z-\zeta))^2 \right)\nonumber\\
&= iM\dc\left(\frac{1}{2} \ln{\vartheta_1(z-\zeta,\tau)} -
\frac{\pi}{\tau_2}(\text{Im}(z-\zeta))^2 \right) =
q{A^{(s)}_{\bar{z}}}^*\,,\\
qA^{(s)}_1 &= q (A^{(s)}_z+A^{(s)}_{\bar{z}})
= -2\pi M(y_2-\eta) + iM\dc_1\ln{\frac{\vartheta_1}{|\vartheta_1|}}\,,\label{singular1}\\
qA^{(s)}_2 &= q (\tau A^{(s)}_z+\bar{\tau} A^{(s)}_{\bar{z}})
= -2\pi M\tau_1 (y_2-\eta) + iM\dc_2\ln{\frac{\vartheta_1}{|\vartheta_1|}}\,.\label{singular2}
\end{align}
Comparing Eq.~\eqref{regular} and Eqs.~\eqref{singular1} and \eqref{singular2} it is clear
that regular and singular gauge fields are related by a singular gauge transformation,
\begin{align}
A^{(r)} = A^{(s)} - \frac{1}{q} d\Lambda\,,
\end{align}
where $\Lambda$ is given by
\begin{align}\label{singularGT}
\Lambda = 2\pi M\eta y_1 - \pi M \tau_1 (y_2 - \eta)^2 + i M
\ln{\frac{\vartheta_1}{|\vartheta_1|}}\,.
\end{align}
The local gauge parameter $\Lambda$ is ill-defined at the singularity
$z=\zeta$ of the Green's function. Away from this point $\Lambda$ is real.
The crucial point of this transformation is that it changes the factor
$|\vartheta_1|^M$ appearing in untwisted wave functions to the
factor ${\vartheta_1^*}^M$ appearing in twisted wave functions,
\begin{align}
\chi \propto |\vartheta_1|^M \quad\longrightarrow\quad \xi = e^{i\Lambda}\chi \propto {\vartheta_1^*}^M\,.
\end{align}
Note that $\vartheta_1^*$ is a function of $\bar{\tau}$,
like the holomorphic part of the zero-mode functions $\xi_j$.

Unwisted wave functions transform trivially under lattice
translations,
\begin{align}
\chi(y_1+1,y_2) = \chi(y_1,y_2)\,,\quad \chi(y_1,y_2+1) =
\chi(y_1,y_2)\,.
\end{align}
The boundary conditions of twisted wave functions are then determined
by the transformation of $\Lambda$ under lattice translations. From
Eqs.~\eqref{singularGT}, \eqref{theta11} and \eqref{theta1tau} one obtains
\begin{equation}
\begin{split}
\Lambda(y_1 + 1,y_2) &= -\pi M (1 - 2\eta) + \Lambda(y_1,y_2)\,,\\
\Lambda(y_1,y_2+1) &=-\pi M (1 + 2\rho - 2y_1) + \Lambda(y_1,y_2)\,.
\end{split}
\end{equation}
This agrees precisely with the twisted boundary conditions
\eqref{twistedbc}, with the identification up to $\text{mod}\ 2$,
\begin{align}
k_1 = M(1 - 2\eta)\,,\quad k_2 = M(1 - 2\rho)\,.
\end{align}
Hence, the location of the singularity at $\zeta = \rho + \tau\eta$
determines the constant Wilson line factors $k_{1,2}$ of the regular
vector field. The generalization of this result to untwisted wave
functions with contributions from different singularities at $\zeta_i = \rho_i + \tau\eta_i$
is obvious. Up to $\text{mod}\ 2$,
$k_1$ and $k_2$ are now given by
\begin{align}
k_1 = \sum_i M_i(1 - 2\eta_i)\,,\quad k_2 = \sum_i M_i (1 - 2\rho_i)\,,
\end{align}
where $-M_i$ are the localized fluxes at the fixed points $\zeta_i$.
This result is indeed consistent with the explicit examples discussed
in the previous section. For $M=1$, $(k_1,k_2)$ is given by $(1,0)$ for $\zeta_2$, $(0,1)$ for $\zeta_3 =
\tau/2$ and $(0,0)$ for $\zeta_4 = (1+\tau)/2$; this agrees with the
list in Eq.~\eqref{SM1}. For $M=2$, with both flux quanta at the same
fixed point, one obviously has $k_1 = k_2 = 0$, which is consistent
with Eq.~\eqref{M2_untwisted}. Finally, for $M=2$, with flux
quanta localized at different fixed points $(\zeta,\zeta')$, $(k_1,k_2)$ is given by 
$(1,0)$ for $(\zeta_2,\zeta_4)$,  $(0,1)$ for $(\zeta_3,\zeta_4)$ and $(1,1)$ for $(\zeta_2,\zeta_3)$.
This is in agreement with Eq.~\eqref{twozeta}. 

Note that the above procedure, mapping singular to regular gauge
fields, is restricted to the bulk, excluding the orbifold fixed
points. It is tempting to conjecture that integer localized fluxes
correspond to localised fermion zero-modes. In this way not only the
$\text{mod} \, 2$ parity but the entire localized flux would be a
physical quantity. It would then influence the fields localized at the
orbifold singularities without modifying the bulk content.
However, despite being of general interest, this question goes beyond the
scope of our investigations.

\section{Summary and Outlook}
\label{sec:Conclusion} 

We have studied in detail $U(1)$ gauge fields on the orbifold
$T^2/\mathbb{Z}_2$. One of the main goals has been to clarify the
quantization condition for magnetic flux. Contrary to the
naive expectation $qF = 2\pi M$, $M\in \mathbb{Z}$, we showed that also
flux values $qF = \pi + 2\pi M$, $M\in \mathbb{Z}$, are allowed,
confirming results in \cite{Abe:2008fi,Abe:2013bca}. This is an effect of the orbifold
fixed points $\zeta_i$. They can have non-trivial Wilson lines $W_i$  around
them, which can be interpreted as localized flux, $qF_i = \pi
(\delta_{(W_i,-1)} + 2 k_i)$, $k_i\in \mathbb{Z}$. The total flux of
bulk and fixed points then satisfies the standard quantization
condition $q(F + \sum_i F_i) \in 2\pi \mathbb{Z}$. To obtain these
results it is crucial to treat the flux background as a vector bundle
on the orbifold.

Localized flux can be used to construct normalized zero-modes of
charged bulk fields \cite{Lee:2003mc}. We used this method to
systematically construct zero-mode wave functions for different flux
densities. The background gauge field is now
singular. It is obtained from torus Green's functions whose
singularities are located at orbifold singularities. The localized
flux densities can vary and are related to the bulk flux density.
The zero-mode wave functions vanish at the fixed points where flux is
localized. Since the Green's function is invariant under lattice
translations and reflection at the origin, the corresponding
untwisted wave functions satisfy trivial boundary conditions.
For comparison, we also recalled the construction of the standard
twisted wave functions for regular background fields. These fields
are not invariant under lattice tranlations and have non-trivial
transition functions. Hence, the corresponding wave functions 
satisfy twisted boundary conditions.

For small values of magnetic flux we showed that there is a one-to-one
correspondence between twisted and untwisted zero-mode
functions, and it is a matter of convenience which basis to use. 
It is satisfactory to see explicitly how untwisted wave functions can
be mapped to twisted wave functions by means of singular gauge transformations.
An advantage of the untwisted wave functions is the geometric origin of the
wave function zeros, which may be phenomenologically interesting.
It appears straightforward to extend the construction of untwisted
wave functions to large magnetic flux as well as to other orbifolds.

Magnetized orbifolds play an important role in compactifications of
type-I string theories. It appears interesting to analyze the role of
localized flux in these constructions and to obtain a better
understanding of the relation to field theory compactifications. This
may be particularly valuable in view of the challenging problem of
supersymmetry breaking.

\section*{Acknowledgments}
We thank Emilian Dudas, Kantaro Ohmori, Makoto Sakamoto, Volker
Schomerus and Joerg Teschner for helpful discussions.
This work was supported by the German Science Foundation (DFG) within
the Collaborative Research Center (SFB) 676 ``Particles, Strings and
the Early Universe''. M.D.'s work is part of the D-ITP consortium, a
program of the Netherlands Organisation for Scientific Research (NWO)
that is funded by the Dutch Ministry of Education, Culture and Science (OCW).
Y.T. is supported in part by Grants-in-Aid for JSPS Overseas Research
Fellow (No.~18J60383) from the 
Ministry of Education, Culture, Sports, Science and Technology in Japan.
}

\appendix

\section{Jacobi theta-functions}

For convenience we list a number of relations among Jacobi
theta-functions which were used in calculations presented in
the previous sections. We follow the conventions of
\cite{polchinski1998string}.

The basic theta-function is given by ($n\in \mathbb{Z}$)
\begin{equation}
\vartheta(z,\tau) = 
\sum_n e^{\pi i \tau n^2 + 2 \pi i n z}\,.
\end{equation}
A useful extension is the theta-function with characteristics,
\begin{equation}
\vartheta\!\begin{bmatrix} \alpha \\ \beta \end{bmatrix}\! (z, \tau) = 
\sum_n e^{\pi i \tau (n + \alpha)^2} e^{2 \pi i  (n + \alpha) (z + \beta) }\,.
\end{equation}
It satisfies the relation
\begin{equation}
\vartheta\!\begin{bmatrix} \alpha \\ \beta \end{bmatrix}\! (z, \tau) = 
\vartheta\!\begin{bmatrix} \alpha \\ 0 \end{bmatrix}\! (z+\beta, \tau) \,
\end{equation}
and is related to the basic theta function by
\begin{equation}
\vartheta\!\begin{bmatrix} \alpha \\ \beta \end{bmatrix}\! (z, \tau) = 
e^{i \pi \tau \alpha^2 + 2 \pi i  \alpha (z + \beta) } \vartheta(z,\tau)\,.
\end{equation}
The theta-function with characteristics 
includes as special cases
\begin{align}
\vartheta(z,\tau) &=\vartheta\!\begin{bmatrix} 0 \\ 0 \end{bmatrix}\! (z,
\tau) =\vartheta\!\begin{bmatrix} 1 \\ 0 \end{bmatrix}\! (z,\tau) \,,\\
-\vartheta_1(z,\tau) &= \vartheta\!\begin{bmatrix} 1/2 \\ 1/2 \end{bmatrix}\! (z, \tau) \,.
\end{align}
An important ``addition formula'' is given by \cite{mumford,Cremades:2004wa}
\begin{equation}\label{addition}
\begin{split}
\vartheta\!\begin{bmatrix} \alpha \\ 0 \end{bmatrix}\! (z_1, M\tau)
&\vartheta\!\begin{bmatrix} \beta \\ 0 \end{bmatrix}\! (z_2, N\tau)=\\
\sum_{m=0}^{M+N-1}&\vartheta\!\begin{bmatrix} (M\alpha +N\beta +mM)/(M+N) \\ 0 \end{bmatrix}\! (z_1+z_2, (M+N)\tau)\\
\times&\vartheta\!\begin{bmatrix}(\alpha-\beta + m)/(M+N) \\ 0 \end{bmatrix}\! (Nz_1-Mz_2, MN(M+N)\tau)\,.
\end{split}
\end{equation}
Further relations read
\begin{align}
\vartheta(z+1,\tau) &= \vartheta(z,\tau)\,,\\
\vartheta(z+\tau,\tau) &= e^{-i\pi\tau -2\pi iz}\vartheta(z,\tau)\,,\\
\vartheta(z,\tau) &= \vartheta(-z,\tau)\,,\\
-\vartheta_1(z,\tau) &= e^{i\pi\tau/4 + i\pi(z+1
/2)} \vartheta(z +(\tau+1)/2,\tau)\,, \\
\vartheta((1+\tau)/2,\tau) &= 0 \,,\\
\vartheta_1(z+1,\tau) &= e^{i\pi} \vartheta_1(z ,\tau) \,,\label{theta11}\\
\vartheta_1(z+\tau,\tau) &= e^{i\pi(1-\tau - 2z)} \vartheta_1(z,\tau) \,,\label{theta1tau}\\
-\vartheta_1(z,\tau)^* &= e^{-i\pi\bar{\tau}/4 - i\pi(\bar{z}+1
/2)} \vartheta(\bar{z} +(\bar{\tau}+1)/2,-\bar{\tau}) \,,
%\label{comcon}
\end{align}

\section{Gamma-matrices}

For completeness, and in order to avoid confusion, we list our
conventions for the 6d gamma-matrices in the following. We start from
the Wess-Bagger conventions in four dimensions \cite{Wess:1992cp},
\begin{equation}
\begin{split}
%\text{diag}(\eta_{\mu\nu}) &= (-1,1,1,1)\,,\quad
\{\gamma_\mu,\gamma_\nu\} &= -2\eta_{\mu\nu}\,,\quad
\eta_{\mu\nu} = \text{diag}(-1,1,1,1)\,,\\
\gamma_5 &= -i\gamma^0\gamma^1\gamma^2\gamma^3\,,\quad
\gamma_5 \psi_L = - \psi_L\,, \quad \gamma_5 \psi_R = \psi_R \,.
\end{split}
\end{equation} 
This is extended to six dimensions using
$\eta_{MN} = \text{diag} (-1,1,1,1,1,1)$ and
\begin{align}
\{\Gamma_M,\Gamma_N\} = -2\eta_{MN} \,.
\end{align}
Explicitly, we use the representation
\begin{align}
\Gamma^\mu = \left(\begin{array}{cc} \gamma^\mu & 0 \\ 0 &
    \gamma^\mu \end{array}\right)\,,\quad
\Gamma^5 = \left(\begin{array}{cc} 0 & i\gamma^5  \\ 
    i\gamma^5 & 0 \end{array}\right)\,,\quad
\Gamma^6 = \left(\begin{array}{cc} 0 & -\gamma^5 \\ 
    \gamma^5 & 0 \end{array}\right)\,,
\end{align}
which implies 
\begin{align}
\Gamma_7 =  -\Gamma^0\Gamma^1\Gamma^2\Gamma^3\Gamma^5\Gamma^6 
= \left(\begin{array}{cc} \gamma^5 & 0 \\ 0 &
    -\gamma^5 \end{array}\right)\,.
\end{align}

% #################################
% #           Bibliography        #
% #################################
\providecommand{\href}[2]{#2}\begingroup\raggedright\endgroup

\end{document}